\documentclass[12pt]{JHEP3}
\usepackage{multicol}
\usepackage[dvips]{epsfig,graphics}
\usepackage{graphicx}

 \newlength{\wth}
\newcommand{\twographst}[2]{%
 \unitlength=1.1in
 \begin{picture}(5.8,2.6)(-0.15,0)
 \put(0,0){\epsfig{file=#1, width=\wth}}
 \put(2.7,0){\epsfig{file=#2, width=\wth}}
 \end{picture}
}

\newcommand{\fourgraphs}[4]{%
\unitlength=1in
\begin{picture}(5.8,5.2)(-0.15,0)
\put(0,0){\epsfig{file=#3, width=\wth}}
\put(2.7,0){\epsfig{file=#4, width=\wth}}
\put(0,2.6){\epsfig{file=#1, width=\wth}}
\put(2.7,2.6){\epsfig{file=#2, width=\wth}}
\end{picture}}

\newcommand{\sixgraphs}[6]{%
\unitlength=1in
\begin{picture}(8.7,7.1)(-0.5,0)
\put(-0.5,4.8){\epsfig{file=#1, width=\wth}}
\put(2.7,4.8){\epsfig{file=#2, width=\wth}}
\put(-0.5,2.3){\epsfig{file=#3, width=\wth}}
\put(2.7,2.3){\epsfig{file=#4, width=\wth}}
\put(-0.5,-0.2){\epsfig{file=#5, width=\wth}}
\put(2.7,-0.2){\epsfig{file=#6, width=\wth}}
\end{picture}}

\newcommand\msq{{m_{\tilde{q}}}}
\newcommand\mlL{{m_{\tilde{e}_L}}}
\newcommand\mlR{{m_{\tilde{e}_R}}}
\newcommand\mnA{{m_{\tilde{\chi}^0_1}}}
\newcommand\mnB{{m_{\tilde{\chi}^0_2}}}
\newcommand\mnC{{m_{\tilde{\chi}^0_3}}}
\newcommand\mnD{{m_{\tilde{\chi}^0_4}}}

\title{Determining SUSY model parameters and masses at the LHC using
cross-sections, kinematic edges and other observables.}

\author{Christopher G. Lester, Michael A. Parker and Martin J. White\\ 
Cavendish Laboratory. Madingley Road. Cambridge CB3 0HE, UK
}

\abstract{We address the problem of mass measurements of
supersymmetric particles at the Large Hadron Collider, using the ATLAS
detector as an example. By using Markov Chain sampling techniques to
combine standard measurements of kinematic edges in the invariant mass
distributions of decay products with a measurement of a missing $p_T$
cross-section, we show that the precision of mass measurements at the
LHC can be dramatically improved, even when we do not assume that we
have measured the kinematic endpoints precisely, or that we have
identified exactly which particles are involved in the decay chain
causing the endpoints. The generality of the technique is demonstrated
in a preliminary investigation of a non-universal SUGRA model, in
which we relax the requirements of mSUGRA by breaking the degeneracy
of the GUT scale gaugino masses. The model studied is compatible with
the WMAP limits on dark matter relic density.}


\preprint{CAV-HEP-2005-15\\ATL-PHYS-PUB-2005-013\\ATL-COM-PHYS-2005-033}

\begin{document}

\section{Introduction}
\subsection{Background and motivation}
The recent data from the WMAP satellite\cite{Spergel:2003cb} have
allowed the matter density of the Universe to be quantified with
greater precision than ever before, whilst also strongly disfavouring
warm dark matter. With the relative\footnote{That is the density
divided by the universe's critical density} matter density $\Omega_m$
constrained by the measurement $\Omega_m h^2 = 0.135^{+0.008}_{-0.009}
$ and the relative baryon density $\Omega_b$ constrained by $\Omega_b
h^2 = 0.0224 \pm 0.0009$, one can infer the following 2-$\sigma$
contraint on the relative density of cold dark matter: $\Omega_{CDM}
h^2 = 0.1126^{+0.0161}_{-0.0181}$, where the reduced Hubble constant
$h$ is measured to be $0.73 \pm 0.03$.

In R-parity conserving supersymmetric (SUSY) models, the lightest
supersymmetric particle (LSP) is stable and is therefore an ideal
candidate for non-baryonic cold dark matter. Past studies in the
context of the minimal supergravity (mSUGRA) model have identified
regions of the five dimensional mSUGRA parameter space in which the
relic density of the LSP (usually the lightest neutralino
$\tilde{\chi}^0_1$) is consistent with dark matter
constraints\cite{Battaglia:2001zp}, and recent studies carried out
post-WMAP have narrowed these regions
further\cite{Battaglia:2003ab}. There has been much recent interest in
examining the phenomenology of SUSY models that are consistent with
the WMAP results in preparation for the arrival of the LHC.
 
The aim of this paper is to use the study of one such model to
demonstrate a new approach to mass measurements at the LHC. In present
analyses, inclusive signatures are rarely used to constrain SUSY
models, despite the fact that they are straightforward to define and
measure at the LHC. This is almost certainly due to the difficulty
associated with calculating the expected values of these signatures at
many points in parameter space, a process that requires a large amount
of computing power. Nevertheless, we demonstrate that inclusive
signatures contain a great deal of information, using as an example
the cross-section of events with missing $p_T$ greater than 500 GeV.

The standard technique for analysis of mSUGRA models is to look for
kinematic endpoints, and use these to measure the masses of particles
involved in cascade decays. These can then be used to obtain the
values of the GUT scale mSUGRA parameters. The problem, however, is
that such an analysis is often loaded with assumptions. Although
endpoint measurements are in principle model independent, it is
usually assumed that one has correctly identified the particles in the
decay chain, giving unreasonably good precision on the measured
masses. Furthermore, it is inevitable that models more general than
mSUGRA will be able to reproduce the endpoints seen in cascade decays,
and hence it is important to develop techniques that allow one to
investigate other possibilities.

Our approach is to combine endpoint measurements with inclusive
signatures through the use of Markov chain sampling techniques, a
method that can in principle be applied to any parameter space, with
any information we happen to have obtained experimentally. The
advantage of Markov chain techniques is their efficiency; a crude scan
of 100 points per axis in a 3 dimensional parameter space would
require one million points, whereas obtaining the same useful
information with our choice of sampling algorithm required only 15,000
points. Even so, in order to evaluate inclusive signatures at many
points in the parameter space within a sensible period of time, it was
necessary to develop an MPI adaptation of the {\tt HERWIG 6.5} Monte
Carlo event generator
\cite{Corcella:2000bw,Corcella:2002jc,Moretti:2002eu} for use on a
supercomputer with parallel processing.

Throughout this paper, we use a particular mSUGRA model as a
description of nature, but it is important to realise that we could in
principle have chosen any SUSY model that fits with current
observations; the techniques described here rely only on the fact that
we have observed endpoints in invariant mass distributions and are
able to measure other well-defined observables. Indeed, given enough
inclusive observables, one would not even need to have observed
endpoints in order to obtain precise results.
  
Section~\ref{sec:lala2} demonstrates the successful application of
kinematic edge analysis to the chosen mSUGRA point before
section~\ref{sec:lala3} reviews Metropolis sampling and applies the
technique to the reconstruction of the masses involved in a squark
decay chain. This differs from current techniques only in the choice
of the method used to fit the masses, as it is assumed in sections 2
and 3 that we have correctly identified the particles in the decay
chain. In section~\ref{sec:lala4}, we introduce a method by which we
can combine the endpoint data with a cross-section measurement in
order to tighten the precision on the masses, using the sampling
techniques reviewed in section~\ref{sec:lala3}. For the sake of
clarity, this is introduced in the familiar context of an mSUGRA
analysis where it is assumed that the particles in the decay chain
have been identified correctly, and we merely wish to fit the
endpoints and obtain masses and mSUGRA parameters.

Finally, in section~\ref{sec:lala5} we admit that we do not know which
particles are in the decay chain, and we also start to relax the
conditions of the mSUGRA model by having non-universal gaugino masses
at the GUT scale. These are both powerful extensions of the current
analysis, and as far as the authors are aware have only rarely been
looked at before (e.g.\ \cite{Lester,Gjelsten:2004ki} for
consideration of particle ambiguity). We also investigate the effect
of a jet energy scale error, in order to demonstrate how one might
include systematic experimental effects in our technique.

The method developed in sections~\ref{sec:lala4} and \ref{sec:lala5}
can easily be generalised to include other inclusive signatures, and
to explore larger parameter spaces, and it can be used in future as a
basis for obtaining precise measurements in general SUSY models.

\subsection{Definition of model}
This paper describes an analysis carried out on a point consistent
with the WMAP data, described by the
following set of mSUGRA parameters:
\label{sec:wmappointdefined}

\begin{center} $m_0=70$ GeV, $m_{1/2}=350$ GeV \

 tan$\beta = 10$, $A_0=0$, $\mu > 0$\
\end{center}
The values of the universal scalar and gaugino masses at the GUT-scale
(respectively $m_0$ and $m_{1/2}$) are chosen such that the point lies
in the coannihilation region in which the LSP's annihilate with
sleptons, thus reducing the LSP relic density to a value within the
range consistent with WMAP. Henceforth we will refer to this model as
the `coannihilation point'.

\DOUBLETABLE[t]
{\begin{tabular}{|c|c|}
\hline
Particle&Mass (GeV)\\
\hline
$\tilde{\chi}_1^0$&137\\
$\tilde{\chi}_2^0$&264\\
$\tilde{e}_L$&255\\
$\tilde{e}_R$&154\\
$\tilde{g}$&832\\
$\tilde{u}_L$&760\\
$\tilde{u}_R$&735\\
$\tilde{d}_L$&764\\
$\tilde{d}_R$&733\\
$\tilde{b}_1$&698\\
$\tilde{b}_2$&723\\
$\tilde{t}_1$&574\\
$\tilde{t}_2$&749\\
$\tilde{\tau}_1$&147\\
$\tilde{\tau}_2$&257\\
$h$&116\\
\hline
\end{tabular}}
{\begin{tabular}{|c|c|}
\hline
Process&Branching Ratio\\
\hline
$\tilde{\chi}_2^0 \rightarrow \tilde{e}_R e$&2\%\\
$\tilde{\chi}_2^0 \rightarrow \tilde{e}_L e$&29\%\\
$\tilde{\chi}_2^0 \rightarrow \tilde{\tau}_1 \tau$&18\%\\
$\tilde{\chi}_2^0 \rightarrow \tilde{\tau}_2 \tau$&2\%\\
$\tilde{\chi}_2^0 \rightarrow \tilde{\chi}_1^0 h$&48\%\\
\hline
Process & Cross-Section \\
\hline
SUSY (Total, {\tt HERWIG}) & 9.3 pb \\
\begin{tabular}{c}
SUSY (After {\tt ATLFAST} \\
missing $p_T > 500$ GeV cut)
\end{tabular} & 2.0 pb \\
\hline 
\end{tabular}
}
{The most important sparticle masses at the coannihilation point.\label{tablemasses}}
{Branching ratios and cross-sections for important processes
at the coannihilation point.\label{branching}}

The masses of the most relevant particles are contained in Table
\ref{tablemasses}, whilst branching ratios for some of the most
significant decay processes are given in Table \ref{branching},
generated using {\tt ISAJET 7.69}.  Cross-sections in
Table~\ref{branching} were calculated with {\tt HERWIG 6.5} and with
fortran {\tt ATLFAST-2.16}. Although similar to the point 5 analysed
in the ATLAS Physics TDR\cite{AtlasTDR}, this particular model differs
by having small mass differences between the $\tilde{\chi}_1^0$ and
the $\tilde{e}_R$ and between the $\tilde{\chi}_2^0$ and the
$\tilde{e}_L$, leading to the production of soft leptons that may be
missed in the detector thereby reducing the efficiency with which we
are able to select relevant SUSY decays.

\section{Kinematic edge analysis}

We begin by demonstrating that standard edge analysis techniques work
(within their limitations) for the chosen coannihiliation point.

\label{sec:lala2}
\subsection{Search for squark decay}
Previous studies (for example \cite{AtlasTDR,Allanach:2000kt,Lester})
have illustrated the procedure of searching for kinematic edges in the
various invariant mass distributions resulting from a given event. By
isolating exclusive decay processes, one can use these kinematic edges
to obtain measurements of the masses of the sparticles that
participate in the decay chain. The procedure is used here in
conjunction with the decay:

\begin{center} $\tilde{q}\rightarrow q\tilde{\chi}^0_2\rightarrow ql^{\pm}_2 \tilde{l}^{\mp}_L \rightarrow ql^{\pm}_2 l^{\mp}_1 \tilde{\chi}^0_1$
\end{center}
This is an excellent starting point for analysis due to the clear
signature provided by the two opposite-sign, same-flavour (OSSF)
leptons. The left-handed slepton is considered here rather than the
right-handed slepton due to the much greater branching ratio
BR($\tilde{\chi}_2^0 \rightarrow \tilde{e}_Le$).  The following
endpoints are expected to be observed in invariant mass spectra
associated with this decay chain
($\tilde{\psi}=m_{\tilde{\chi}^0_2}^2,
\tilde{q}=m_{\tilde{q}}^2,\tilde{l}=m_{\tilde{e}_L}^2,\tilde{\chi}=m_{\tilde{\chi}^0_1}^2)$:

\begin{equation}\label{lledge}
(m_{ll}^2)^{\mbox{\tiny edge}}=\frac{(\tilde{\psi}-\tilde{l})(\tilde{l}-\tilde{\chi})}{\tilde{l}}
\end{equation}

\begin{equation}\label{llqedge}
(m_{llq}^2)^{\mbox{\tiny edge}}=\left\{ \begin{array}{l} \mbox{max}
                            \left [
                            \frac{(\tilde{q}-\tilde{\psi})(\tilde{\psi}-\tilde{\chi})}{\tilde{\psi}},\frac{(\tilde{q}-\tilde{l})(\tilde{l}-\tilde{\chi})}{\tilde{l}},\frac{(\tilde{q}\tilde{l}-\tilde{\psi}\tilde{\chi})(\tilde{\psi}-\tilde{l})}{\tilde{\psi}\tilde{l}}
                            \right] \\ \mbox{except when } \tilde{l}^2
                            < \tilde{q}\tilde{\chi} < \tilde{\psi}^2
                            \mbox{and } \tilde{\psi}^2\tilde{\chi} <
                            \tilde{q}\tilde{l}^2 \\ \mbox{where one
                            must use}
                            (m_{\tilde{q}}-m_{\tilde{\chi}_1^0})^2.
                            \end{array} \right.
\end{equation}

\begin{equation}\label{lqhighedge}
(m_{lq}^2)^{\mbox{\tiny edge}}_{\mbox{\tiny max}}=\mbox{max}
\left[\frac{(\tilde{q}-\tilde{\psi})(\tilde{\psi}-\tilde{l})}{\tilde{\psi}},\frac{(\tilde{q}-\tilde{\psi})(\tilde{l}-\tilde{\chi})}{\tilde{l}}
\right]
\end{equation}

\begin{equation}\label{lqlowedge}
(m_{lq}^2)^{\mbox{\tiny edge}}_{\mbox{\tiny min}}=\mbox{min}
\left[\frac{(\tilde{q}-\tilde{\psi})(\tilde{\psi}-\tilde{l})}{\tilde{\psi}},\frac{(\tilde{q}-\tilde{\psi})(\tilde{l}-\tilde{\chi})}{(2\tilde{l}-\tilde{\chi})}
\right]
\end{equation}

\begin{equation}\label{llqthreshold}
(m_{llq}^2)^{\mbox{\tiny thres}}=\frac{2\tilde{l}(\tilde{q}-\tilde{\psi})(\tilde{\psi}-\tilde{\chi})+(\tilde{q}+\tilde{\psi})(\tilde{\psi}-\tilde{l})(\tilde{l}-\tilde{\chi})-(\tilde{q}-\tilde{\psi})\sqrt{(\tilde{\psi}+\tilde{l})^2(\tilde{l}+\tilde{\chi})^2-16\tilde{\psi}\tilde{l}^2\tilde{\chi}}}{4\tilde{l}\tilde{\psi}}
\end{equation}
where ``min'' and ``max'' refer to minimising and maximising with
respect to the choice of lepton. In addition, ``thres'' refers to the
threshold that appears in the $m_{llq}$ distribution when events are
chosen such that $m_{ll}^{\mbox{\tiny edge}}/\sqrt2 < m_{ll} <
m_{ll}^{\mbox{\tiny edge}}$, corresponding to the angle between the
two lepton momenta exceeding $\pi/2$ in the slepton rest frame (see
\cite{Lester}).

\subsection{Monte Carlo event simulation}
Monte Carlo simulations of SUSY production at the above mass point
have been performed using {\tt HERWIG
6.5}\cite{Corcella:2000bw,Corcella:2002jc,Moretti:2002eu}, with the
particles subsequently passed through the fortran {\tt ATLFAST-2.16} detector
simulation \cite{ATLFAST}. A {\tt HERWIG} input file was generated
using \tt ISAJET v7.69 \rm\cite{Paige:2003mg} in conjunction with the Herwig-Isajet interface {\tt ISAWIG} which converts the {\tt ISAJET} output into {\tt HERWIG} input format. A sample corresponding to $100 \mbox{fb}^{-1}$ has been
generated (being one year of design luminosity in the high luminosity
mode).

\subsection{Invariant mass distributions}
\subsubsection{Cuts}
\label{sec:lala231}
In order to see the above edges clearly, one must apply various cuts
to the event data in order to isolate a clean sample of the squark
decay chain. Here, one can select events with the OSSF lepton
signature described above, and one can also exploit the fact that the
required events have a large amount of missing energy (due to the
departure from the detector of two invisible
$\tilde{\chi}_1^0$'s). Furthermore, one expects to obtain hard jets in
SUSY events, resulting from the decay of gluinos and squarks. All
plots are obtained through the use of the following cuts:

\begin{itemize}
\item
$E_{T}^{\mbox{\tiny miss}} > 300 \mbox{ GeV;}$
\item
exactly two opposite-sign leptons with $p_T > 5 \mbox{ GeV}$ and $|\eta| < 2.5$;
\item
at least two jets with $p_T > 150 \mbox{ GeV}$;
\end{itemize}
Although the cuts chosen are similar to those used for point 5 in the
ATLAS Physics TDR, there are some exceptions. For example, one needs
to impose a $p_T$ cut on the leptons in the event due to the fact that
{\tt ATLFAST} is not parametrised properly for low $p_T$ leptons,
and yet it is essential to pick up soft leptons due to the small mass
differences that crop up in the decay chain. Hence, a compromise
between these two factors must be chosen. Some plots are the result of
additional cuts, and these are given below.

The SM background for dilepton processes is generally negligible once
the missing $p_T$ cut has been applied, though the OSSF lepton
signature can be produced by SUSY processes other than the decay of
the $\tilde{\chi}_2^0$. One would expect these to produce equal
amounts of opposite-sign opposite-flavour (OSOF) leptons and hence one
can often subtract the dilepton background by producing ``flavour
subtracted plots'' in which one plots the combination
$e^+e^-+\mu^+\mu^--e^+\mu^--e^-\mu^+$. This is only useful in cases
where there are sufficient statistics, and was not done for every
plot below.

\subsubsection{$m_{ll}$ plot}

As seen in figure~\ref{mllpic}, a sharp edge is produced in the
spectrum at $\approx$ 58 GeV, and this is a very clear signature. The
second edge visible at $\approx$ 98 GeV results from the right-handed
selectron. In practise, it will be very difficult to assign these two
edges correctly, and this problem is revisited in
section~\ref{sec:lala5}.

\subsubsection{$m_{llq}$ plot}
\label{sec:mllqfirstdefined}
This is produced by selecting events with exactly two leptons, and
forming the invariant mass $m_{llq}$ first with the jet $q_1$ with the
largest transverse momentum, and then with the jet $q_2$ with the
second largest transverse momentum. As the hardest jets in events
containing cascade decays of squarks and gluinos usually come from the
processes $\tilde{q} \rightarrow \tilde{\chi}_2^0q$ and $\tilde{q}
\rightarrow \tilde{\chi}_1^0q$, the lower of the two invariant masses
formed in this way should lie below the llq edge defined by
equation~(\ref{llqedge}), and so $m_{llq}$ is defined by $m_{llq} =
\min(m_{llq_1}, m_{llq_2})$.  Figure~\ref{mllq-pic} shows a clear
endpoint in the $m_{llq}$ distribution at $\approx$ 600 GeV.

\DOUBLEFIGURE[t]{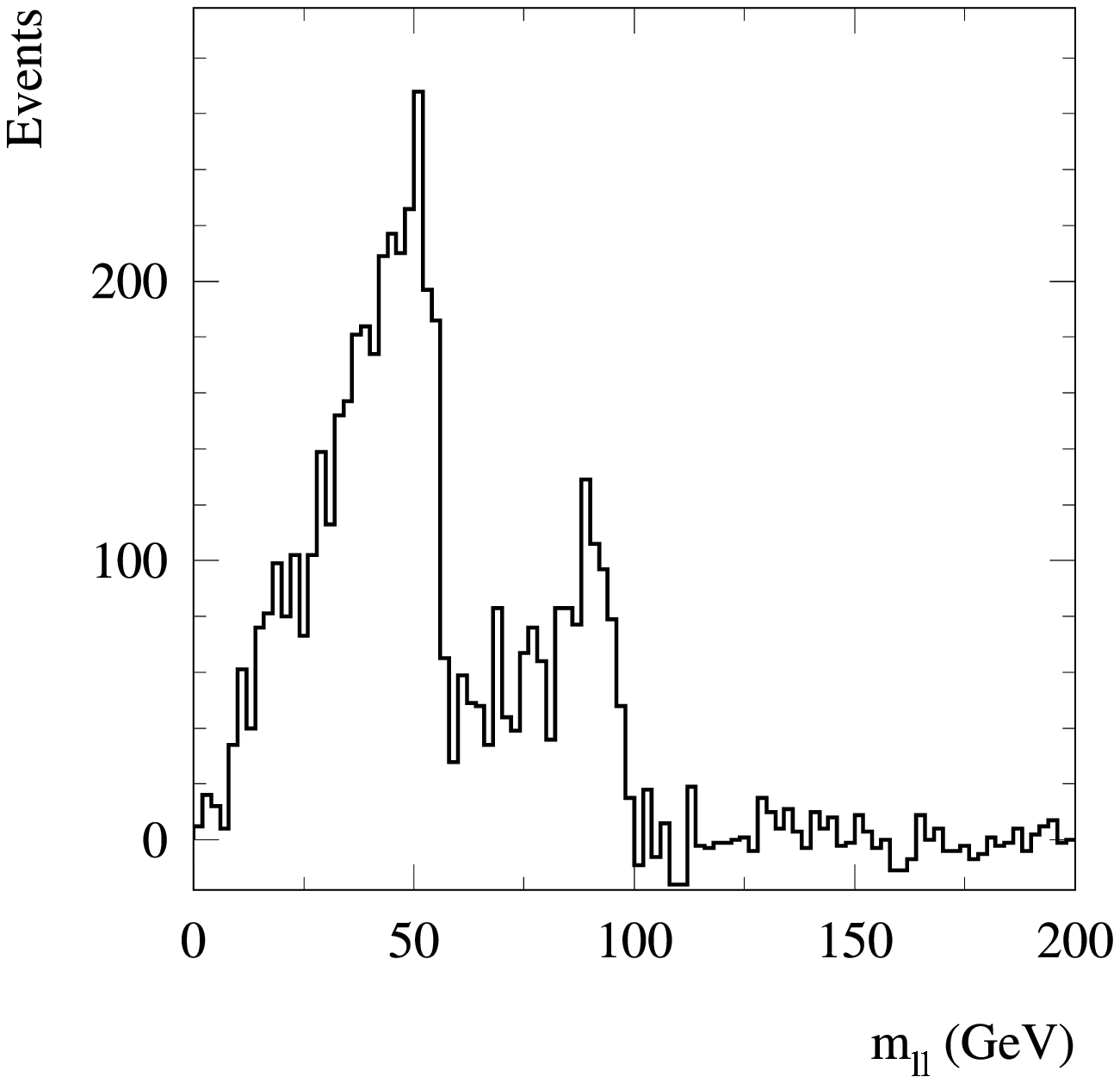,width=2.6in}{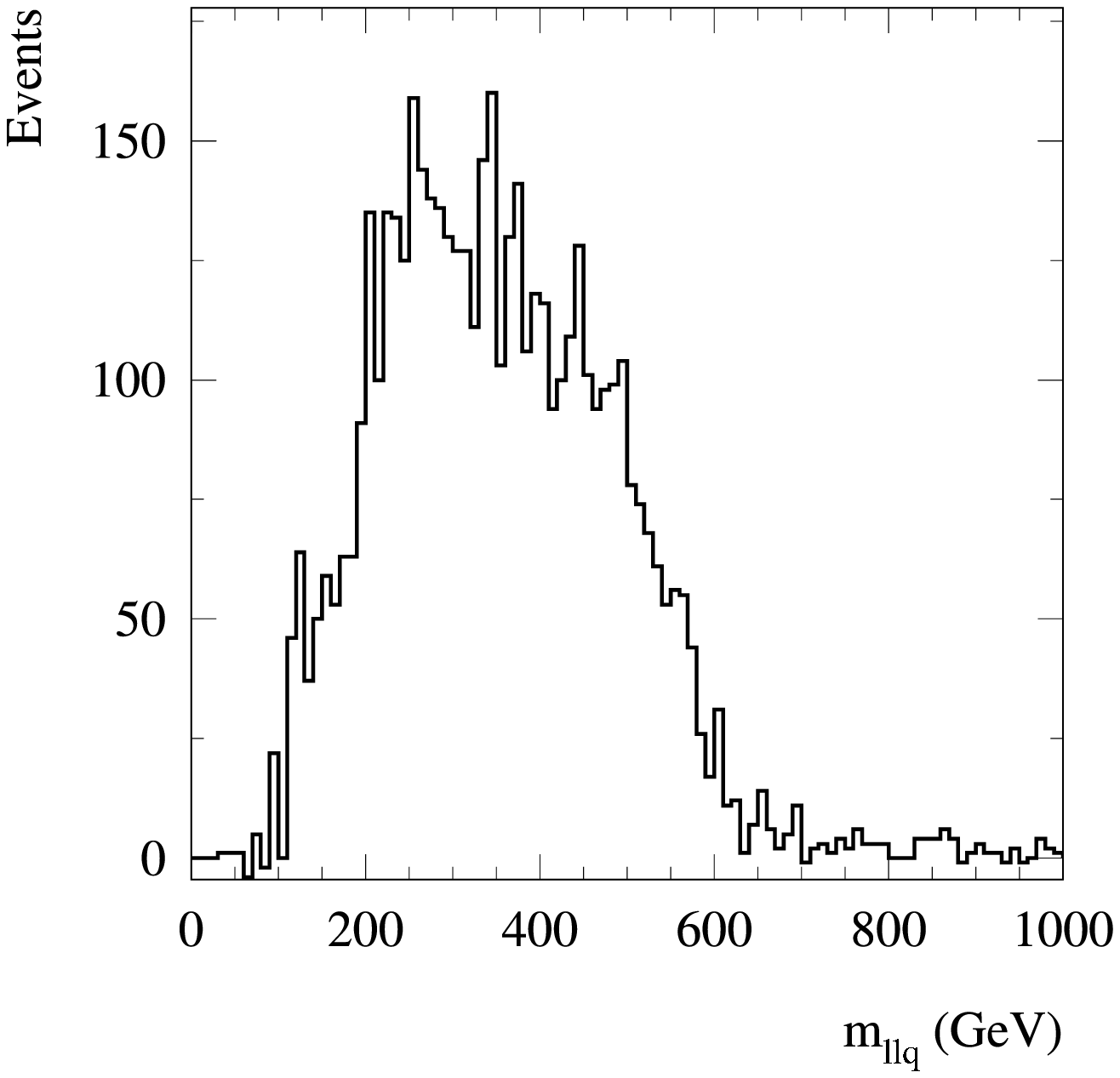,width=2.6in} {The
flavour-subtracted dilepton invariant mass \label{mllpic} plotted with
the cuts described in the text.}{The llq invariant mass
\label{mllq-pic} plot.}

\subsubsection{$m_{llq}^{\mbox{\tiny thres}}$ plot}
The $m_{llq}$ variable plotted in order to measure
$m_{llq}^{\mbox{\tiny thres}}$ is defined almost in the same way as
the $m_{llq}$ variable defined in section~\ref{sec:mllqfirstdefined}.
The two differences are that this time (1) $m_{llq} = \max(m_{llq_1},
m_{llq_2})$ (because a threshold\footnote{The terms ``endpoint'' and
``threshold'' are used to refer the the extremal values of a random
variable or observable at respectively high and low mass values.  The
term ``edge'' describes the shape of the distribution of that variable
near its endpoint or threshold.} is expected rather than an endpoint)
and (2) events must satisfy an additional constraint that $m_{ll}$
must exceed $m_{ll}^{\mbox{\tiny max}}/\sqrt2$.  The resulting
$m_{llq}$ distribution may be seen in figure~\ref{mllq-thres}.  This
plot is not flavour-subtracted.

A threshold is clearly observed a little above 100 GeV, though it is
difficult to state its precise position due to uncertainty in the
expected shape of the edge and the manner in which it is modified by
detector effects. This is discussed further below.
 
\FIGURE{\epsfig{file=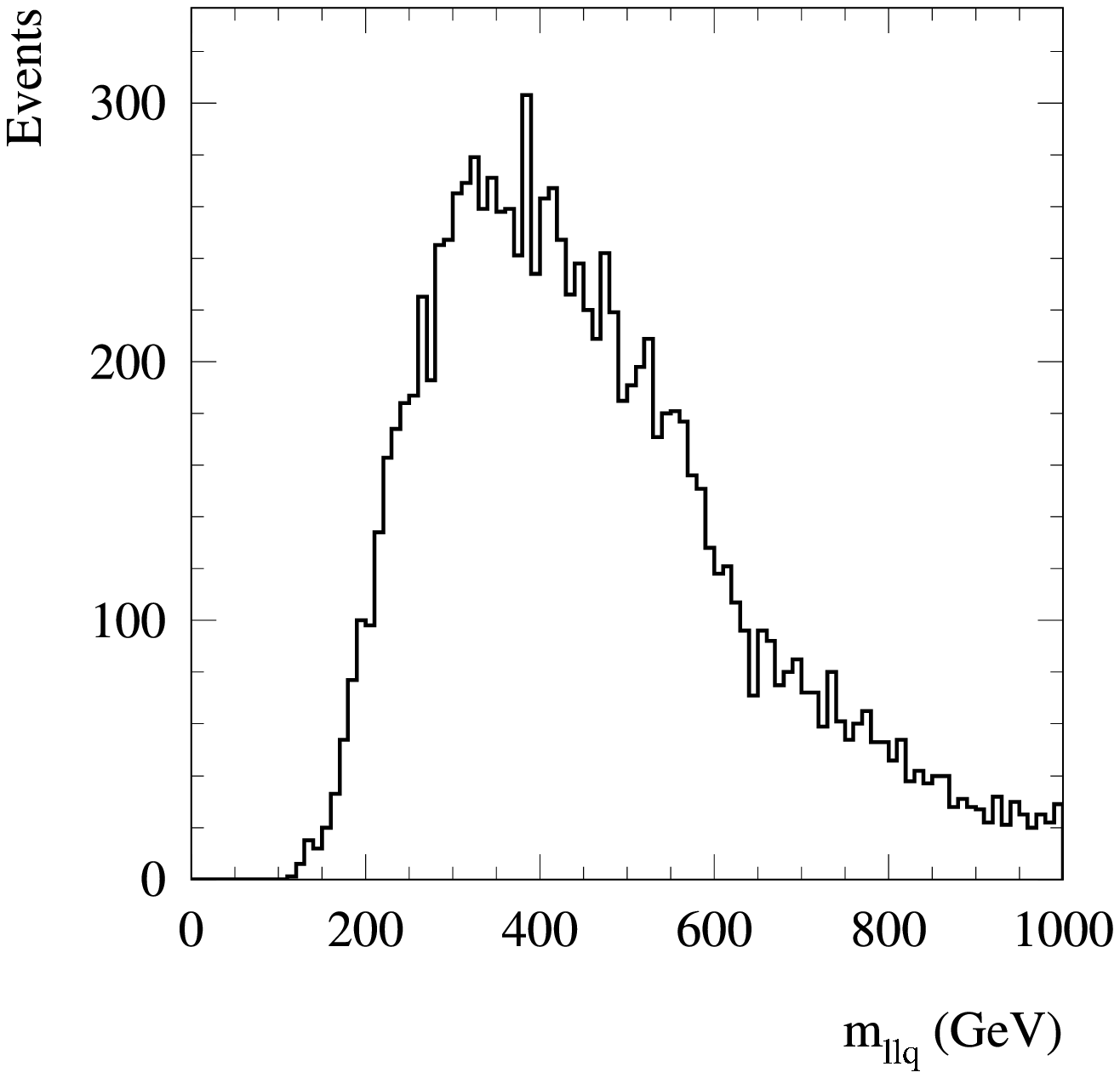,width=2.6in}
        \caption{The llq invariant mass threshold plot.}
	\label{mllq-thres}}

\subsubsection{$m_{lq}^{\mbox{\tiny max}}$ and $m_{lq}^{\mbox{\tiny
      min}}$  plots}
In creating the $m_{lq}^{\mbox{\tiny max}}$ and $m_{lq}^{\mbox{\tiny
      min}}$  plots the following steps are taken.  First, one of the
two hardest jets in the event is selected by the same method used in
section~\ref{sec:mllqfirstdefined}, i.e.~by looking for the
combination yielding the lower value of $m_{llq}$.  Having identified
this jet (call it $q$), the quantities $m_{l_1q}$ and $m_{l_2q}$ are
formed.  The larger of these two combinations $m_{lq}^{\tiny high} =
\max{(m_{l_1q}, m{l_2q})}$ and the lower of them $m_{lq}^{\tiny low} =
\min{(m_{l_1q}, m{l_2q})}$ are identified.  
The distribution of $m_{lq}^{\tiny high}$
 is plotted in figure~\ref{mlqhigh} and the endpoint located
 therein is identified as being
$m_{lq}^{\mbox{\tiny max}}$.  The distribution of $m_{lq}^{\tiny low}$
 is plotted in figure~\ref{mlqlow1} and the endpoint located
 therein is identified as being
$m_{lq}^{\mbox{\tiny min}}$. 

For the $m_{lq}^{\mbox{\tiny max}}$ plot (figure~\ref{mlqhigh}) events
were subject to the additional constraint that one of the llq
invariant masses formed with the two hardest jets must be above the
llq endpoint, and the other must be below.

The $m_{lq}^{\mbox{\tiny min}}$ plot (figure~\ref{mlqlow1}) has one
additional cut: the dilepton invariant mass must be less than the
value of $m_{ll}^{\mbox{\tiny max}}$ observed in figure~\ref{mllpic}.

Both plots exhibit endpoints, and the edge is particularly abrupt in
the $m_{lq}^{\mbox{\tiny max}}$ histogram. Although there are events
beyond the endpoint in the $m_{lq}^{\mbox{\tiny min}}$ plot (due to
SUSY background processes), there is nevertheless a convincing edge at
$\approx$ 180 GeV.

\DOUBLEFIGURE[t]{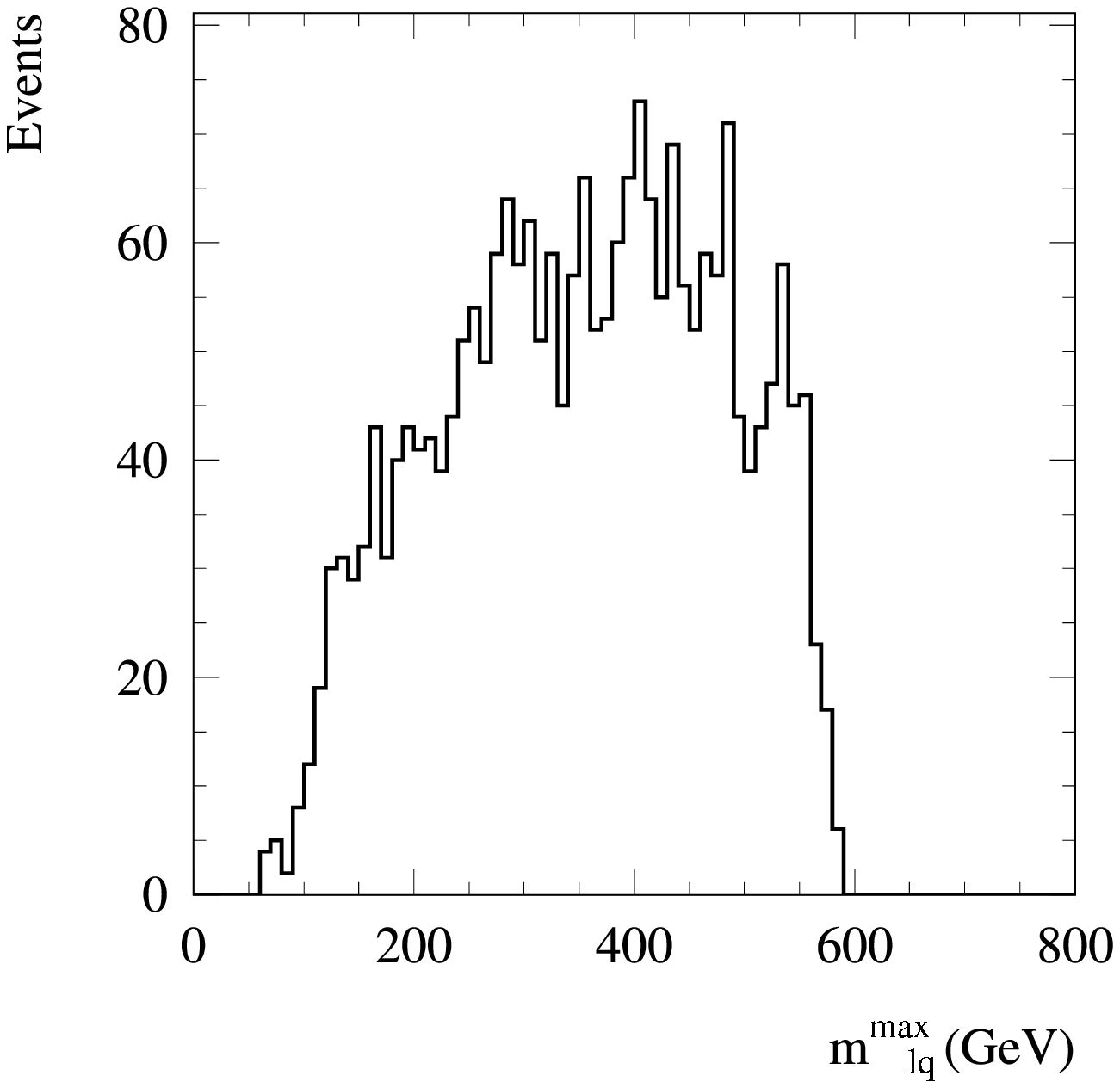,width=2.6in}{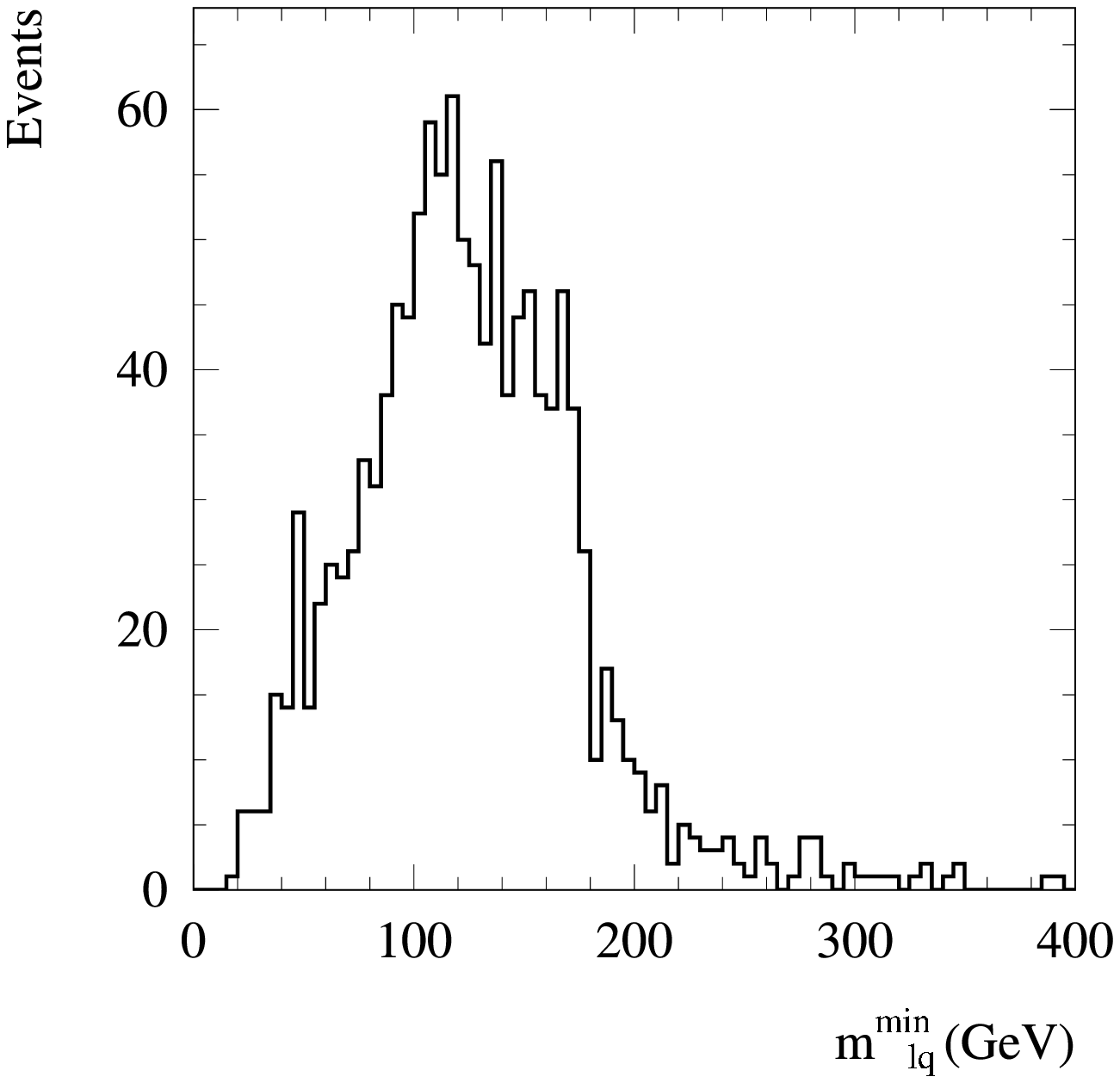,width=2.6in}
{The lq max invariant \label{mlqhigh} mass plot.}{The lq min invariant
mass \label{mlqlow1} threshold plot.}

\subsection{Comparison of observed and predicted edges}
The edges predicted by equations~(\ref{lledge}) to
(\ref{llqthreshold}) are summarised in Table \ref{edgepredictions},
where the spread of in the squark masses has been ignored,
approximating them at a common value of 750 GeV, and all other masses
are taken from Table \ref{tablemasses}. The observed positions of the
endpoints are also given.

\TABLE{
\begin{tabular}{|c|c|c|}
\hline
Edge&Predicted (GeV)&Observed (GeV)\\
\hline
ll edge&57.64&57.5$\pm$2.5\\
llq edge&600.1&600$\pm$10\\
llq threshold&134.0&150$\pm$30\\
lq max edge&592.1&590$\pm$10\\
lq min edge&181.7&180$\pm$10\\
\hline
\end{tabular}
\caption{Predicted and observed edge positions for the mSUGRA mass
point described in the text. Error estimates have been obtained `by
eye', and reflect lack of information regarding the precise shapes of
the endpoints.}
\label{edgepredictions}
}

\label{sec:explainingbyeye}
It is common when extracting the observed edges from plots such as
those above to fit a function to the endpoint in order to determine
both the precision and the accurate position. For the purposes of this
analysis, the edges, and their estimated errors, have been determined
`by eye' for several reasons. Firstly, not all edges can be fitted
with functions (in the case of the llq threshold, for example, the
correct shape is not known). Indeed, recent work in
\cite{Gjelsten:2004ki} highlights the need for caution in applying
these functions too readily without first investigating the
theoretical shape of the distribution, as endpoints can often exhibit
tails or `feet' that will be confused with SUSY background and hence
may lead to inaccurate measurements. The shapes of the endpoints for
distributions involving quarks vary significantly over the parameter
space, introducing a model dependence into the precision with which
one may realistically measure endpoint positions and hence
masses. Given that the purpose of this note is primarily to use an
arbitrary example in order to demonstrate our use of Markov chain
sampling techniques, a full investigation of how to resolve this model
dependence is considered to be beyond the scope of this paper, and we
will use the conservative errors given in table
\ref{edgepredictions}. For those interested, the fitting of endpoint
functions has been done in work leading to \cite{GianlucaTalk} which
contains estimates of the precision expected if one were to take a
more optimistic view.


\section{Mass reconstruction}
\label{sec:lala3}
\subsection{Background}
Having obtained measurements of kinematic edges, the next step is to
attempt to reconstruct the masses involved in the squark cascade
decay. This has been done using a Markov Chain Monte Carlo method,
details of which may be found in appendix A. The technique is an
excellent way of efficiently exploring high dimensional parameter
spaces, and it is in section~\ref{sec:lala4} that the full advantages
of the technique become apparent.
\subsection{Application of Metropolis Algorithm}
We now apply the sampling techniques described in the appendix to our
mass reconstruction problem. The five endpoints observed in the
previous section essentially provide an (over-constrained) set of
simultaneous equations in the four unknowns $m_{\tilde{q}}$,
$m_{\tilde{e}_L}$, $m_{\tilde{\chi}_2^0}$ and $m_{\tilde{\chi}_1^0}$,
and these can be solved to determine the masses. Given a set of
observed edges $\textbf{e}^{obs}$, and a set of postulated masses
\textbf{m}, the ultimate goal is to evaluate
$p(\textbf{m}|\textbf{e}^{obs})$ and thus to find the regions of
parameter space favoured by the data. The best way of doing this is to
sample masses $\textbf{m}$ from $p(\textbf{m}|\textbf{e}^{obs})$,
subsequently histogramming the samples to reveal directly the shape
of the probability distribution.

Using Bayes' Theorem we know that
\begin{eqnarray}
p(\textbf{m}|\textbf{e}^{obs})
\propto p(\textbf{e}^{obs}|\textbf{m})p(\textbf{m}). \label{initbayes}
\end{eqnarray}
We choose the prior $p(\textbf{m})$ to be uniform\footnote{Some points
$\textbf{m}$ in mass space do not satisfy the hierarchy $m_{\tilde{q}}
> m_{\tilde{\chi}_2^0} > m_{\tilde{e}_L} > m_{\tilde{\chi}_1^0} > 0$
required by our decay chain.  Under our model, then, these points
yield $p(\textbf{e}^{obs}|\textbf{m})=0$ and veto the selection of
such points.  While this veto is technically \label{secoftechnopte}
part of the likelihood (given our model) it simplifies later
discussion of the likelihood in more complicated scenarios if we pull
the veto out of the likelihood and move it into the prior
$p(\textbf{m})$.  In practise then, our effective prior is uniform
over all of the region of mass space in which the required hierarchy
is present, and zero elsewhere.  The effect is the same as if we had
left the veto {\em in} the likelihood, but the likelihoods will be
simpler to describe and define.} over the mass space considered.  This
choice seems a good as any other, and has the added benefit that plots
of our posterior distribution $p(\textbf{m}|\textbf{e}^{obs})$ are
also just plots of the likelihood $p(\textbf{e}^{obs}|\textbf{m})$,
permitting the effects of other priors $p(\textbf{m})$ to be easily
imagined.

One can sample from $p(\textbf{m}|\textbf{e}^{obs})$ using the
Metropolis Method as follows.  First a mass point \textbf{m} is
chosen, and $p(\textbf{m}|\textbf{e}^{obs})$ is evaluated using
equation~(\ref{initbayes}).  For the edges $e_1$, $e_2$, $e_3$, $e_4$,
and $e_5$, the likelihood $p(\textbf{e}^{obs}|\textbf{m})$ is given by
the product
\begin{eqnarray}
p(\textbf{e}^{obs}|\textbf{m}) = \prod_{i = 1}^5 {
p(e_i^{obs}|\textbf{m})  }, \label{eq:productything}
\end{eqnarray}
where
\begin{equation}\label{pof}
p(e_i^{obs}|\textbf{m}) \approx \frac{1}{\sqrt{2\pi\sigma^2_i}}\exp\left(-\frac{(e_i^{obs}-e_i^{pred}(\textbf{m}))^2}{2
\sigma^2_i}\right)\label{thingwithedgediffin}
\end{equation}
in which $\sigma_i$ is the statistical and fit error associated with
the edge measurement of edge $e_i$, and where $e_i^{obs}$ and
$e_i^{pred}(\textbf{m})$ are respectively the observed and predicted
positions of the edge.  This probability distribution assigns a weight
$p(\textbf{m}|\textbf{e}^{obs})$ to each point $\textbf{m}$ in mass
space, including the errors associated with the endpoint
measurements. Note that $p(\textbf{m}|\textbf{e}^{obs})$ is the
equivalent of the $P^*(\textbf{x})$ defined later on in equation~(\ref{probdist}), as it is defined only up to an unknown normalisation
constant.

So, in order to plot the probability distribution, one follows the
following steps of the Metropolis Algorithm:\footnote{See
Appendix~\ref{app:metropmethod} for discussion of the motivations
behind each of these steps, and for definitions of ``proposal
functions'' and the decision mechanism.}
\label{sec:stepbystep} 

\begin{enumerate}
\item
A new mass point $\textbf{m}^{proposal}$ is suggested on the basis of
the current point $\textbf{m}^{current}$.  The mass-space proposal
distribution for the Metropolis Algorithm was chosen to be a
4-dimensional Gaussian whose width in each dimension was 5 GeV and
whose centre was the position of the current point
$\textbf{m}^{current}$.  The widths were chosen for the efficiency
reasons outlined in section~\ref{sec:whyQwaswhatitis} and will not
effect the results once convergence has occurred.
\item
$p(\textbf{m}^{proposal}|\textbf{e}^{obs})$ is evaluated at the proposed point.
\item
A decision is made on whether to jump to the new point, or remain at
the current point on the basis (see equation~(\ref{eq:aisdefined})) of
the ratio of $p(\textbf{m}^{proposal}|\textbf{e}^{obs})$ to
$p(\textbf{m}^{current}|\textbf{e}^{obs})$.
\item
If a decision to {\em not jump} is made, then the next point in the
chain $\textbf{m}^{next}$ is again set equal to
$\textbf{m}^{current}$, otherwise it is set equal to
$\textbf{m}^{proposal}$.  When proposals are rejected, therefore,
successive points in the chain are duplicates of each other.
\item
All steps are repeated until the sampler has sufficiently explored the
interesting regions of parameter space.
\end{enumerate}
It is noted that in the real ATLAS detector, one might have a
systematic shift of the endpoints due to the jet energy scale error,
and this is considered in section~\ref{sec:lala5}.
\subsection{Mass space plots}
The Metropolis sampler ensures that points which are more likely are
sampled more often. One can observe the shape of the probability
distribution by simply histogramming the sampled points. This is a 4
dimensional shape in mass space, which can be viewed as a projection
onto each pair of axes. This is done in figure~\ref{n1n2}, revealing
that a lengthy region of parameter space is compatible with the edge
data, and extra information is required to constrain this
further. Note that the endpoint equations discussed previously are
sensitive principally to mass differences, and hence one observes
lines in each plane of the mass space, constraining each mass only in
terms of the others. Given that the endpoint data does not set the
overall mass scale, the system is able to wander to high masses
without affecting the position of the endpoints provided that the mass
differences remain consistent. In the next section, we show that one
can use other measurements from the LHC that are sensitive to the mass
scale to constrain these regions further.

Finally, it is noted that the lines are broader in the plots featuring
squark masses, and this is due to the fact that the end points were
calculated using an average squark mass, whilst the Monte Carlo events
feature a range of squark masses. Hence the resolution is smeared
somewhat relative to the other masses.

\FIGURE{
\sixgraphs{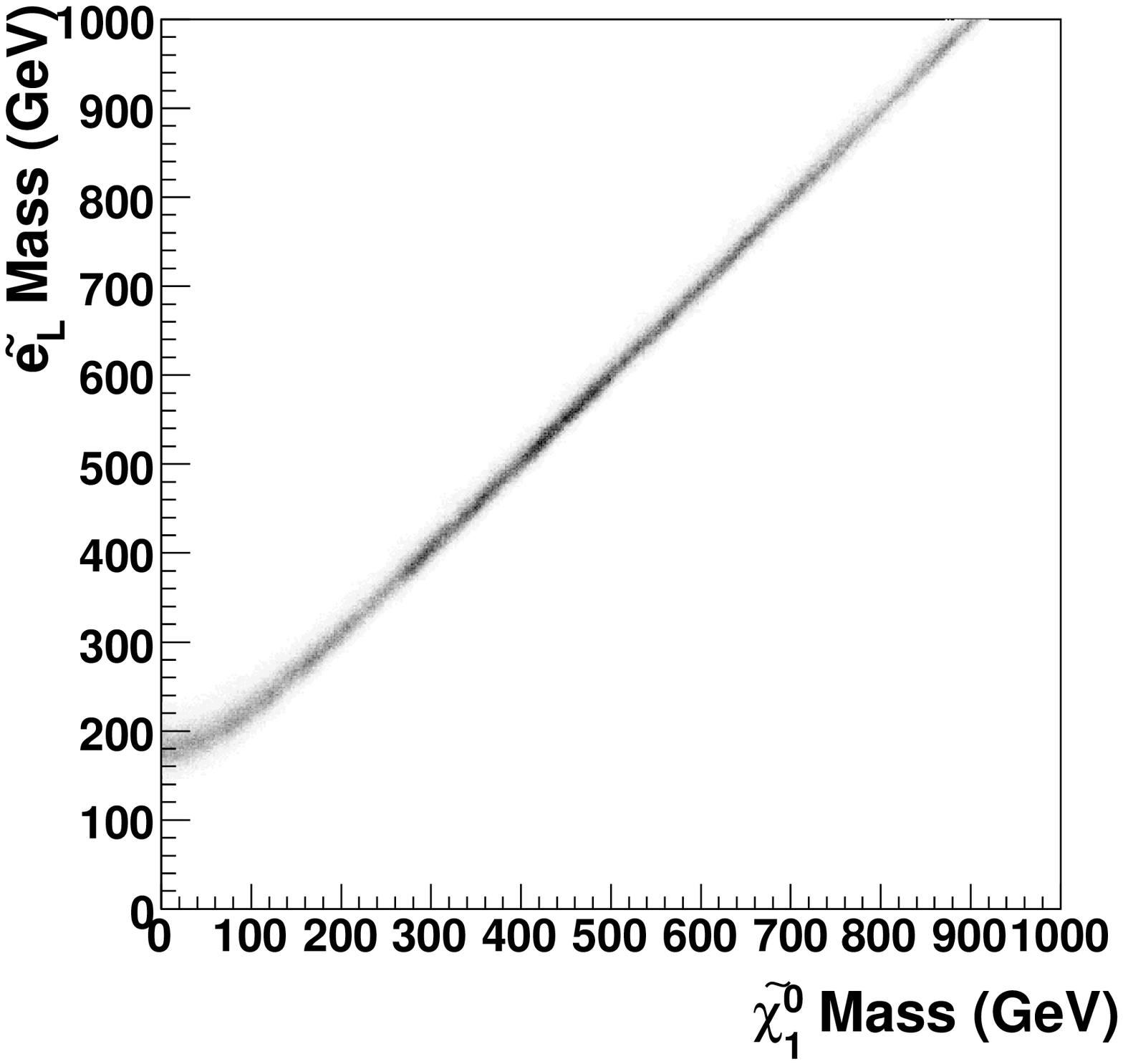}{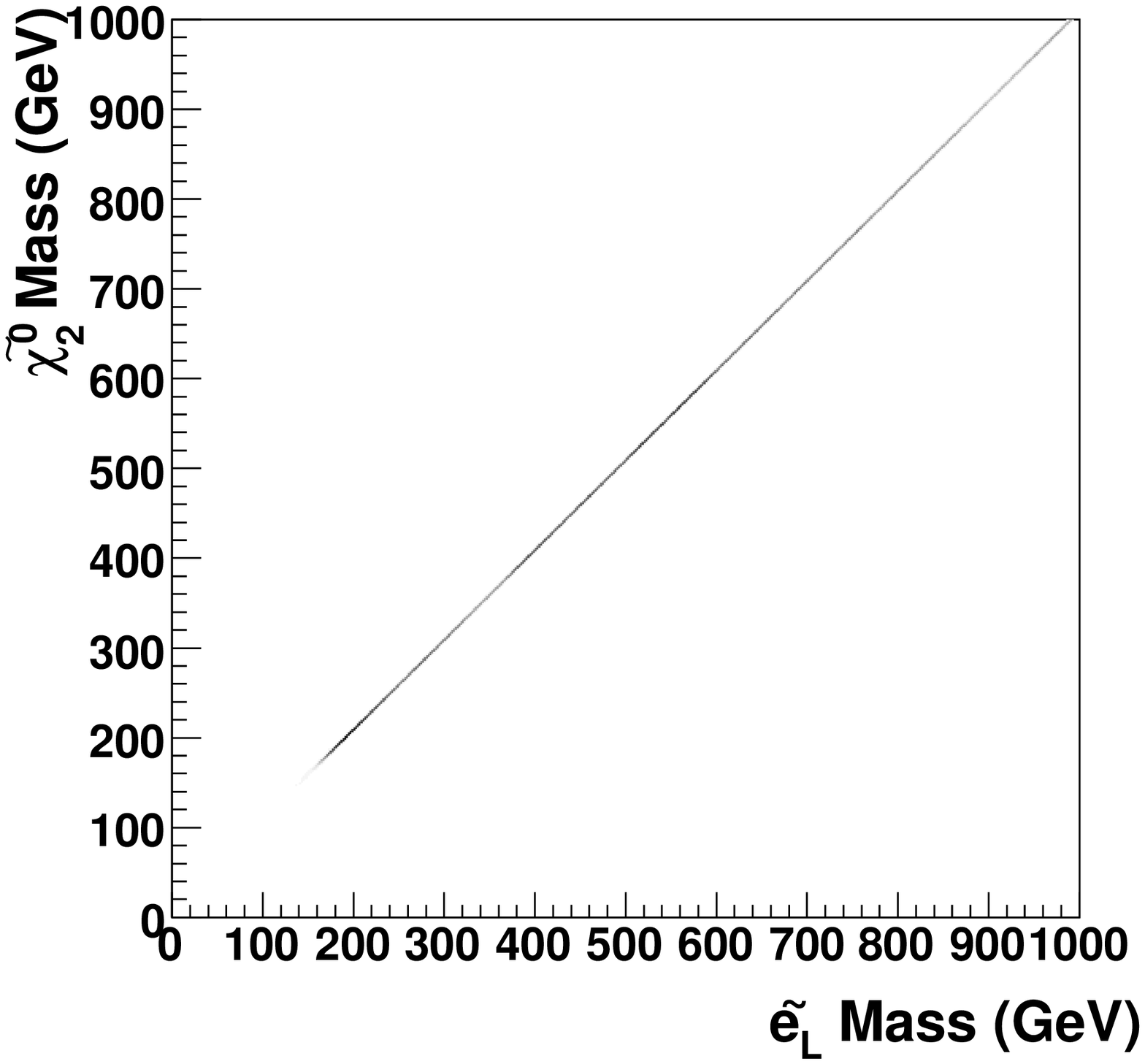}{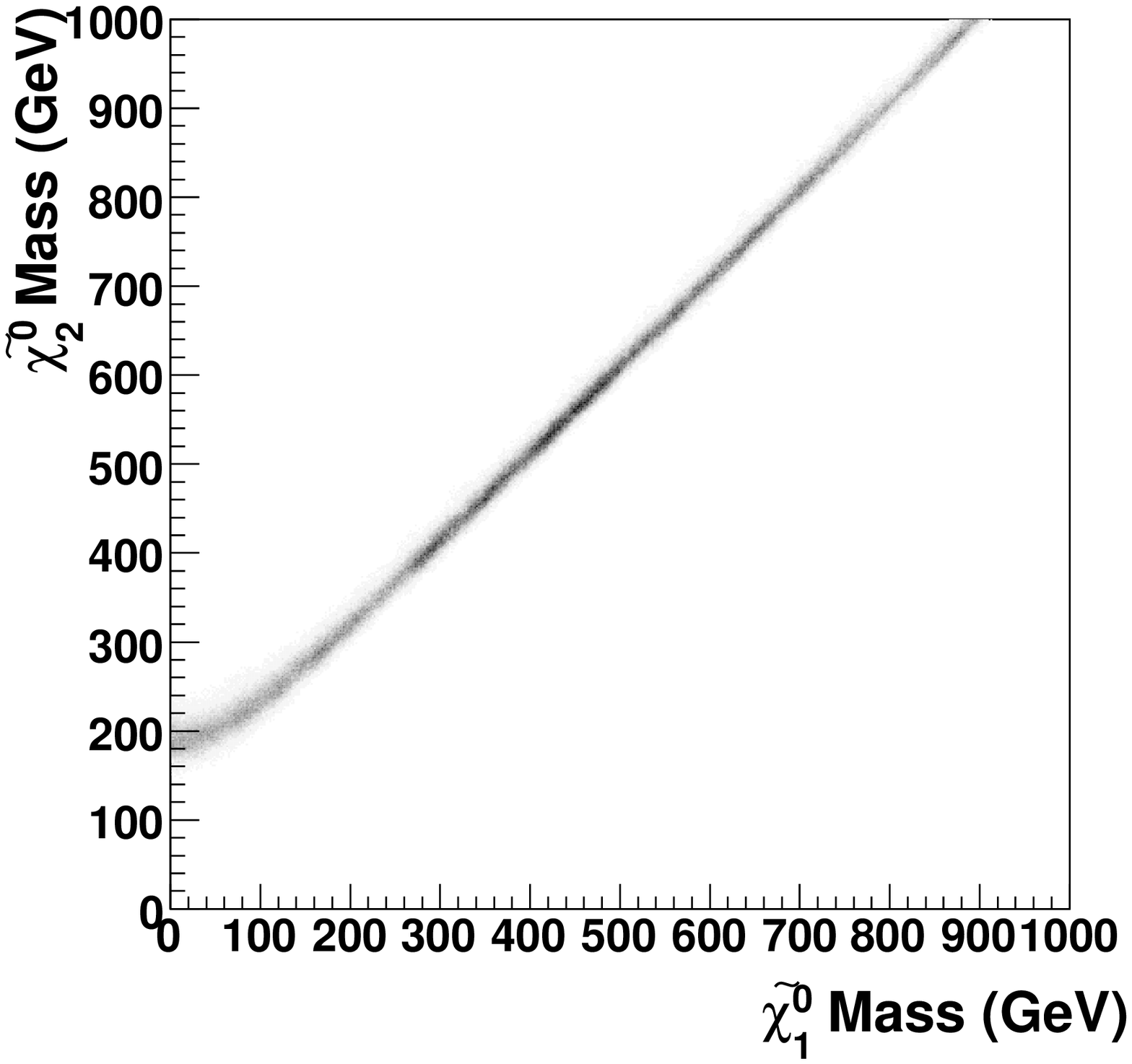}{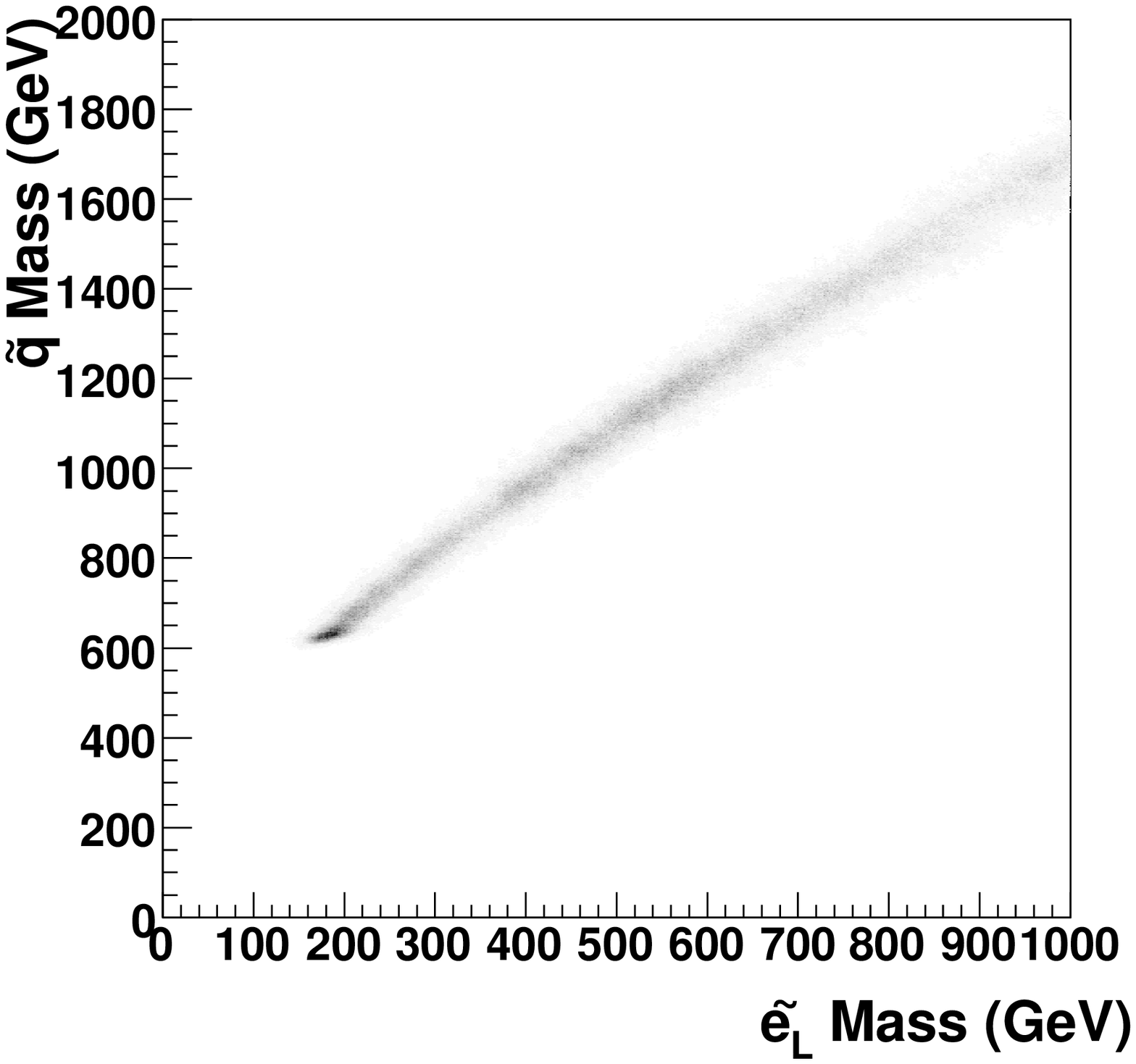}{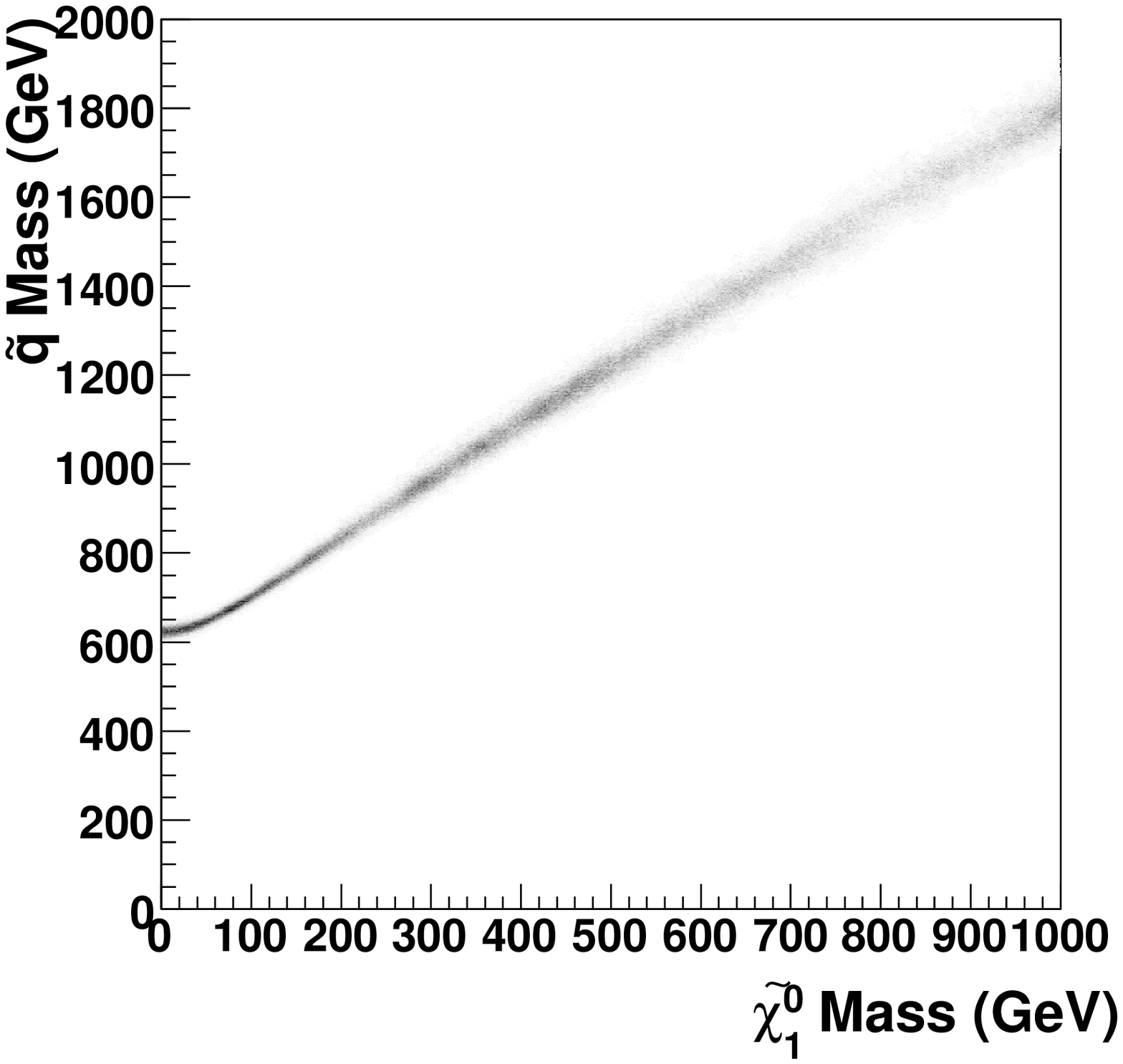}{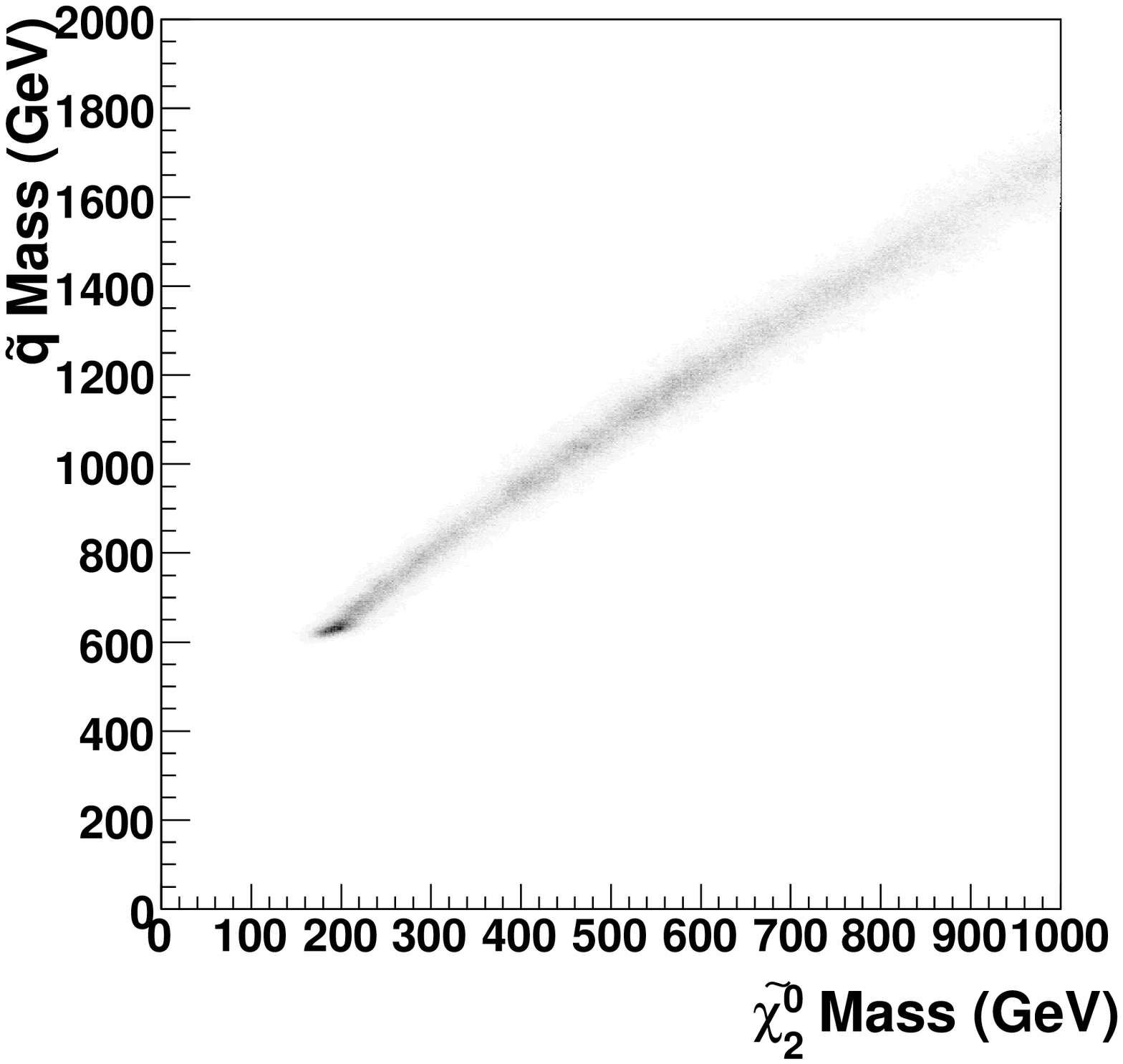}
\caption{The region of mass space consistent with the kinematic edge
measurements described in the text, obtained using a Markov chain
sampler.}
\label{n1n2}

}

\section{Cross-section information in mSUGRA space}
\label{sec:lala4}
\subsection{Background}
In principle, any piece of relevant information may be used to further
constrain the regions consistent with the kinematic edge analysis
presented in the previous section. This may be in the form of further
kinematic edges, which will provide a direct constraint on the weak
scale sparticle masses, or in the form of constraints at the SUSY
scale. The greater the number of relevant pieces of information one is
able to obtain, the better the precision of the mass measurements.

One example is given here, and developed further in this section. It
should be possible to measure the cross-section of events with missing
$p_T$ greater than 500 GeV in the ATLAS detector to a reasonable
precision. As the masses of sparticles increase, the missing $p_T$
will increase, but the total production cross-section will decrease
and hence the high mass solutions encountered in the previous section
will lead to missing $p_T$ cross-sections that are lower than the
value obtained at the coannihilation point. Thus, the cross-section
information can be added to the definition of the probability function
for the Markov Chain to give a tighter constraint on the SUSY masses.

It should be noted that up to now we have performed a model
independent analysis but, from here on in, some model will have to be
assumed in order to draw conclusions from our measurements. This is
because endpoint data can be analysed purely in the mass space
$S_{mass}$ (hereafter ``$M$'') defined by the weak scale masses, but
inclusive measurements must be compared to a given scenario (through
the use of a suitable Monte Carlo generator) before any conclusions
can be drawn, and therefore must be analysed in the space of
parameters, $S_{model}$ of that model. In section~\ref{sec:lala4}, we
investigate the constraints imposed by a cross-section measurement on
the parameter space $S_{mSUGRA}$ (hereafter ``$P$'') of a particular
model, mSUGRA, in order to introduce the technique in a familiar
context.  The limitations of this approach will become apparent by the
end of section~\ref{sec:lala4} and will be tackled in
section~\ref{sec:lala5}.

In view of this change of the constrained-space, (from the space of
weak-scale masses $\textbf{m}\in M$ to the space of mSUGRA models
$\textbf{p}\in P$) the description of the Metropolis algorithm in
section~\ref{sec:stepbystep} must, in section~\ref{sec:lala4}, be
considered re-written in terms of $p(\textbf{p}|\textbf{e}^{obs})$
rather than $p(\textbf{m}|\textbf{e}^{obs})$.  This is made more
explicit in section~\ref{sec:absoluncerjetesc} when a {\em further}
enlargement of the constrained-space is made to accommodate
uncertainty in the absolute jet energy scale.

\subsection{Cross-section measurement}
\subsubsection{Implementation}
It is assumed in this study that the cross-section of events with
missing $p_T$ greater than 500 GeV can be measured at ATLAS. One can
then pick points in the mSUGRA parameter space $S_{mSUGRA}$, work out the mass
spectrum, generate Monte Carlo events and work out the cross-section
of events passing this cut. Only certain points in the parameter space
are consistent with this measurement, and these will give a range of
masses that are consistent. Naively, the overlap of this region of the
mass space with the region consistent with the edge data will give the
new region of mass space that is compatible with the ATLAS data. In
fact, since the end points are not entirely independent of the
cross-section measurement, one needs to include both sets of
information in the final analysis in order to obtain the required
region. The `overlap' picture is approximately true, however, given
that the measurement of the cross-section is not strongly related to the measurements of the edge positions,
and is a useful guide for preliminary investigation before the final
analysis is carried out.

A plot of the missing $p_T$ cross-section in the $m_0$, $m_{1/2}$
plane for fixed tan$\beta$ and positive $\mu$ is shown in
figure~\ref{crossplot}. As can be seen, there is a lot of variation
over the parameter space and a measurement of even modest precision
will be able to rule out large areas.

\FIGURE{
\epsfig{file=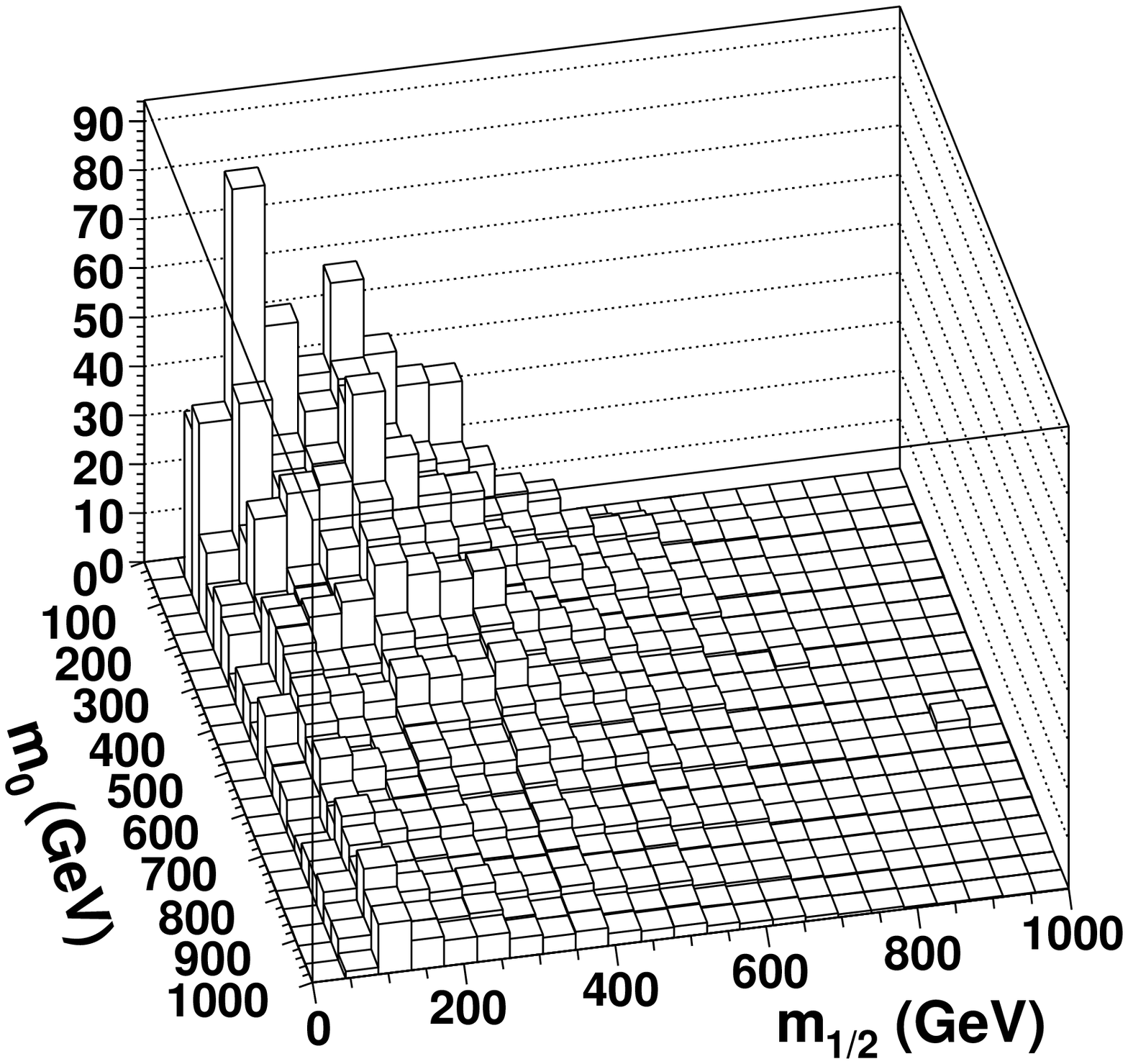,width=2.6in}
\caption{The cross-section in picobarns for events passing a missing
$p_T$ cut of 500 GeV, for tan$\beta$ = 10 and positive $\mu$, obtained
using {\tt HERWIG}. The value at the coannihilation point is 2.03
pb.  The irregularity in the plot comes from the statistical error
from having only simulated 1000 events at each point, in keeping with
the method used in the sampler.}
\label{crossplot}
}

The full process of picking mSUGRA points and obtaining the
cross-section that pass the missing $p_T$ cut has been accomplished by
successively running {\tt ISAJET}, {\tt HERWIG} and {\tt ATLFAST},
with 1000 events being generated at each point.  This is rather time
consuming, however, and a simple scan of the mSUGRA parameter space is
unfeasible if any more than two of the mSUGRA parameters are
varied. For this reason, we again use the Metropolis sampling
technique introduced in the previous section and, indeed, it is here
that the power of the method becomes apparent. The algorithm has been
used to map the interesting region of the parameter space with fewer
points than would be required in a scan in order to obtain similar
performance.

To demonstrate this, consider the following. There are four and a half
parameters in the mSUGRA parameter space, though we have held $A_0$
constant for simplicity.\footnote{In retrospect there was no
compelling reason to hold $A_0$ fixed, and in a later study we expect
to look at the effect of allowing $A_0$ to vary and be measured by the
data along with all the other model parameters.} Of the remaining
parameters, one is simply a sign (the sign of $\mu$), and hence one
sampling run was performed with this sign positive, and another with
it negative. In any one application of the software, then, three
parameters are varied -- $m_0$, $m_{1/2}$ and tan$\beta$ -- and even a
coarse scan requiring 100 points along each axis would require one
million points for each sign of $\mu$. The Metropolis algorithm maps
the interesting regions of the space in approximately 15,000 points
per sign of $\mu$, a dramatic improvement without which the analysis
would have taken many months, if not years.

Even with this improvement, it was still necessary to reduce the run
time of {\tt HERWIG} significantly through the use of a parallel
computer. Although the Metropolis algorithm itself cannot be
parallelised, we have adapted {\tt HERWIG} to run on a parallel
machine with the use of {\tt MPI} code, thereby substantially reducing
the run time per point.

\subsubsection{Definition of Metropolis quantities for cross-section}
\label{sec:simplecrossseccase}
We now define the Metropolis algorithm for use with (only) the
cross-section data.  As in the previous section, we require the
definition of the probability distribution
$p(\textbf{p}|\sigma^{obs})$ from which samples are to be taken, in
which $\sigma^{obs}$ represents the cross-section supposedly
``observed'' or measured by the experiment.  Lacking real data, we
take $\sigma^{obs}$ to be 2.04 pb, the value predicted by a {\tt
HERWIG} and {\tt ATLFAST} simulatuion of the coannihiliation point of
section~\ref{sec:wmappointdefined}.  The evaluation of
$p(\textbf{p}|\sigma^{obs})$ necessitates the definition of a suitable
prior $p(\textbf{p})$ on the model space $P$ which again we take to be
flat (but equal to zero for invalid values of any of the model
parameters $p_i \in \textbf{p}$).  Finally the Metropolis sampler's
proposal distribution must be modified to act on the model space $P$
rather than on the mass space $M$.  The proposal distribution was
again chosen to be a multi-dimensional Gaussian centred on the current
point $\textbf{p} \in P$.  The widths of the proposal distribution in
$m_0$, $m_{1/2}$ and tan$\beta$ were respectively usually 25 GeV, 25
GeV and 2 GeV, except when both cross-section {\em and} edge
constraints were in use simultaneously (only in
sections~\ref{sec:furtheranalsec} and beyond) in which case a smaller
set of widths was used (5 GeV, 5 GeV and 2 GeV).  The widths were
chosen for the efficiency reasons outlined in
section~\ref{sec:whyQwaswhatitis} and will not affect the results once
convergence has occurred.

The sampled probability distribution $p(\textbf{p}|\sigma^{obs})$
follows a similar definition to that encountered previously for
$p(\textbf{m}|\textbf{e}^{obs})$.  The analogue of
equation~(\ref{eq:productything}) is then just the single term
$p(\sigma^{obs}|\textbf{p})$ quantifying the cross-section likelihood
according to:
\begin{equation}\label{eq:newproductything}
p(\sigma^{obs}|\textbf{p})\approx\frac{1}{\sqrt{2\pi\sigma^2_{err}}}\exp\left(-\frac{   (\sigma^{obs}-\sigma^{pred}(\textbf{p}))^2  }{2 \sigma^2_{err}}\right),
\end{equation}
where $\sigma_{err}$ is the error associated with the cross-section
measurement $\sigma^{obs}$, and $\sigma^{pred}(\textbf{p})$ is the
value of the cross-section expected at the point $\textbf{p}$ in
mSUGRA parameter space $P$ as again predicted by a {\tt HERWIG} and
{\tt ATLFAST} simulation.

The error $\sigma_{err}$ on the observed cross-section $\sigma_{obs}$
was taken to be ten per cent.  This figure was chosen somewhat
arbitrarily, for similar reasons to those given when explaining the
sizes of the errors assumed for the enpoint measurments (see
section~\ref{sec:explainingbyeye}): this paper is designed to
illustrate a {\em method}, not to claim that a particular measurement
can be made with a certain precision.  In contrast, if we had access
to {\em real} data, it would be of vital importance to make the
estimation of the cross-section error as accurate as possible.  The
eventual precision of the final answer will be strongly correlated
with the error attributed to the cross-section.  In retrospect, the
chosen value of ten per cent probably underestimates the combination
of (1) the statistical error, (2) luminosity error, (3) the
theoretical uncertainty on the signal cross-section, and (4) the
combined experimental and theoretical uncertainty on the prediction
for the number of standard model events likely to pass the signal
cuts.  If we were to be granted further time on the supercomputer and
were able to start the analysis again from scratch, we would probably
re-simulate with a larger and more realistic error of thirty percent.
Further work (beyond the scope of this paper) should be done to
investigate the expected size of this error, and to confirm that the
effect of increasing this error estimate is just to enlarge the size
of the final regions.  Within this article, however, the cross section
error will be taken to be the stated ten per cent -- and this will be
sufficient for the purpose of demonstrating how the proposed method
can be used in practice.

Certain regions of mSUGRA parameter space $P$ are known to be
unphysical -- for example there may be no electroweak symmetry
breaking or there may be a charged LSP.\ \ In both cases, \tt ISAJET
\rm will detect this and fail to run.  Furthermore there are points
$\textbf{p}$ for which {\tt HERWIG} will not run. When any of these
problems occur we take the point $\textbf{p}$ to be unphysical and
multiply the likelihood by zero (as unphysical points {\em cannot}
have generated the observed data!).

\subsubsection{Results in mSUGRA space (for cross-section information alone)}
The results of the Markov Chain in mass space for positive $\mu$ can
be seen in figure~\ref{plusmu}, with those for negative $\mu$
presented in figure~\ref{minusmu}. The distributions look very similar
in the $m_0, m_{1/2}$ plane, reflecting a lack of sensitivity to the
sign of $\mu$. The tan$\beta$ distribution is approximately flat for
negative $\mu$, whilst there is some insignificant preference for the
`correct' value of $\tan\beta=10$ in the positive $\mu$ case.

\FIGURE{
\twographst{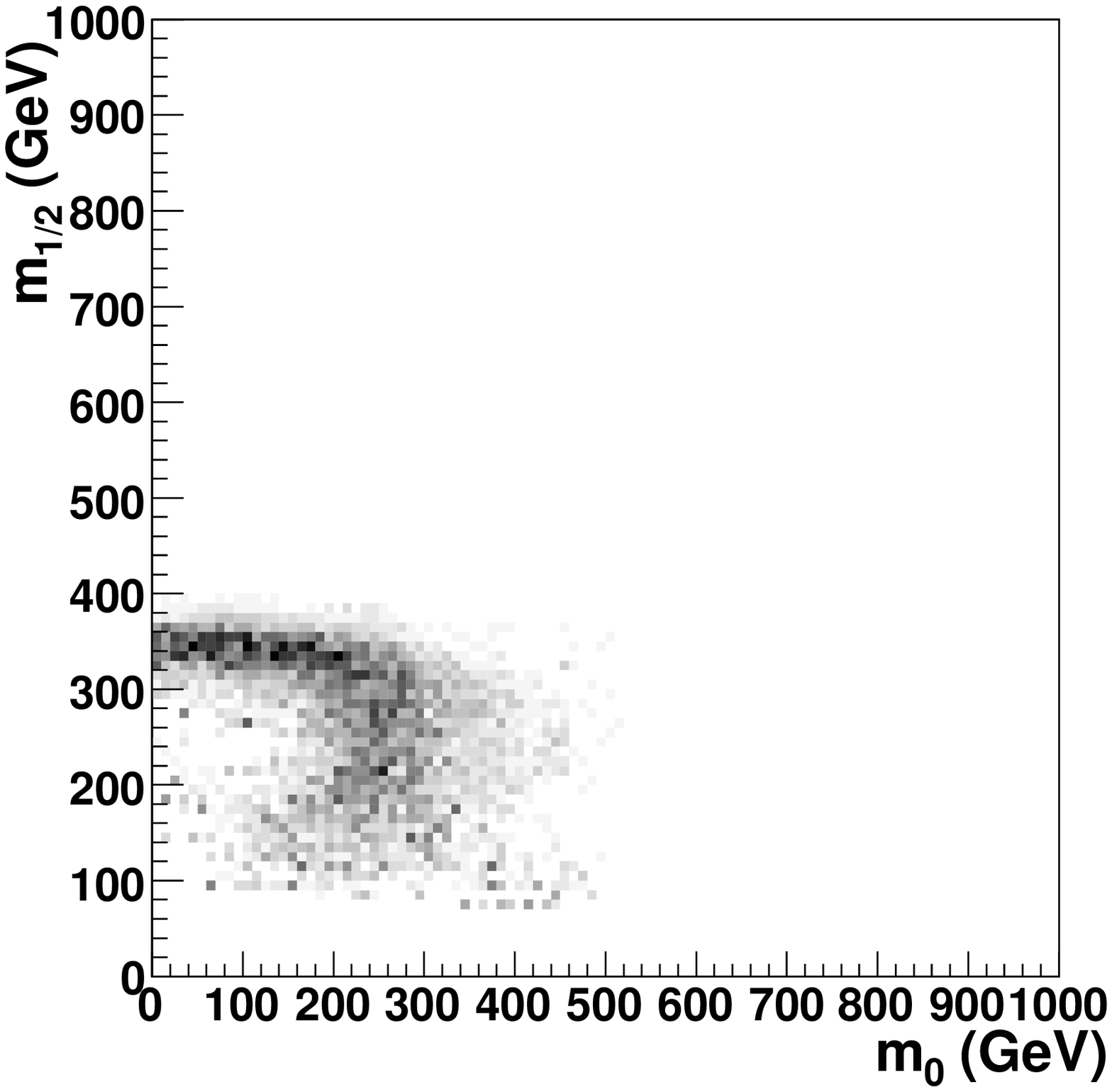}{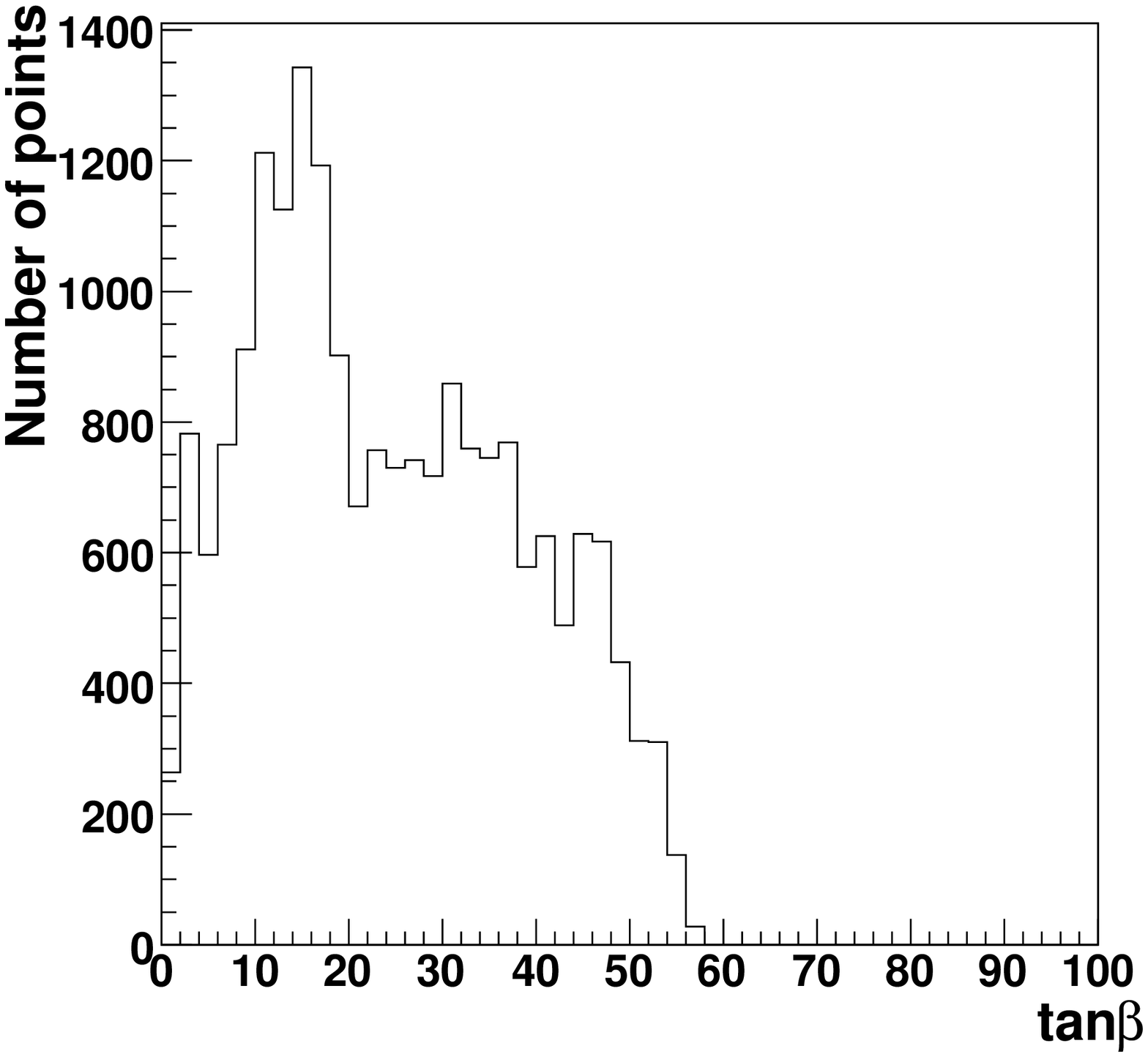}
\caption{The region of mSUGRA parameter space consistent with the
measurement of the cross-section of events with missing $p_T$ greater
than 500 GeV, for positive $\mu$.}
\label{plusmu}
}

\FIGURE{
\twographst{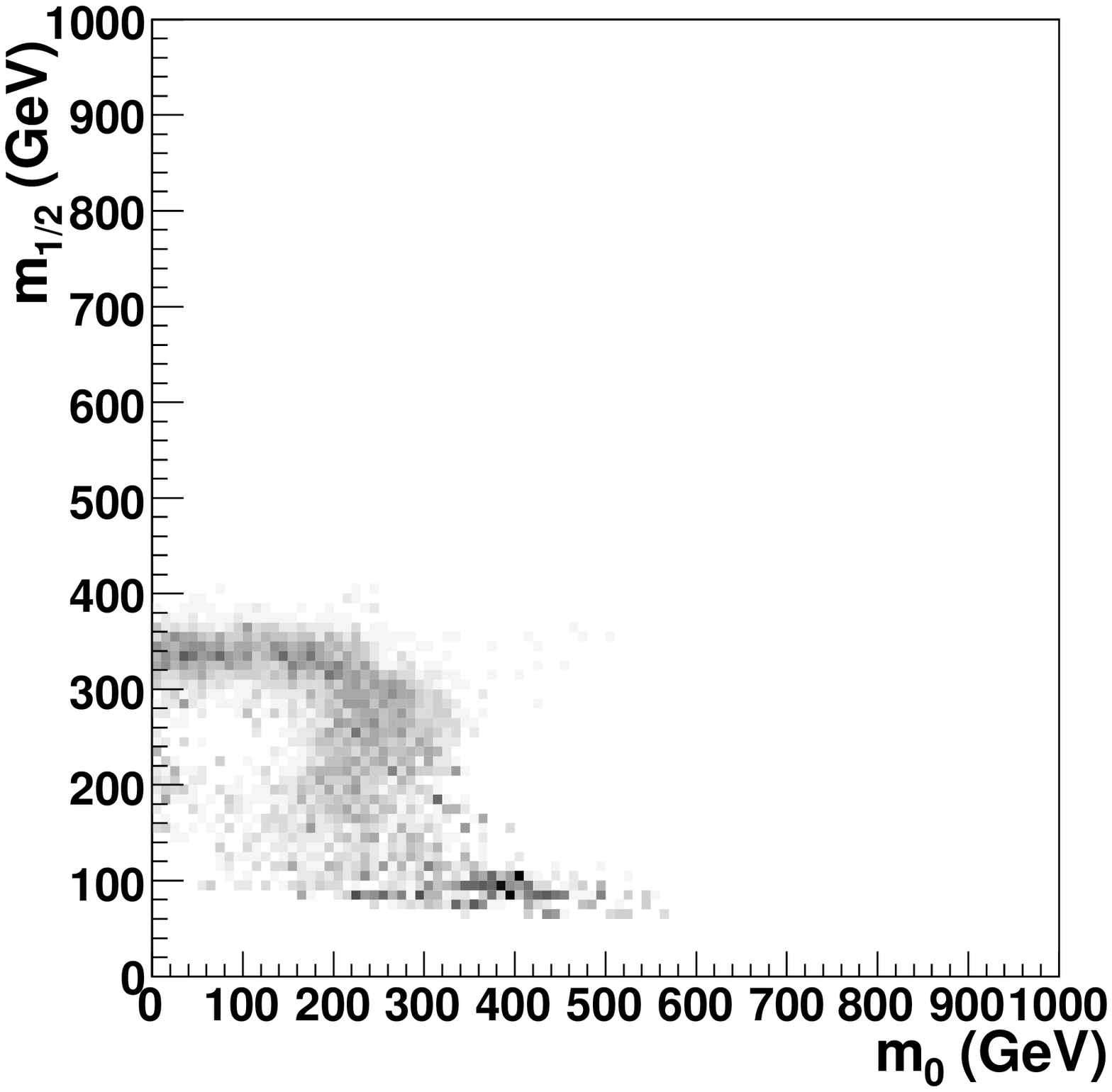}{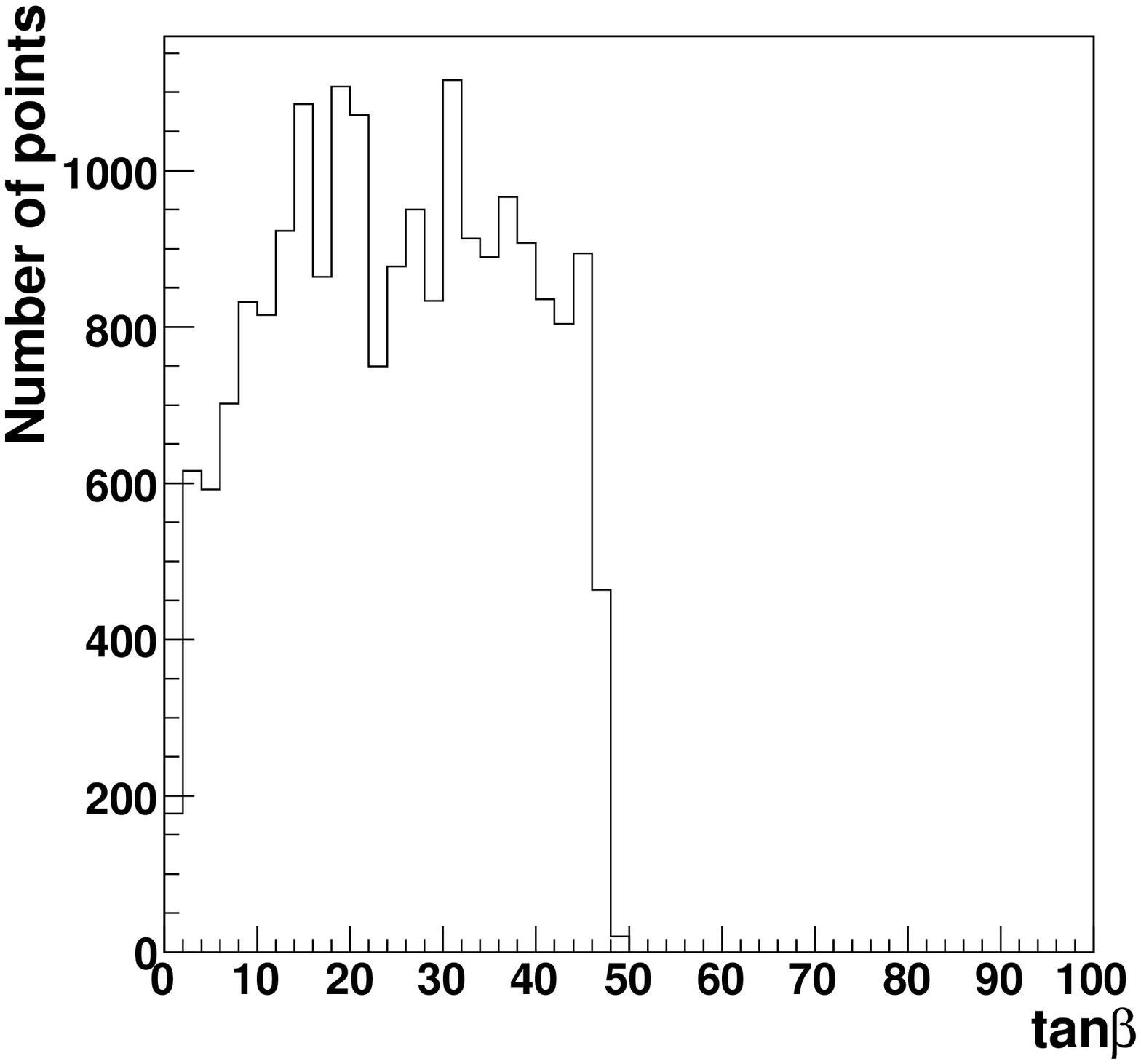}
\caption{The region of mSUGRA parameter space consistent with the
measurement of the cross-section of events with missing $p_T$ greater
than 500 GeV, for negative $mu$.}
\label{minusmu}
}

\subsubsection{Results in mass space (for cross-section information alone)}
We now relate the results in figures~\ref{plusmu} and \ref{minusmu} to
the weak scale mass space in which we have already observed the
regions consistent with the kinematic edge analysis. The positive
$\mu$ and negative $\mu$ data sets presented previously have been
evolved to the weak scale using {\tt ISAJET}and combined into a single
data set by weighting each of the two sets by the average likelihood
of that set. The region obtained in mass space is shown in
figure~\ref{masscross}, and is dramatically different from that
obtained using the edge analysis. The overlap between the regions
found by the two methods (figures~\ref{n1n2} and \ref{masscross}) is
shown in figure~\ref{overlap}, and was obtained by multiplying the
previous data sets.

The overlap of the two regions has produced much tighter constraints
on the particle masses, even with a relatively conservative estimate
of the precision of the endpoint measurements. It is worth noting that
the projections of the region of overlap on each pair of axes give
different size regions in each plane, with the smallest being that in
the plane of the neutralino masses. This could be used to remove some
of the area shown in the other planes, although the strictly correct
procedure (followed in section~\ref{sec:furtheranalsec}) is to run a
Markov Chain with the edge and cross-section information implemented
at the same time.

\FIGURE{
\sixgraphs
{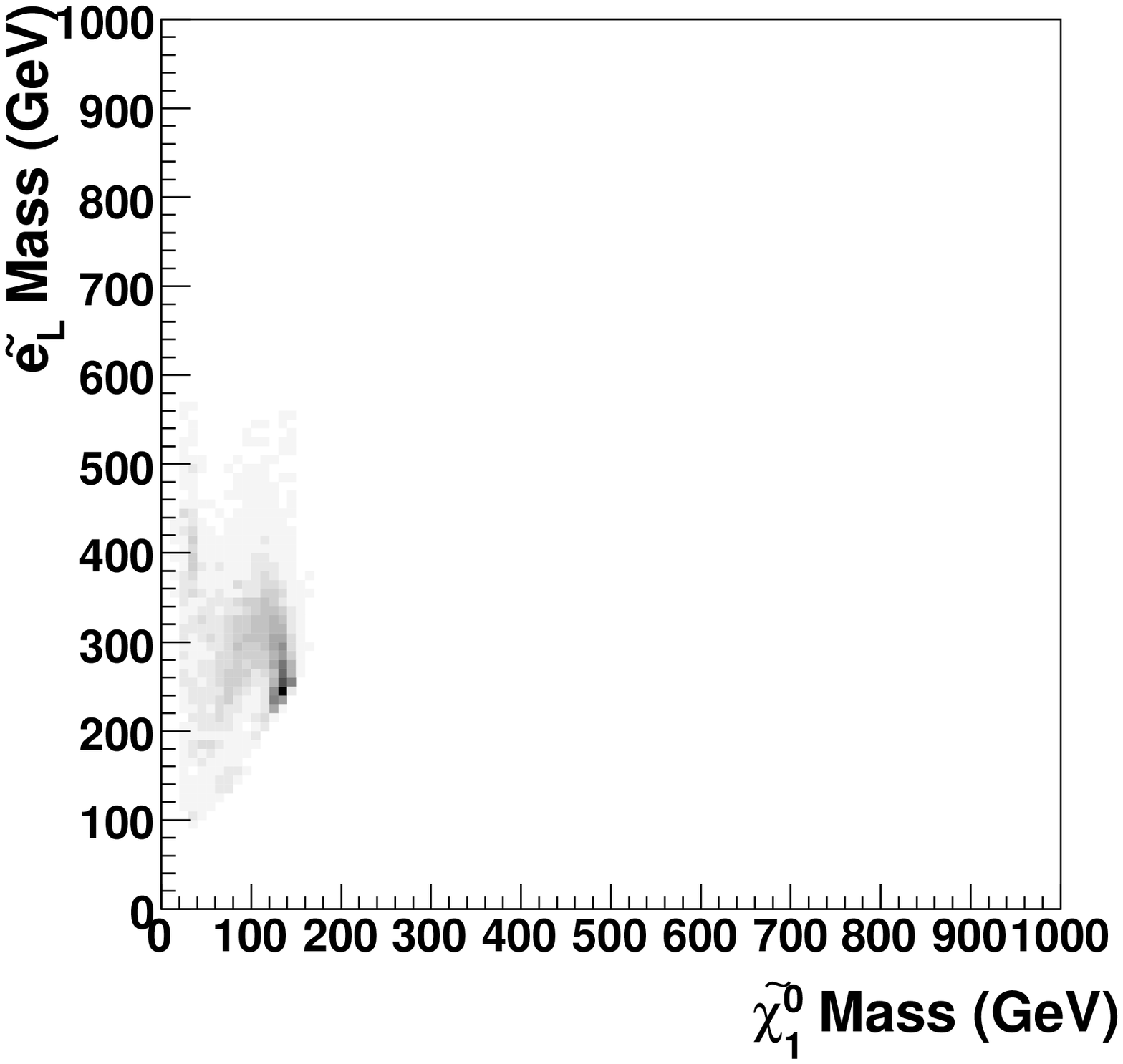}{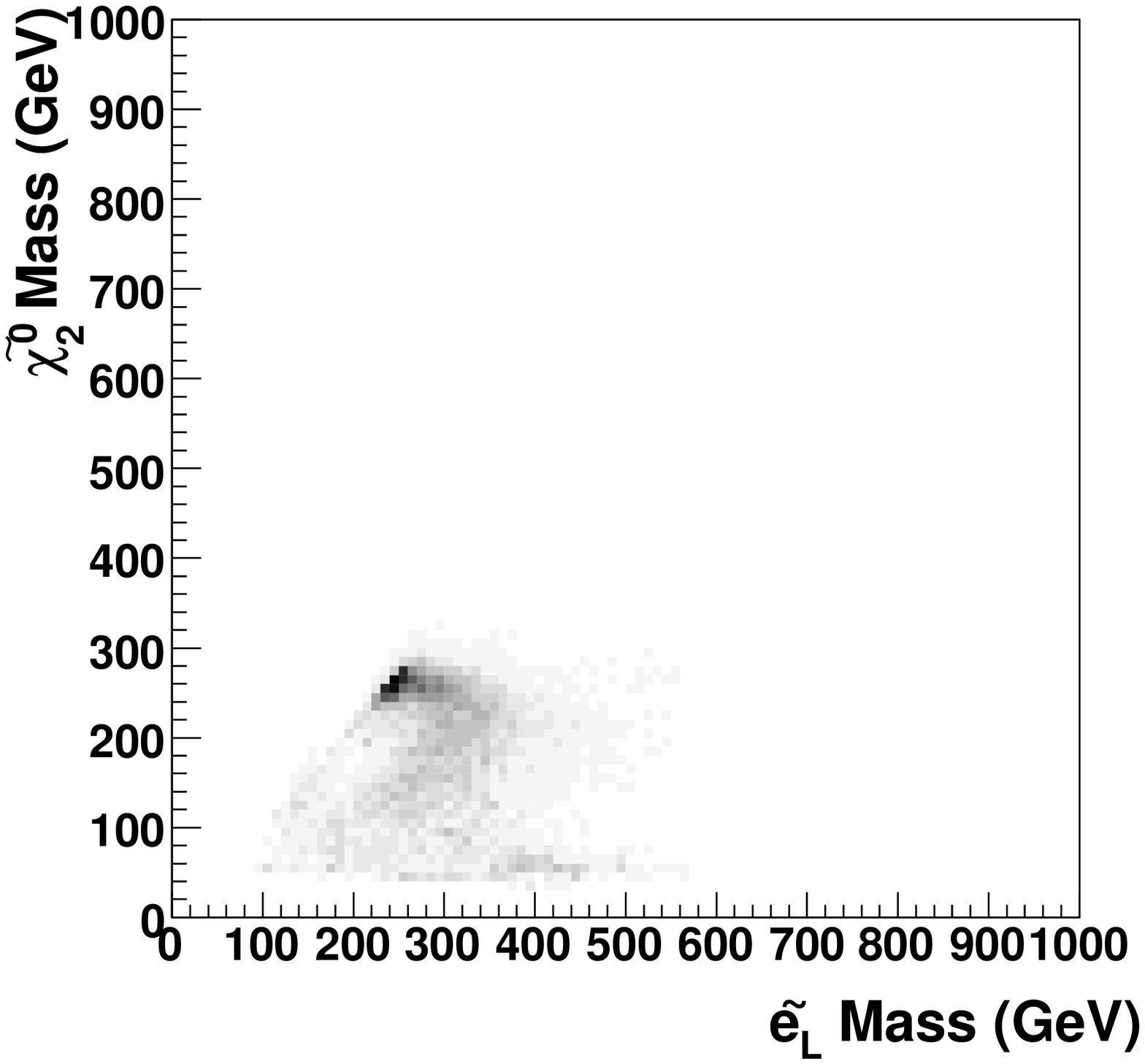}
{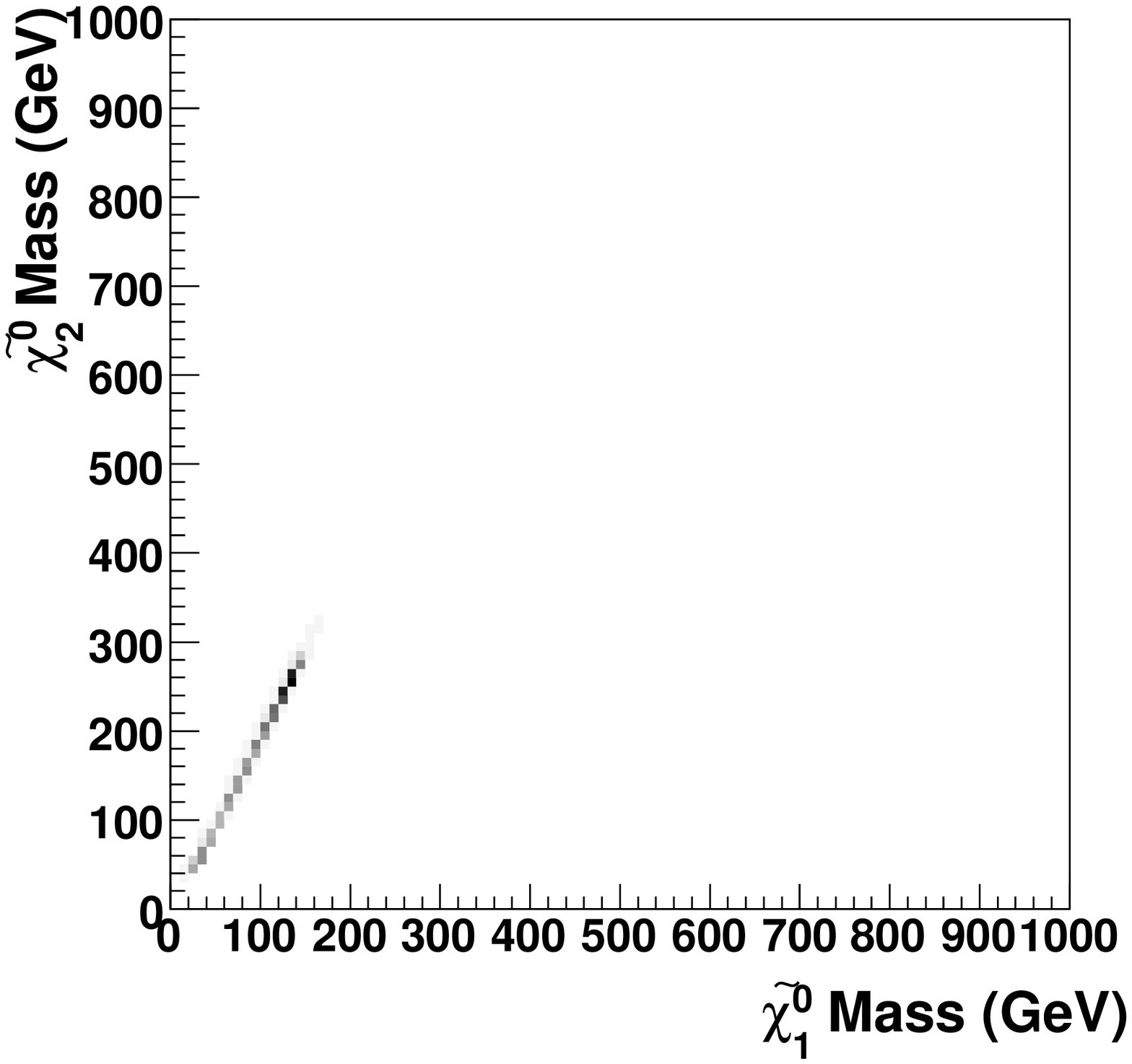}{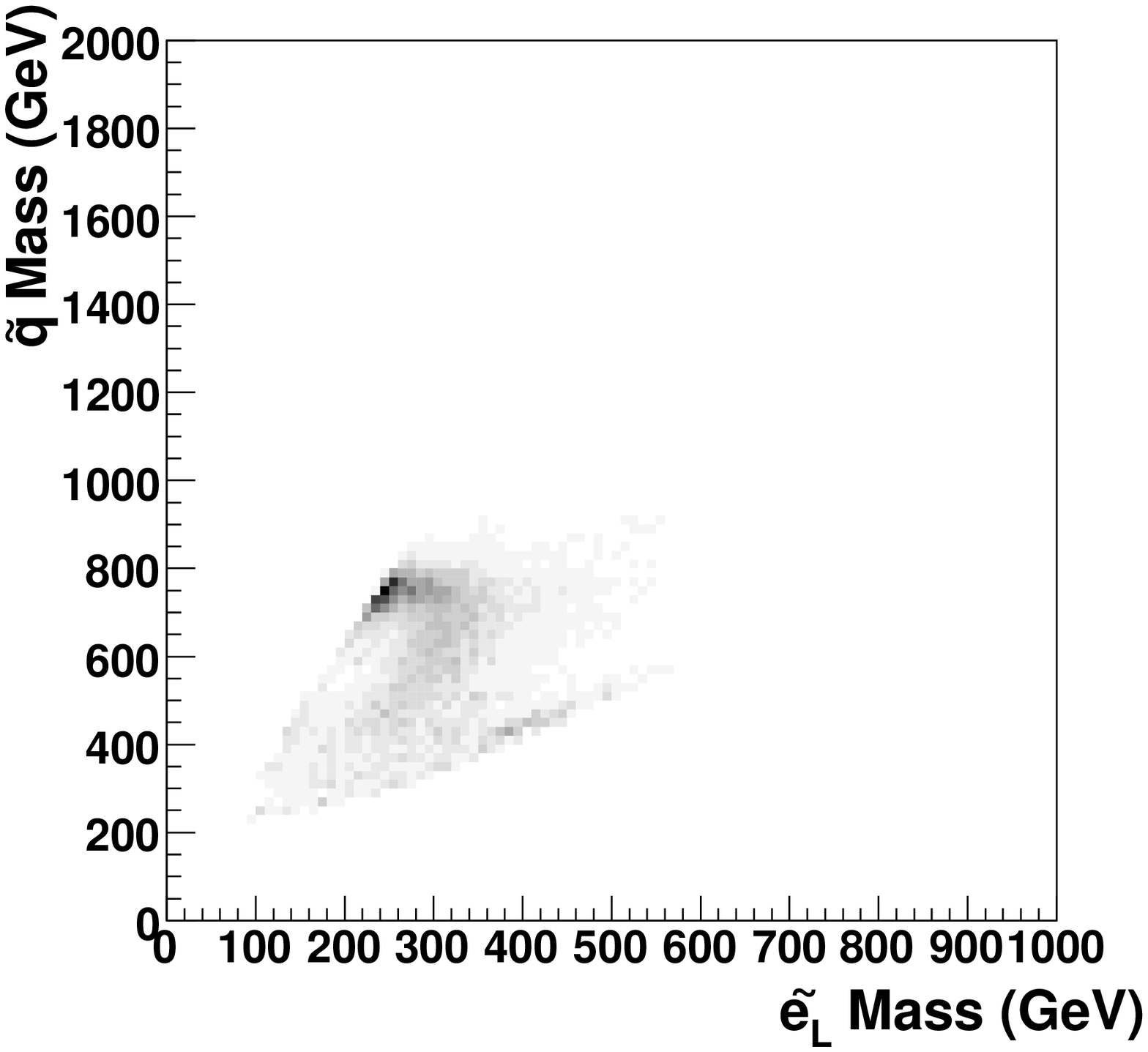}
{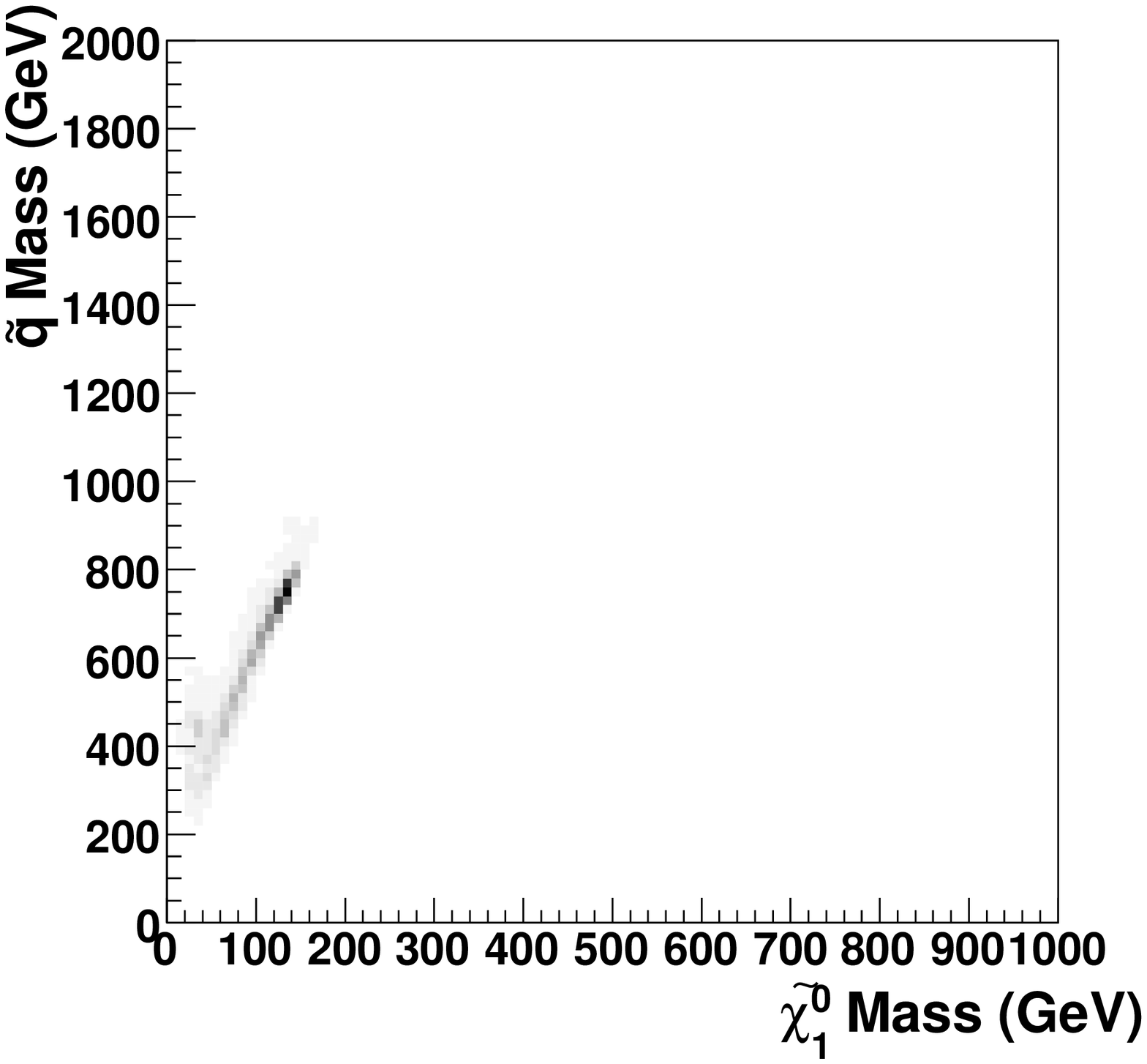}{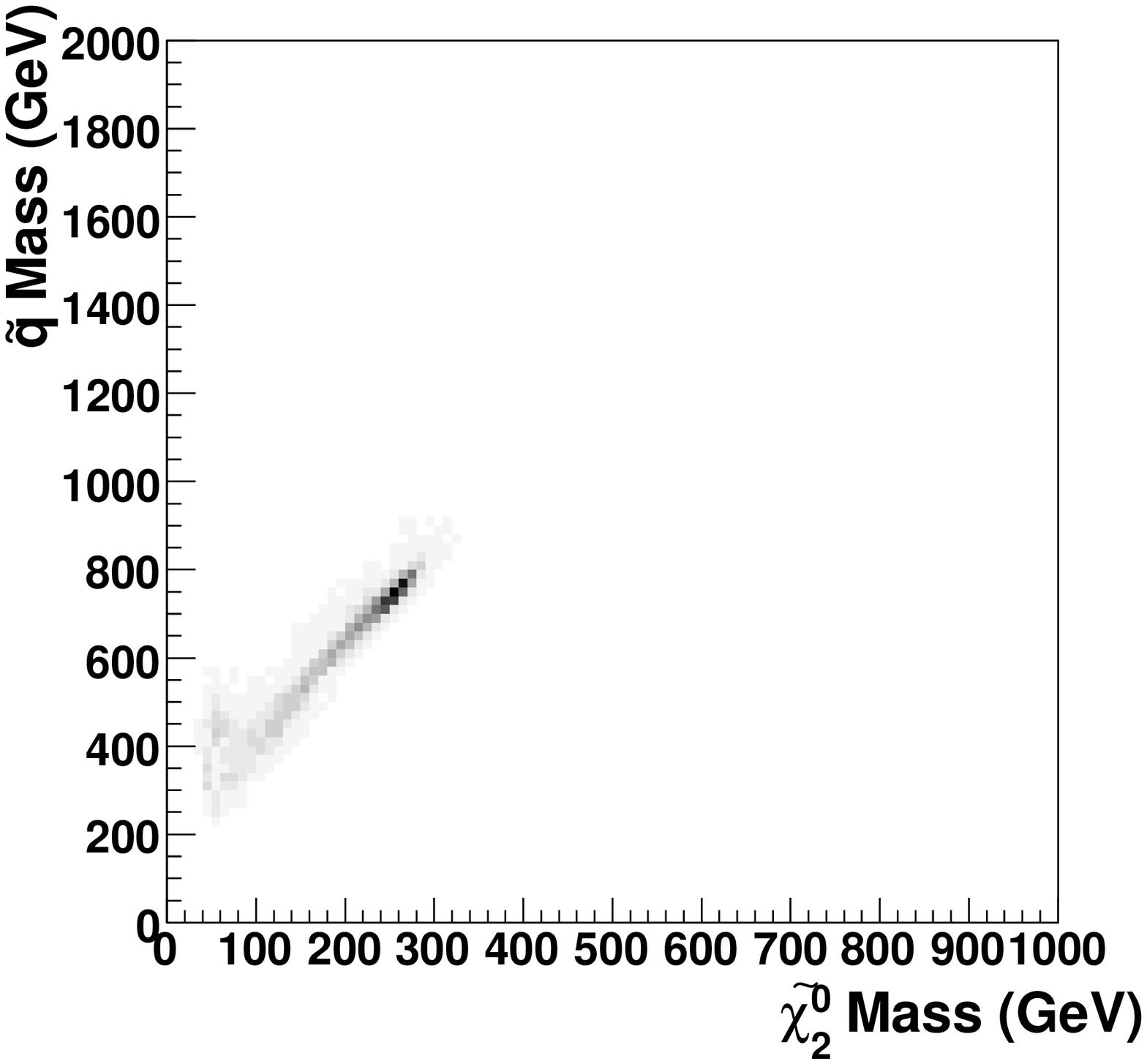}
\caption{The region of mass space consistent with a measurement at
10\% precision of the cross-section of events with missing $p_T$
greater than 500 GeV, obtained using a Markov chain sampler.}
\label{masscross}
}

\FIGURE{
\sixgraphs
{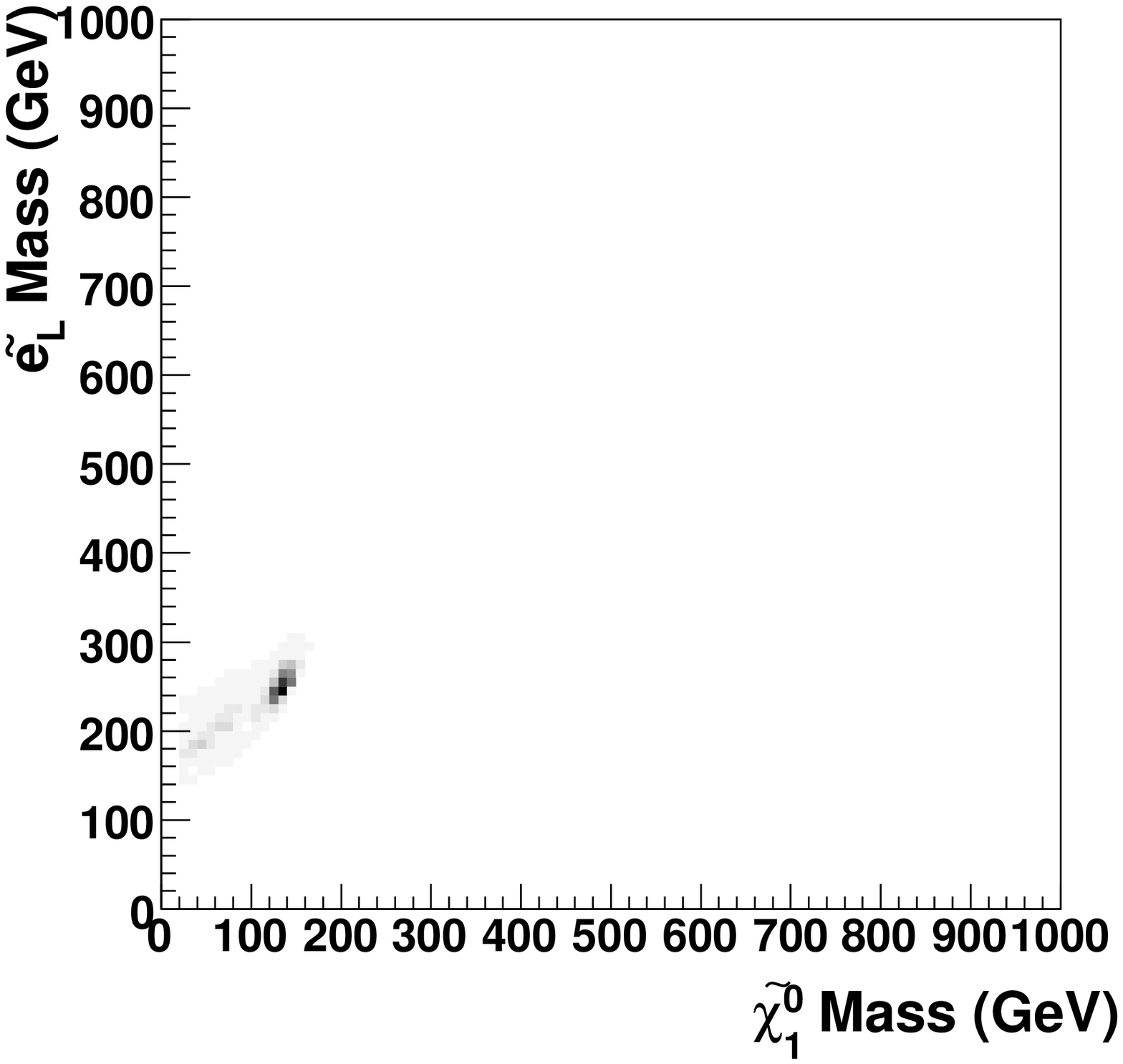}{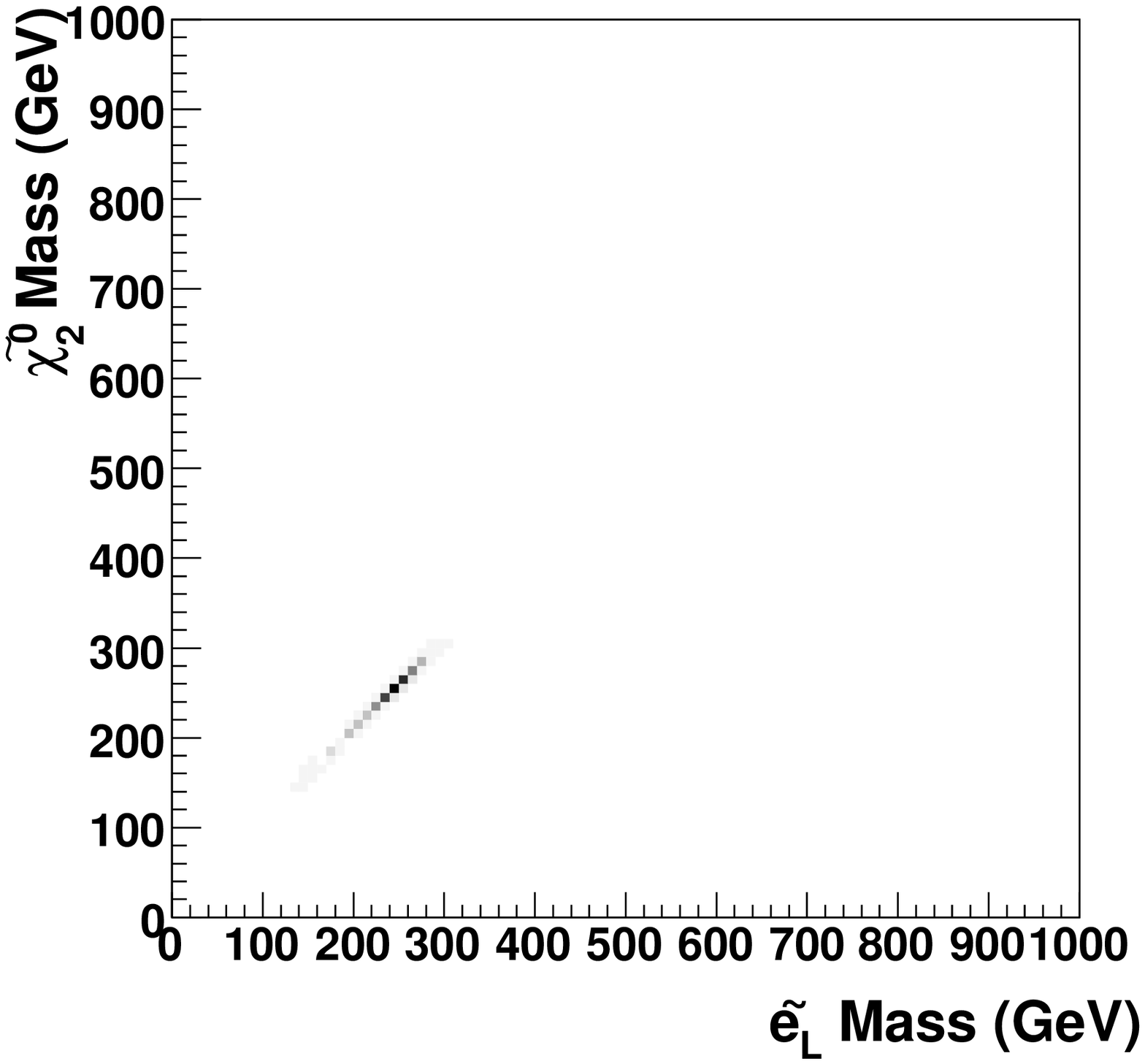}
{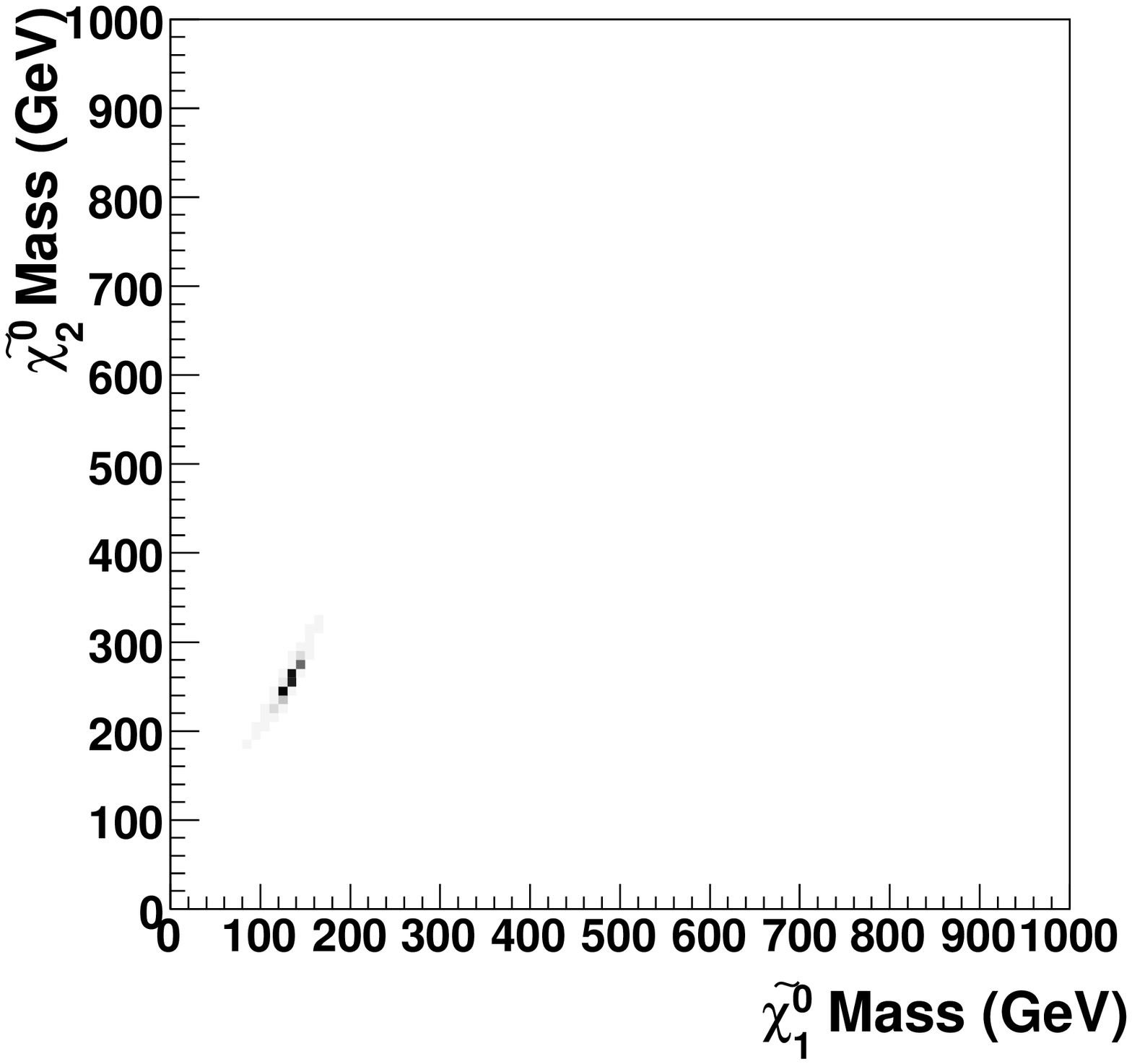}{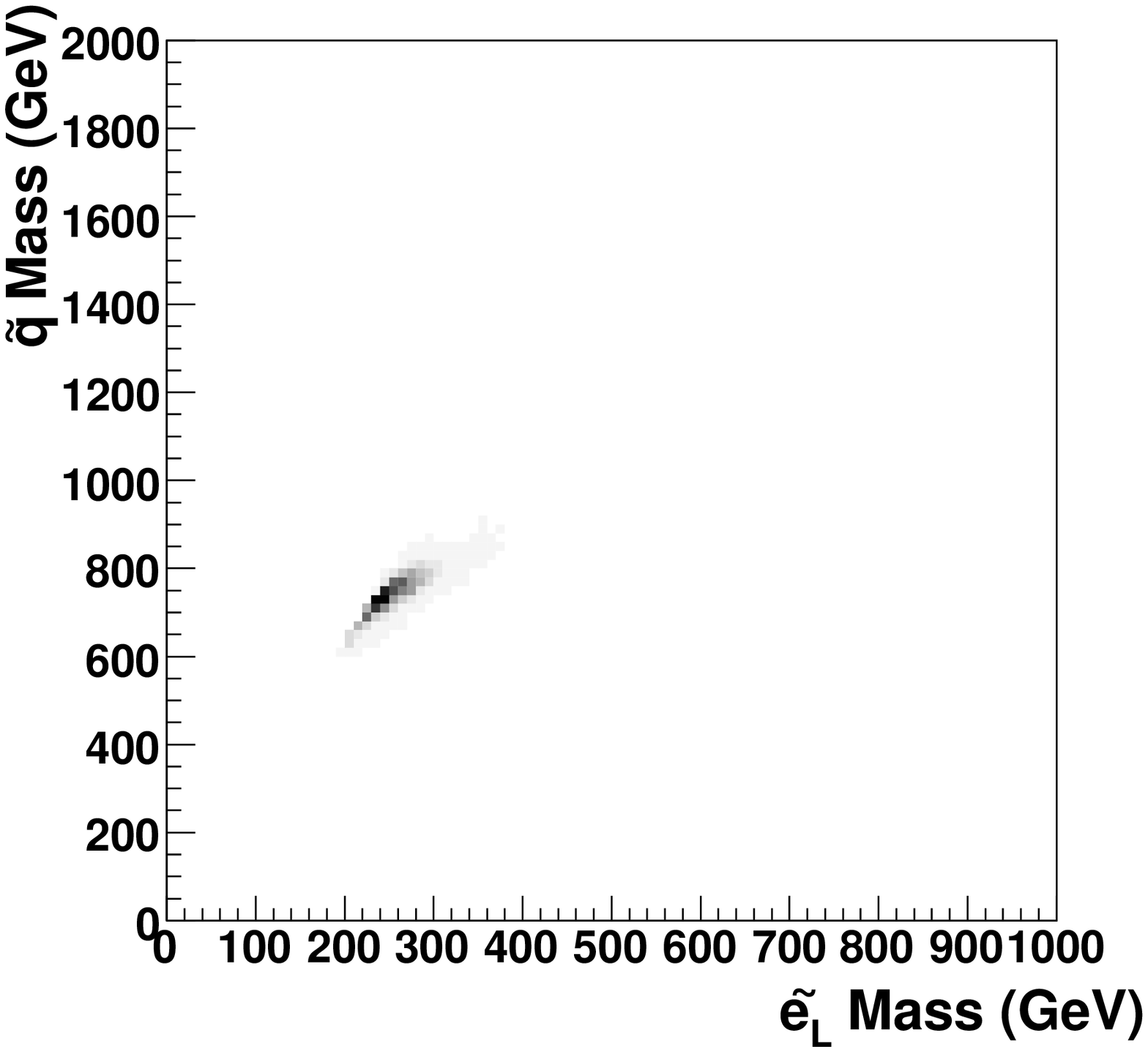}
{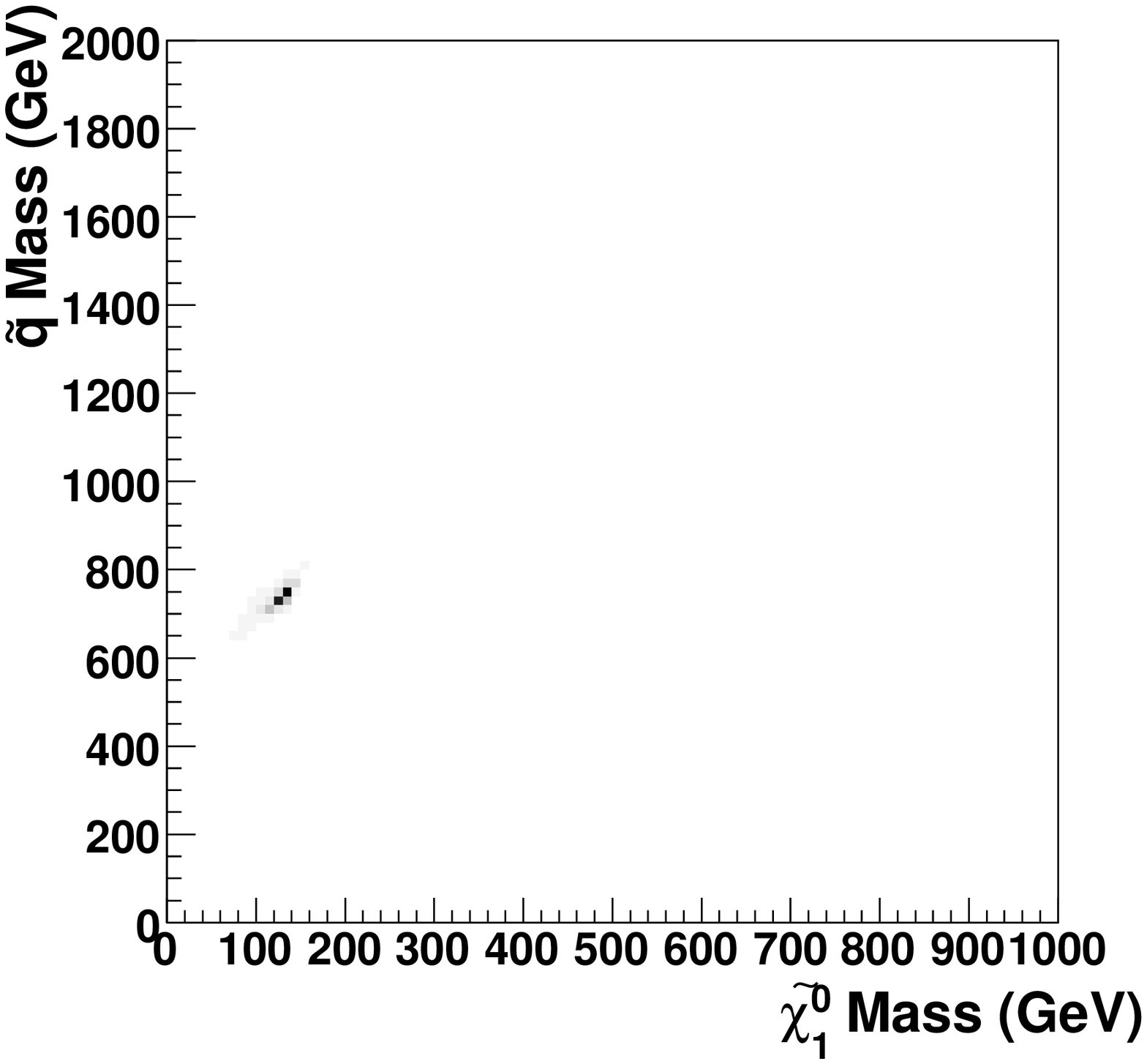}{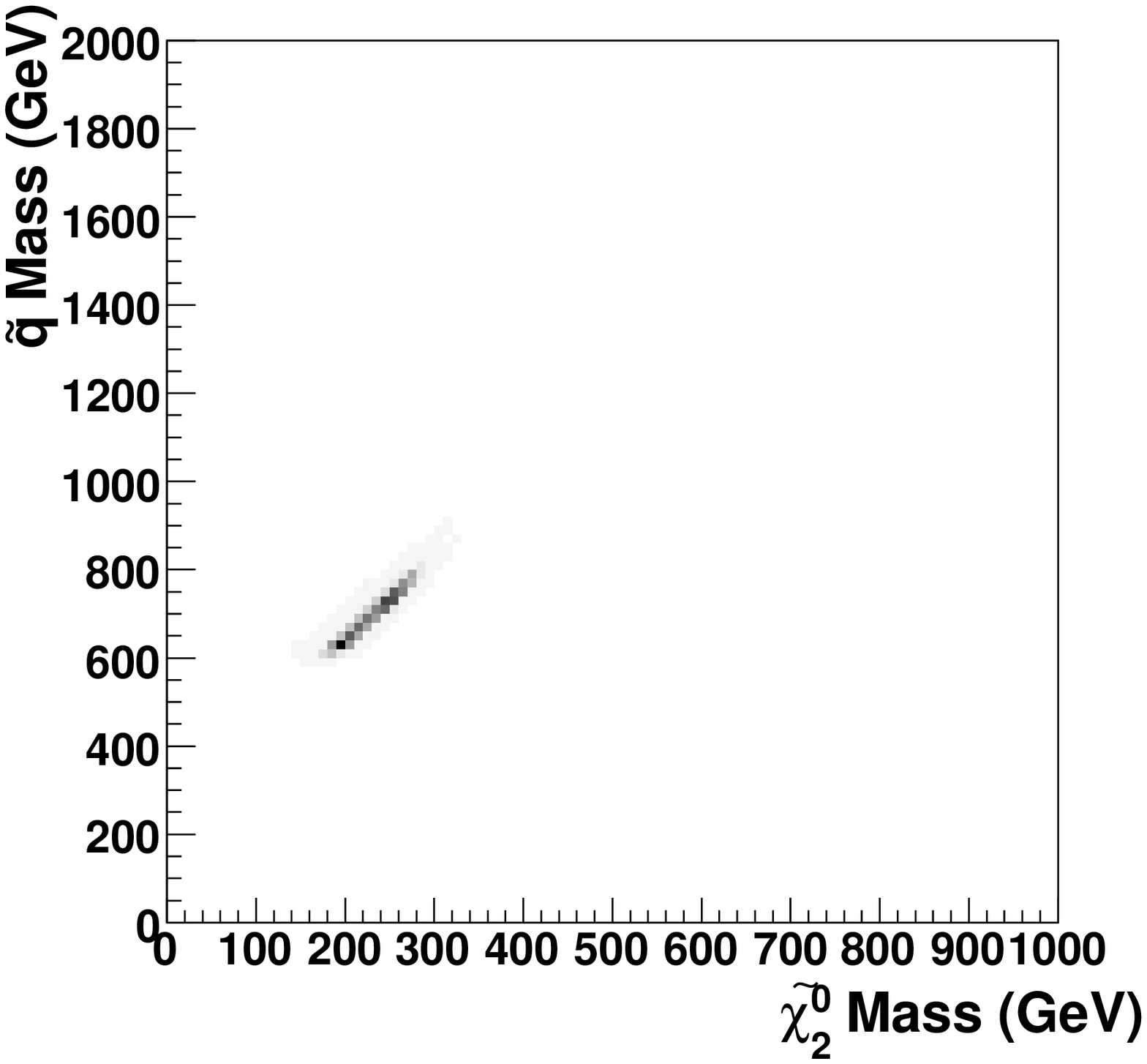}
\caption{The region of mass space consistent with a measurement at
10\% precision of the cross-section of events with missing $p_T$
greater than 500 GeV, overlapped with a measurement of the squark
decay kinematic endpoints obtained in section~\ref{sec:lala2}.}
\label{overlap}
}

\subsection{Further analysis}
\label{sec:furtheranalsec}
The overlap plots presented in the previous subsection give a rough
idea of what to expect from the combination of edge and cross-section
information, but the approach is only approximately valid given that
the cross-section measurement is not independent of the kinematic edge
positions. In order to be fully rigorous, one must run a Markov Chain
whose probability density function combines both the cross-section and
the edge information at the same time -- in other words one must
sample this time from $p(\textbf{p} | \textbf{e}^{obs},
\sigma^{obs})$.

Accordingly, a Metropolis sampler of $p(\textbf{p} | \textbf{e}^{obs},
\sigma^{obs})$ was set to explore the mSUGRA parameter space $P$.

At each point $\textbf{p}\in P$ the number of events passing the
missing $p_T$ cut was obtained from the {\tt ATLFAST} output whilst
the {\tt ISAJET} mass spectrum was used to find the expected
position of the endpoints. This information was then compared to the
`measured' information (in this case, the endpoints shown earlier, and
the cross-section obtained through Monte Carlo simulation of the
coannihilation point) in the definition of the probability
weight for each point $\textbf{p}\in P$.  The likelihood
$p(\textbf{e}^{obs}, \sigma^{obs}|\textbf{p})$, the analogue of
equations~(\ref{eq:productything}) and (\ref{eq:newproductything}), is
this time the product of the pair of them:
\begin{eqnarray}
p(\textbf{e}^{obs}, \sigma^{obs}|\textbf{p}) = p(\sigma^{obs}|\textbf{p}) \prod_{i = 1}^5 {
p(e_i^{obs}|\textbf{m}(\textbf{p}))  }. \label{eq:combinedproductything}
\end{eqnarray}

The same flat prior $p(\textbf{p})$ on mSUGRA space $P$ was used as in
section~\ref{sec:simplecrossseccase}.  The likelihood was multiplied
by zero if the sparticle masses $\textbf{m}(\textbf{p})$ obtained at a
point $\textbf{p}$ were not consistent with the mass hierarchy
required for the squark decay chain to exist.  The Metropolis
algorithm's proposal distribution was the same as that used previously
in section~\ref{sec:simplecrossseccase}.  Chains were run separately
for positive and negative $\mu$.

\subsubsection{Results for cross-section and edge
measurements {\em together}}

The mSUGRA space results for cross-section {\em and} edge measurements
are shown in figures~\ref{plusmu_final} and \ref{negativemu_final},
with the results in mass space shown in figure~\ref{mass-final}. Note
that inclusion of the cross-section information greatly improves the
precision in the $m_0, m_{1/2}$ plane.

We would like to emphasise at this stage that the majority of the
apparent improvement is {\em not} the result of the inclusion of the
cross-section measurement -- but is rather a well known side-effect of
the fit taking place in a model space which is more tightly
constrained (masses depend primarily on just the two parameters $m_0$
and $m_{1/2}$) than the original mass space (four free masses).  Many
points in mSUGRA space are now rejected as they give the wrong mass
hierarchy for the decay chain. This leads to a jump in mass-space
precision, at the expense of incorporating some model dependence.
If we are prepared to accept the model dependence introduced by moving
to the space of a particular model (in this case mSUGRA) we are forced
to accept ``uncomfortably'' tight constraints on the compatible
regions of parameter space.  Why uncomfortable?  Uncomfortable because
the choice of mSUGRA was somewhat arbitrary, and made without a strong
degree of belief that mSUGRA is an effective theory of Nature.  Given
this lack of confidence in mSUGRA itself, there seems little use in
being able to quote tiny errors on the parts of it which are
compatible with the data -- especially when even small departures from
the mSUGRA model might lead to significant changes in the sparticle
spectra or properties.

However, this very distaste is now the motivation for recognising that
we are no longer restricted to looking at overly constrained models
like mSUGRA, and suggests that we can now look at a wider class of
models in which we hope to have a higher degree of faith.  In this way
we can lose some of the unpleasant model dependence just introduced,
and can for the first time actually put the cross-section measurement
in a position in which it can play an {\em active} role in
constraining competing theories.  We thus hope to illustrate the power
of our technique.

In section~\ref{sec:genrlisingsec} we go on to increase the
dimensionality of the parameter space in exactly this way (by relaxing
the conditions that impose, for example, universal gaugino masses at
the GUT scale, etc) and still maintain good precision by using the
endpoint data together with the cross-section measurement.  Inclusive
and exclusive data is combined to explore more general SUSY models in
order to learn yet more about the SUSY Lagrangian.


Finally, it is crucial to note that there is another limitation in the
analysis so far in that it has been assumed that one has established
that the particles involved in the decay chain are the two lightest
neutralinos and the left-handed slepton. In practise, one could just
as easily fit endpoints using, for example, the heaviest two
neutralinos and the right-handed slepton. This ambiguity ought to be
reflected in the analysis, and has only rarely been considered before
(see for example \cite{Lester,Gjelsten:2004ki}). This is also
considered in section~\ref{sec:genrlisingsec}.

\FIGURE{
\twographst{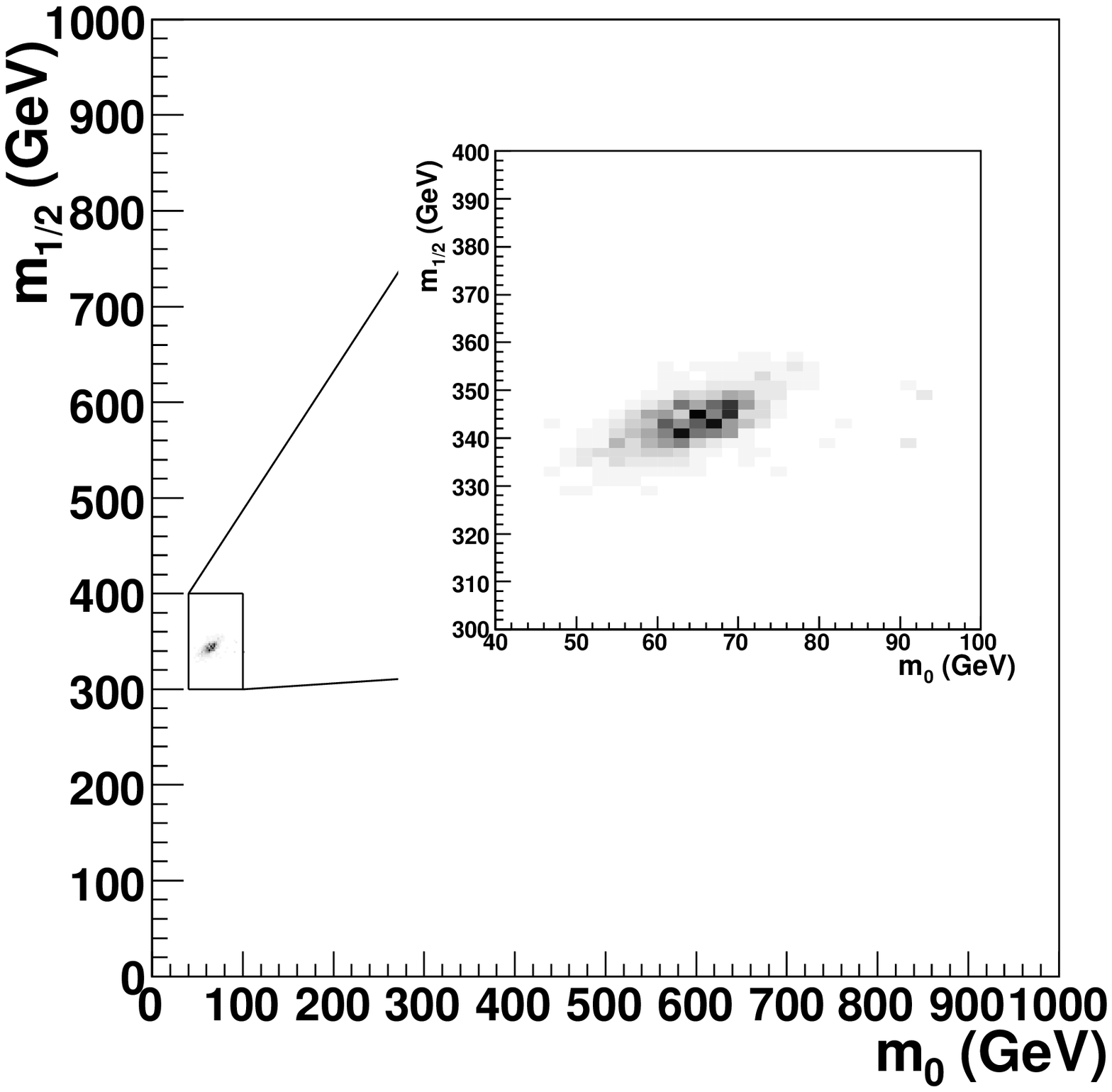}{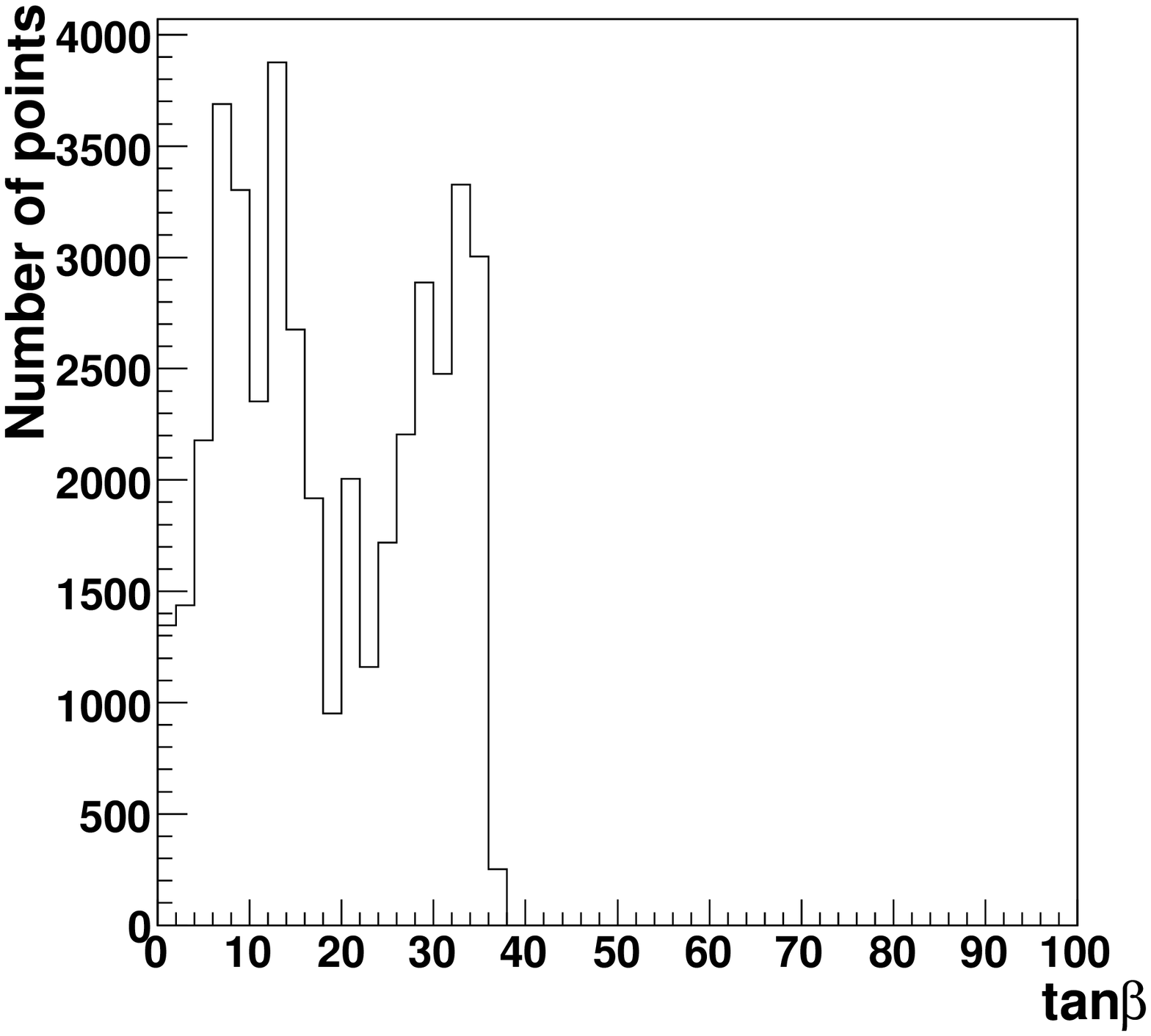}
\caption{The region of mSUGRA parameter space consistent with the
measurement of the cross-section of events with missing $p_T$ greater
than 500 GeV and with the endpoint measurements obtained in section~\ref{sec:lala2},
for positive $\mu$.}
\label{plusmu_final}
}

\FIGURE{
\twographst{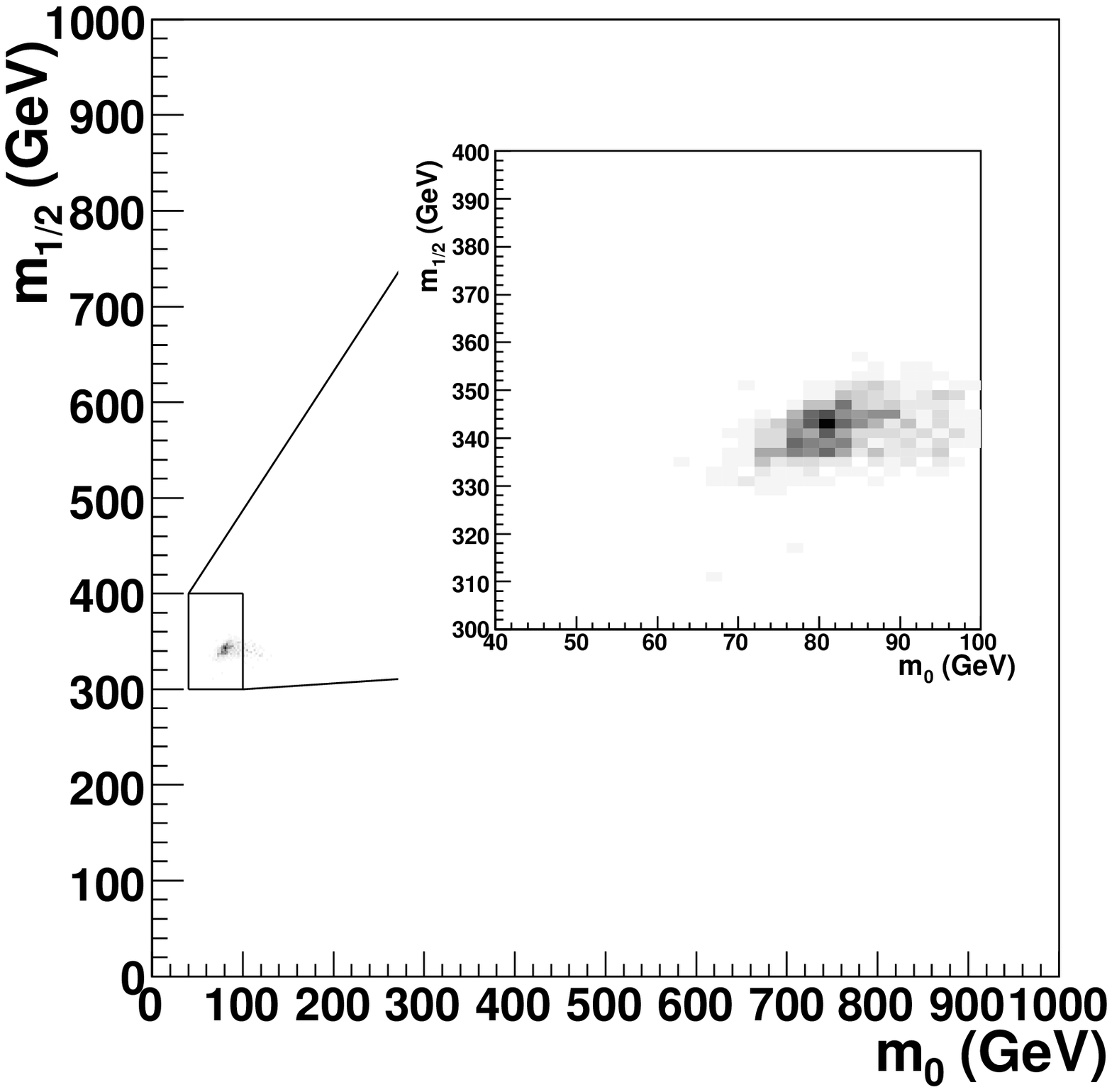}{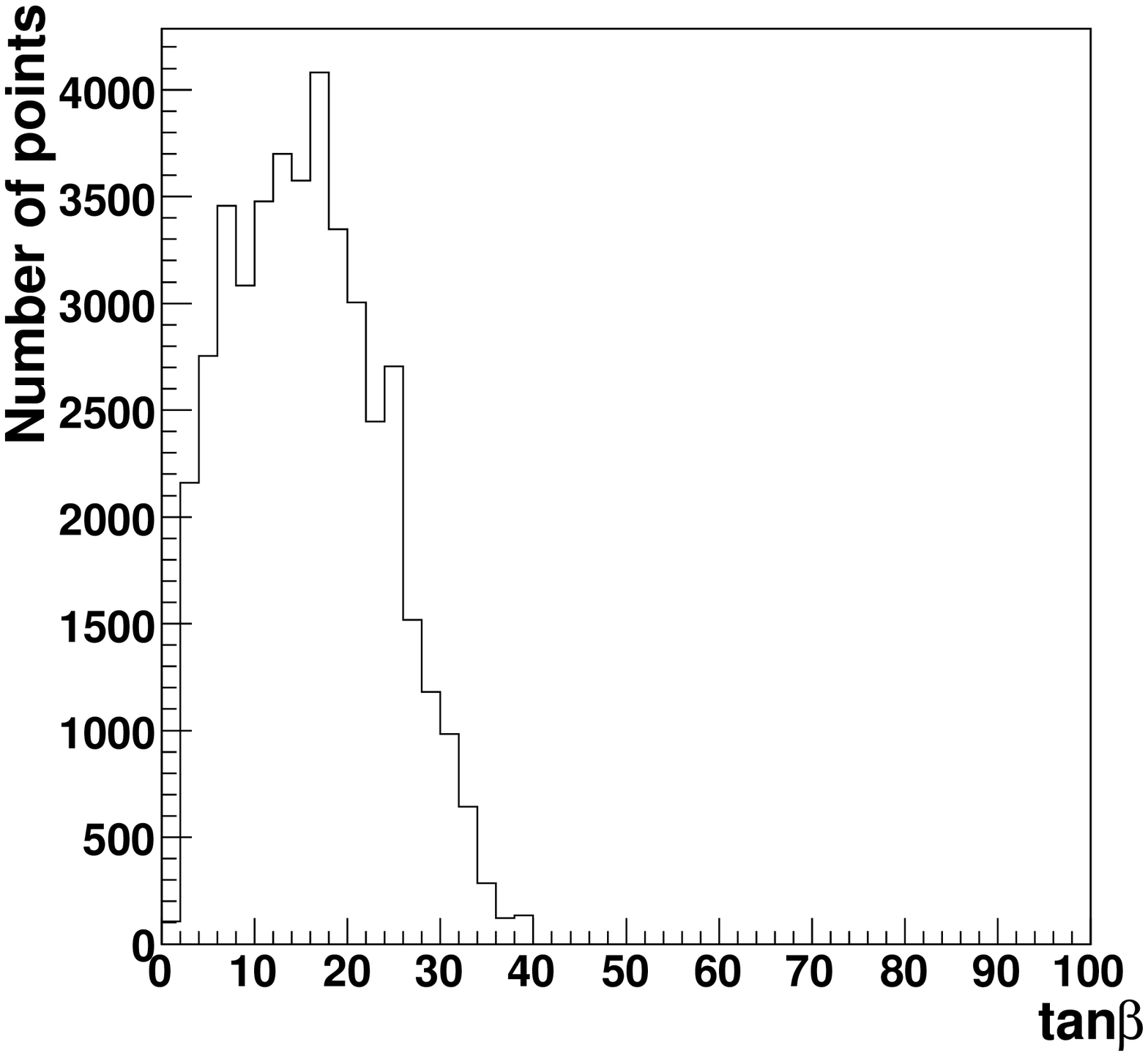}
\caption{The region of mSUGRA parameter space consistent with the
measurement of the cross-section of events with missing $p_T$ greater
than 500 GeV and with the endpoint measurements obtained in section~\ref{sec:lala2},
for negative $\mu$.}
\label{negativemu_final}
}

\FIGURE{
\sixgraphs
{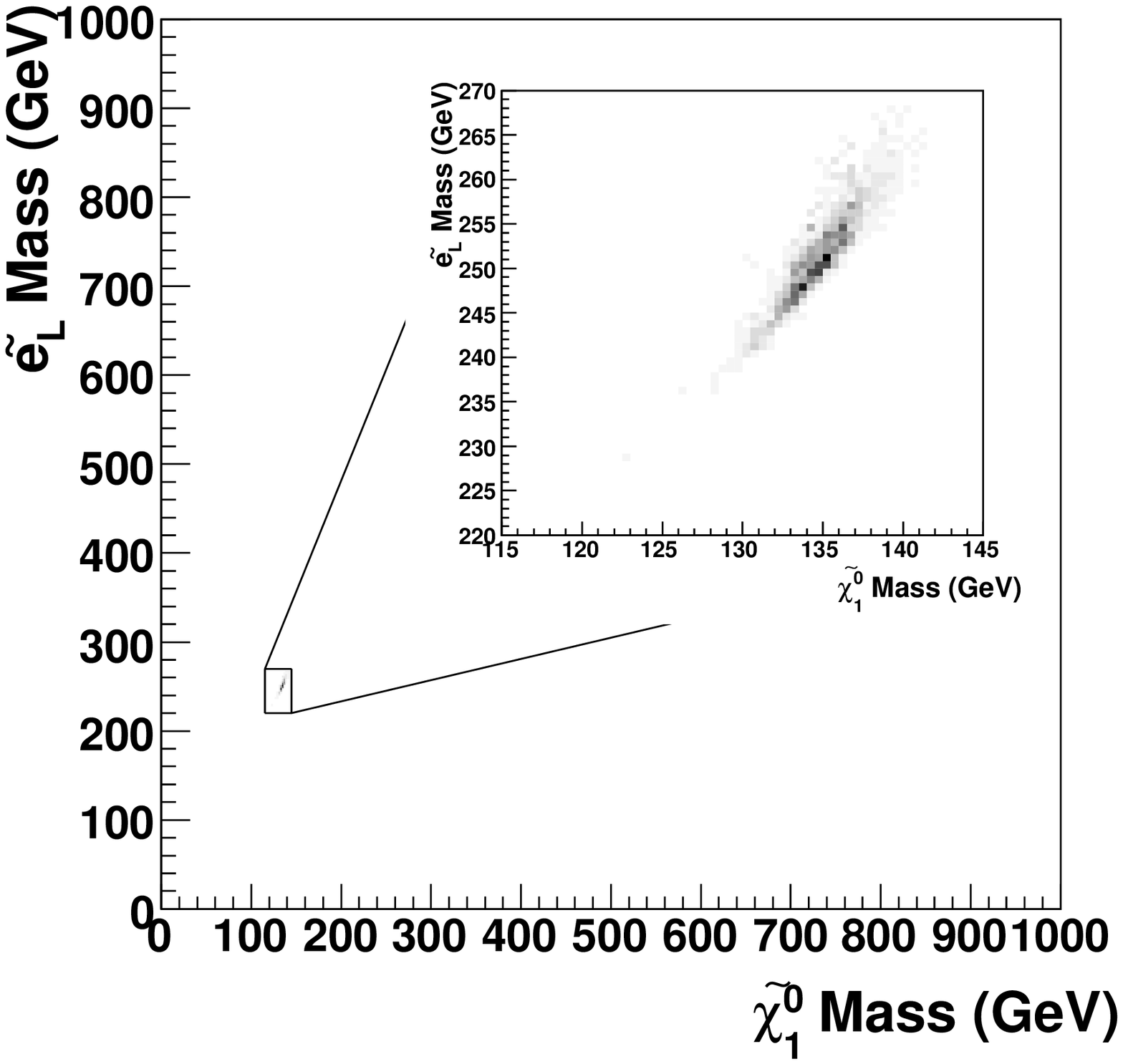}{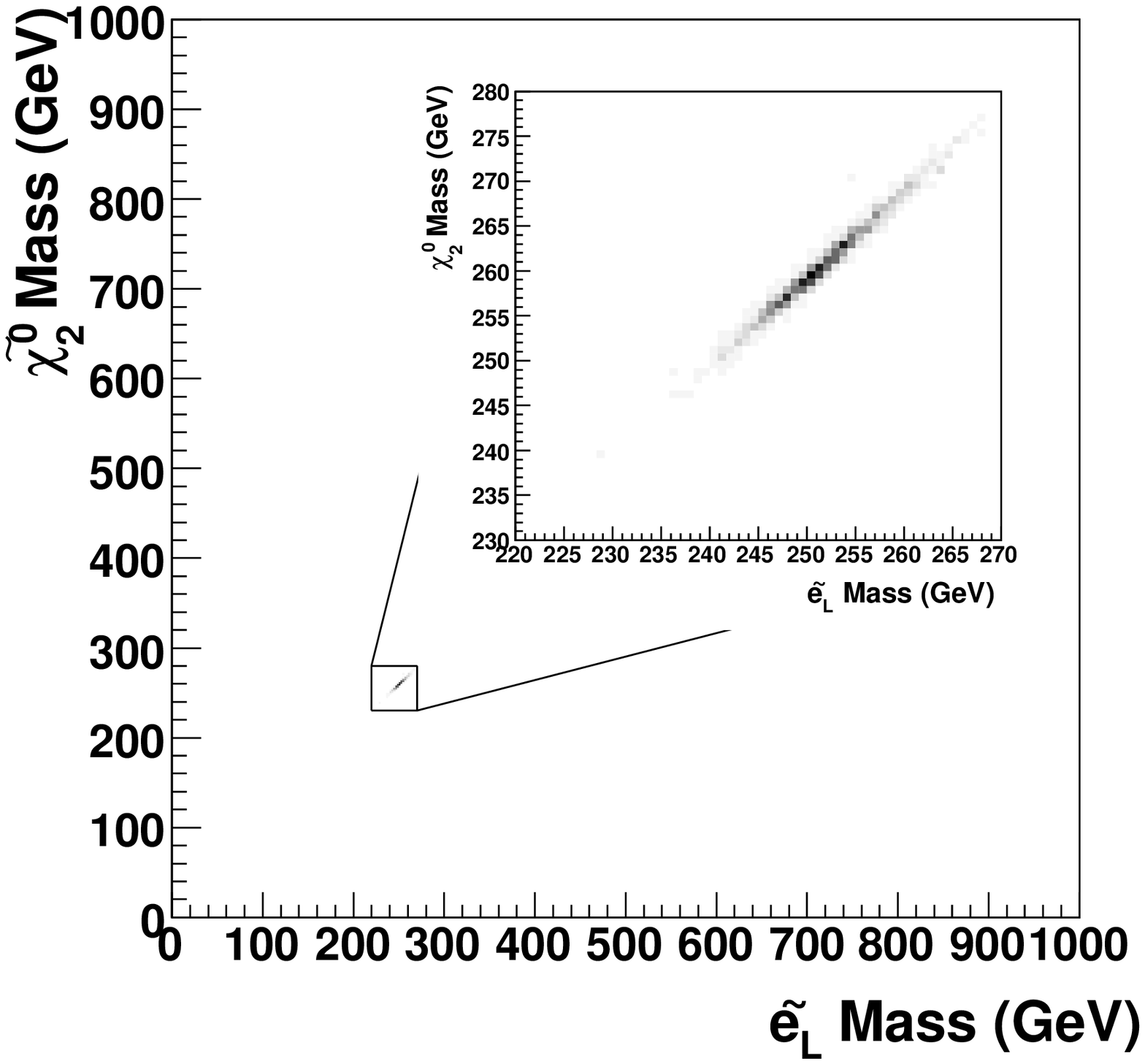}
{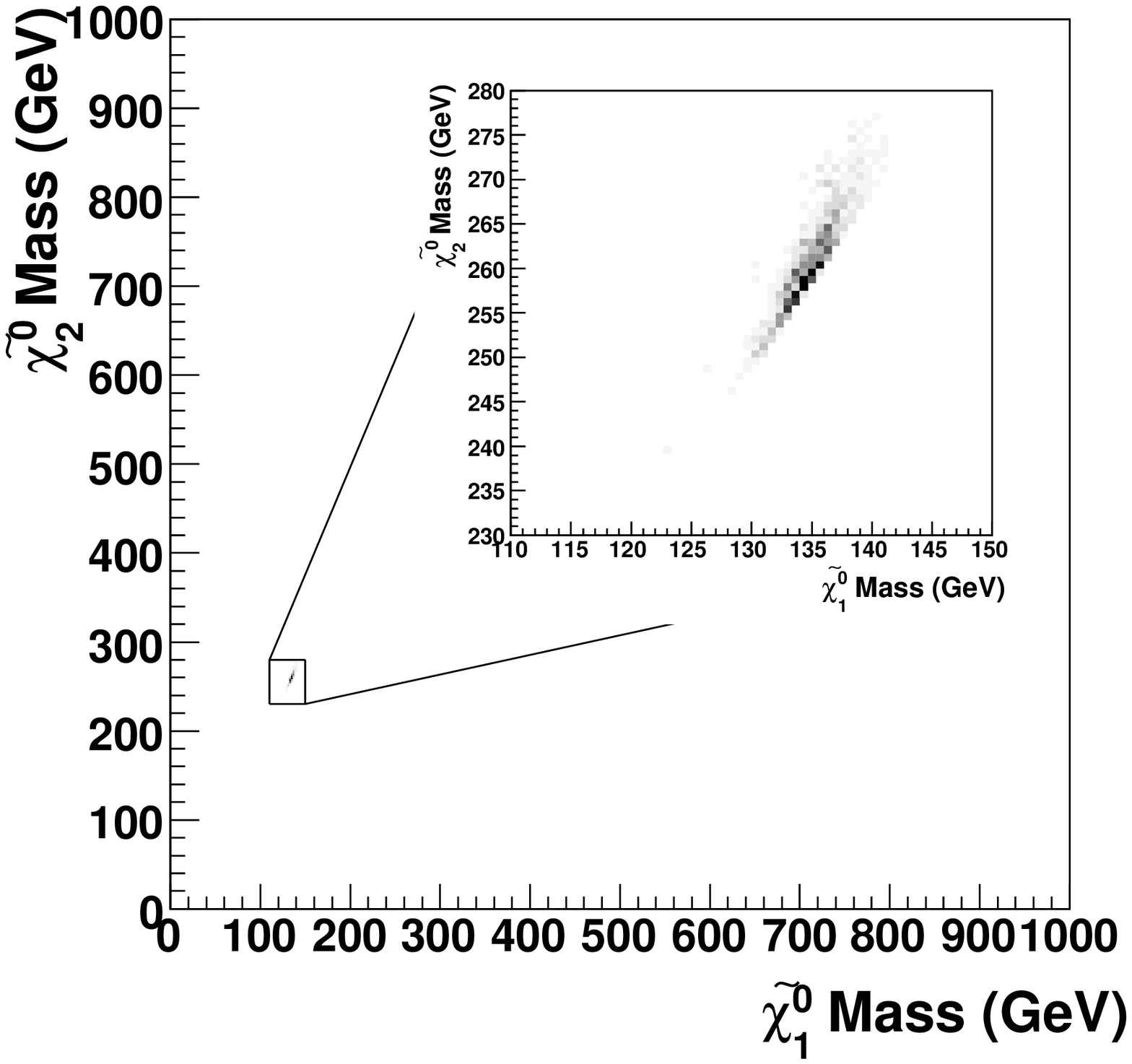}{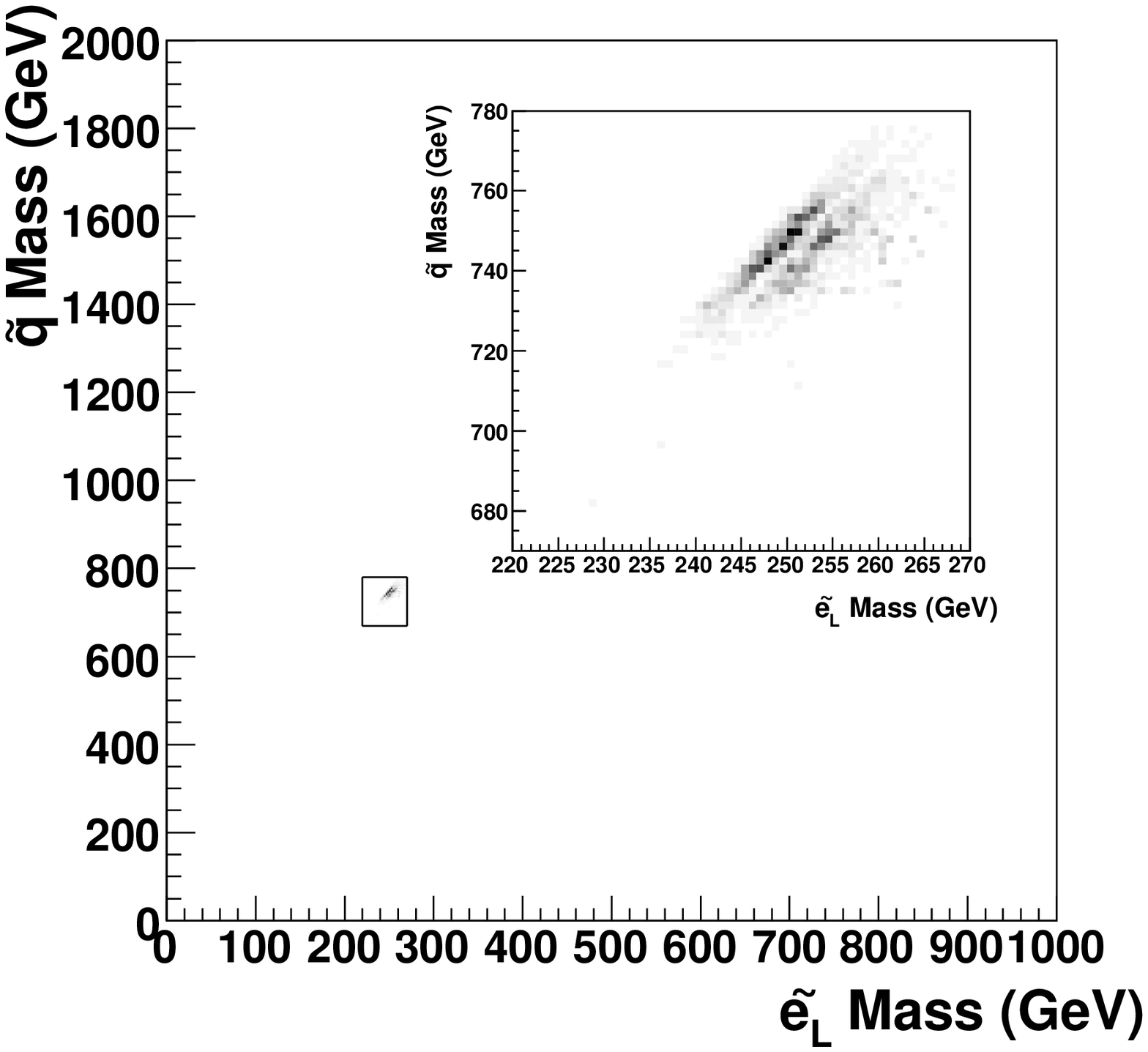}
{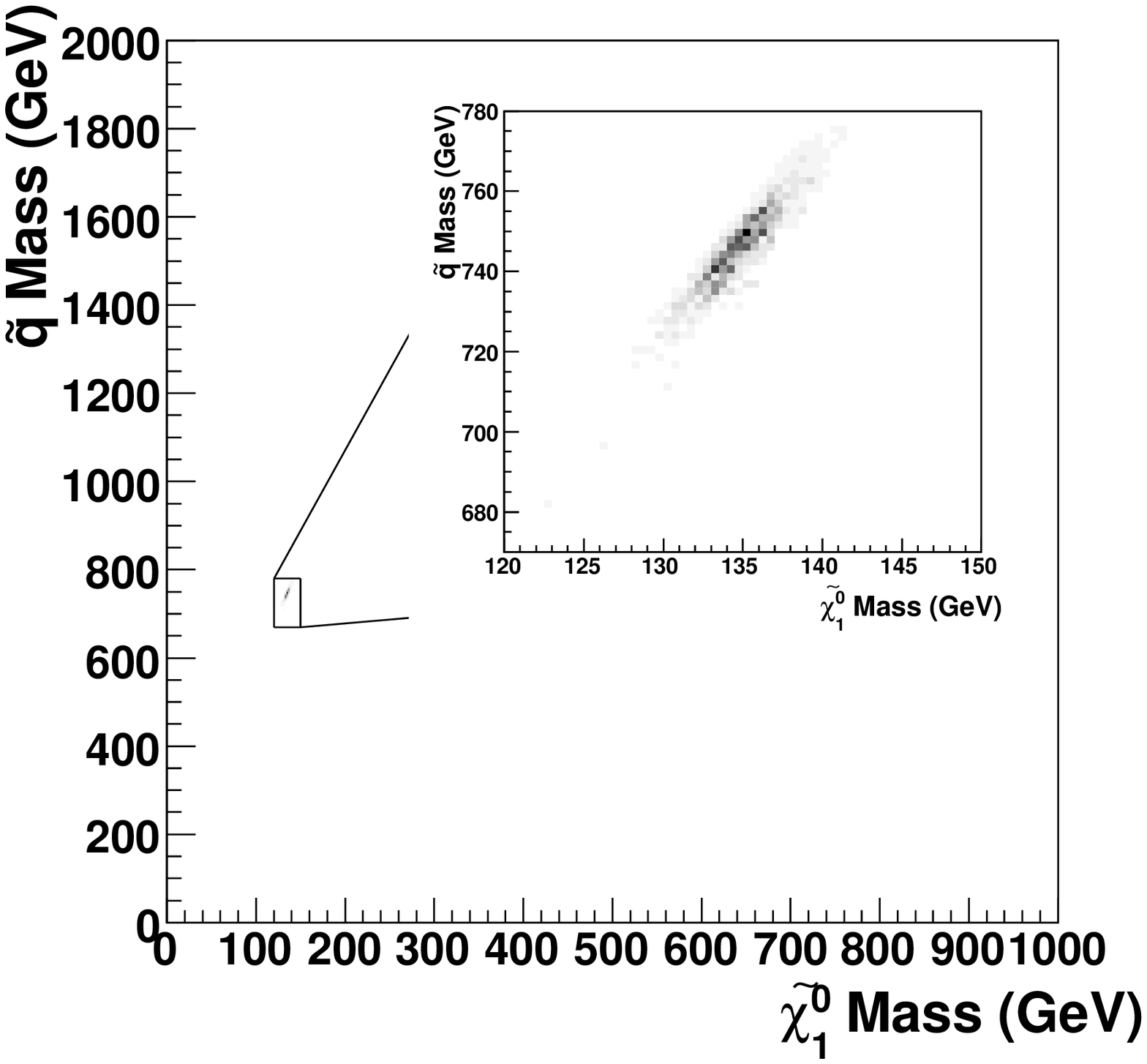}{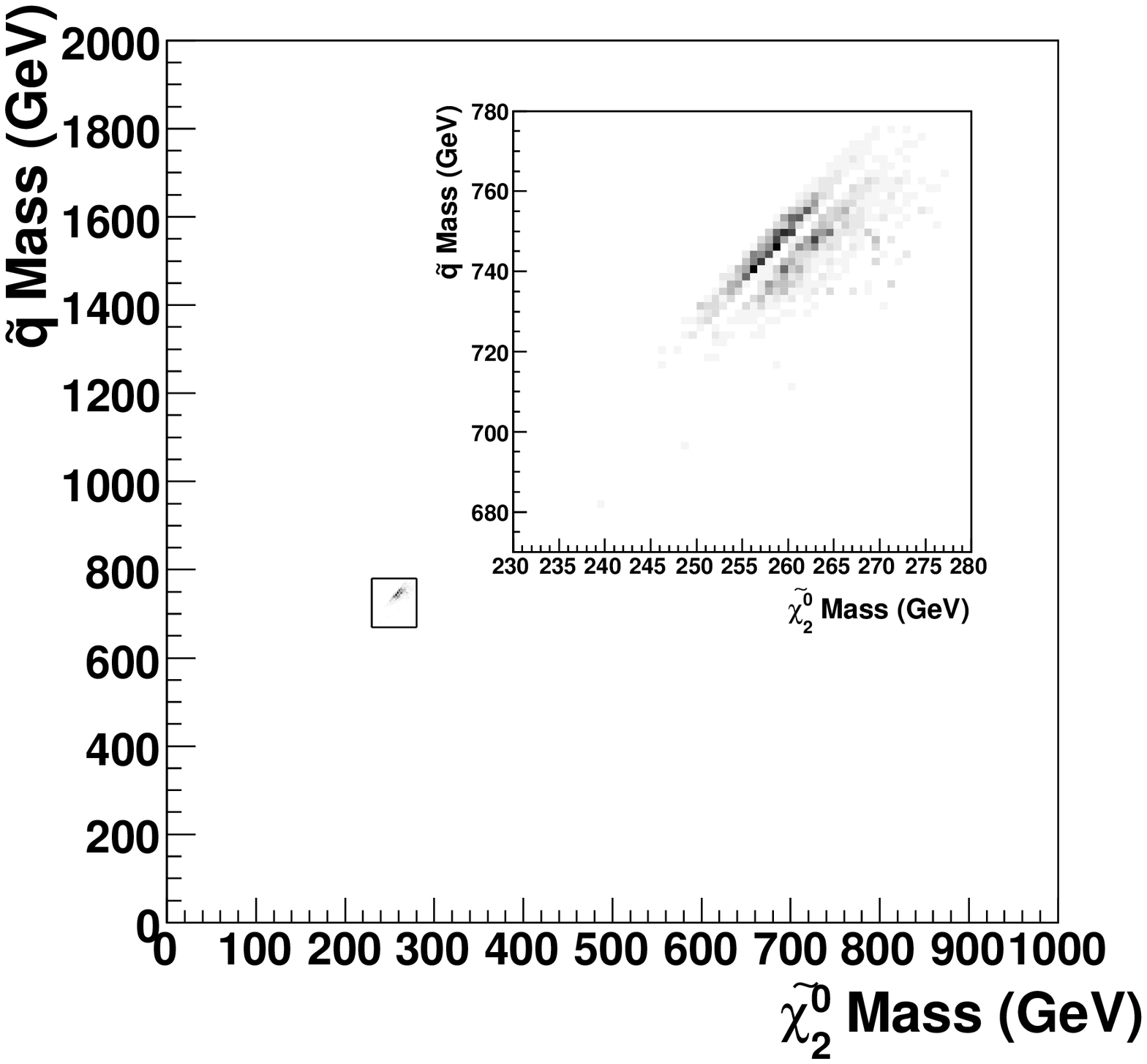}
\caption{The region of mass space consistent with a measurement at
10\% precision of the cross-section of events with missing $p_T$
greater than 500 GeV combined with the endpoints measured in
section~\ref{sec:lala2}, obtained using a Markov chain sampler in
mSUGRA space.}
\label{mass-final}
}

\section{Going beyond mSUGRA}
\label{sec:lala5}
\label{sec:genrlisingsec}
We have seen thus far that one can sample from the mSUGRA parameter
space using both kinematic endpoint data and a simple cross-section
measurement. Endpoint data alone gives more than adequate precision
within the framework of mSUGRA, provided one samples the mSUGRA
parameter space and assumes that one has identified the particles in
the decay chain correctly. The aim of this section is to generalise
this analysis to include both ambiguity in the decay chain and more
general assumptions about the mechanism of SUSY breaking. We will also
consider the effect of the jet energy scale error on the endpoint
positions, thus demonstrating how one would include correlated
experimental effects in our analysis.

\subsection{Effect of a jet energy scale error}
\label{sec:jetenscalesec}
Any detector such as ATLAS does not measure the energy of jets
perfectly, but instead has some energy scale error. Given that most of
the endpoints feature a quark jet, it is worth investigating the
effect of the energy scale error on the positions of the endpoints,
and the subsequent effect on our precision in the mSUGRA parameter
space.

Firstly, it is noted that for jets whose energy exceeds 70 GeV (the
likely energy of the jet in our endpoints given the relatively large
mass difference between the squarks and the neutralinos), the energy
scale error is expected to be of the order of 1 per cent
\cite{AtlasTDR}. This is much lower than the errors we have already
attributed to the endpoints that arise from mismeasurement, and hence
the effect will not cause a discernible difference to our results. We
have nevertheless included the effect in our analysis as an example of
how one can incorporate experimental effects in our analysis.

To accommodate the effect of an unknown shift $s$ in the absolute jet
\label{sec:absoluncerjetesc} energy scale, we add $s$ to the parameter
set explored by the sampler.  In other words, the sampler now wanders
around the extended space $Q = P \otimes S$ defined as the product of
the mSUGRA parameter space $P = \{\textbf{p}\}$ with the set $S$ of
possible values of $s$.  At each point $\textbf{q} = (\textbf{p}, s)
\in Q$ we work out the masses $\textbf{m}(\textbf{p})$ of the
particles in the decay chain.  We then calculate the ``idealised''
positions of the edges corresponding to these masses (as before) but
we then {\em move} the positions of these edges by the amount
predicted by the current hypothesis $s$ for the the absolute jet
energy scale correction.  The resulting modified edge positions
$\textbf{e}^{pred} = \textbf{e}^{pred}(\textbf{q}) =
\textbf{e}^{pred}(\textbf{m}(\textbf{p}), s)$, which now depend on
$s$, are the values which are used in the new version of
equation~(\ref{thingwithedgediffin}).

Having extended $P$ to the larger space $Q$, our goal is now to sample
not from $p(\textbf{p}|\textbf{e}^{obs})$ but from
$p(\textbf{q}|\textbf{e}^{obs})$.  The latter is proportional to
$p(\textbf{e}^{obs}|\textbf{q})p(\textbf{q})$.  The first term
$p(\textbf{e}^{obs}|\textbf{q})$ may be calculated almost exactly as
before in equation~(\ref{thingwithedgediffin}) but with the new
modified edge positions $\textbf{e}^{pred}(\textbf{m}(\textbf{p}), s)$
described above.  The last term $p(\textbf{q})$ may be decomposed by
independence into two parts: $p(\textbf{p})p(s)$.  The first of these,
$p(\textbf{p})$, is the mSUGRA-space prior which we have seen
before,\footnote{We must remember that, as in
earlier sections, the likelihood
$p(\textbf{e}^{obs}|\textbf{q})$ will be zero (given our model) at
points where the masses of the particles in the chain do not obey the
necessary mass hierarchy.  It was computationally easier for us to
place this veto into the prior $p(\textbf{p})$ as before.} while the
other, $p(s)$, is the expected distribution of the final uncertainty
in the absolute jet energy scale.  Following \cite{AtlasTDR} we take
$p(s)$ to be a Gaussian of width 1\%.

In order to determine the particular amounts $\delta_i$ by which the
$i^{th}$ endpoint should be shifted for a given jet energy scale
correction factor $s$, we run a toy Monte Carlo simulation at that
point and for that edge.\footnote{Strictly speaking the toy Monte
Carlo simulation is only needed for the $llq$ edge and the $llq$
threshold as the shifts in the edge positions for the {\em other}
edges are linear in $\sqrt{s}$ and may be calculated analytically.}
This is done once with and once without the correction factor $s$
multiplying the jet energies.  The positions of the endpoints are
compared in the two cases.  Different endpoints are thus shifted by
different fractions of the energy scale error $s$.

The results including uncertainty in the jet energy scale are shown in
figures~\ref{plusmu_shift} and \ref{minusmu_shift} for positive and
negative $\mu$ respectively and are comparable to those obtained
previously (figures~\ref{plusmu_final} and \ref{negativemu_final}) when
uncertainty in the jet energy scale was not considered.

\FIGURE{
\twographst{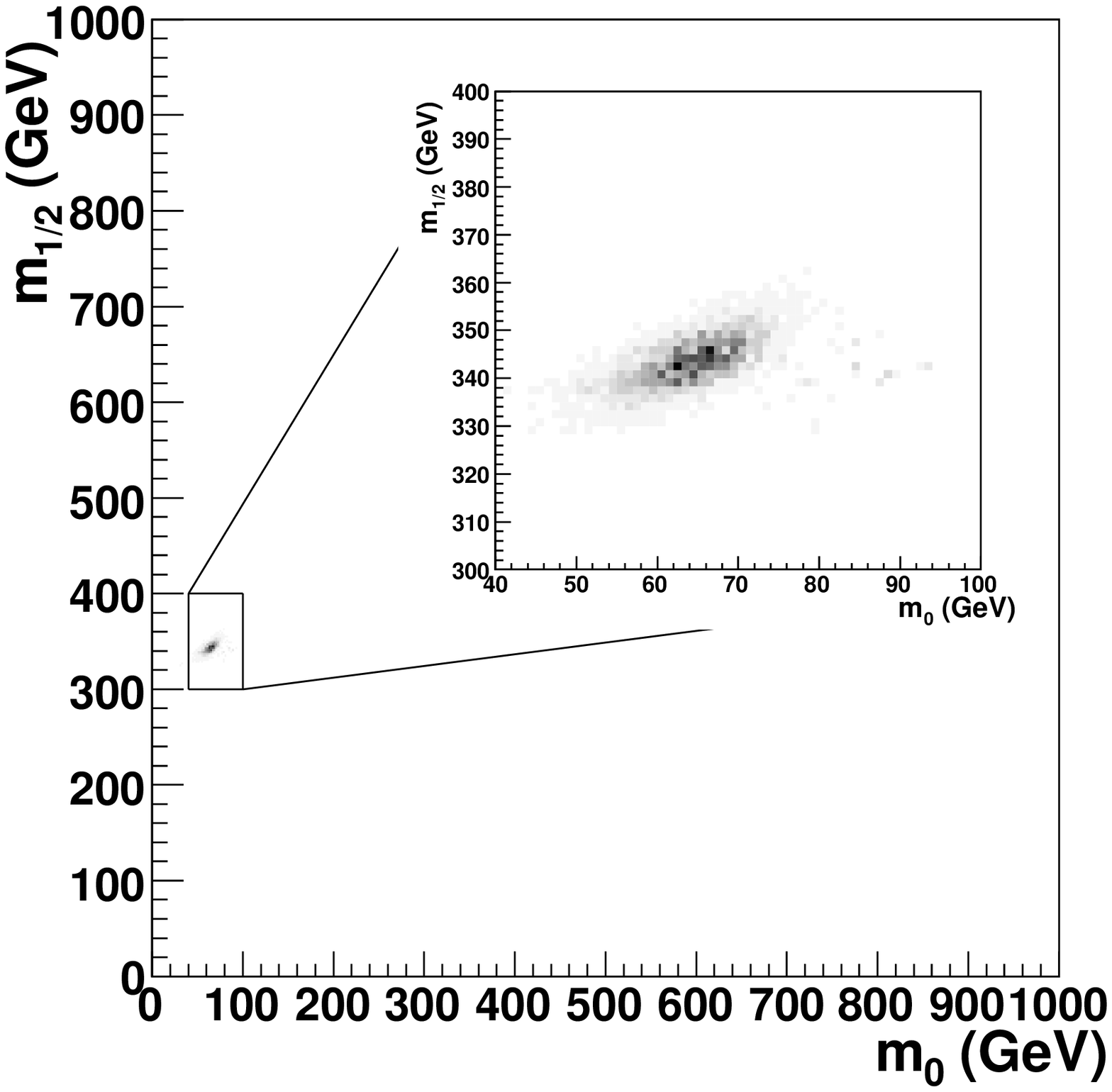}{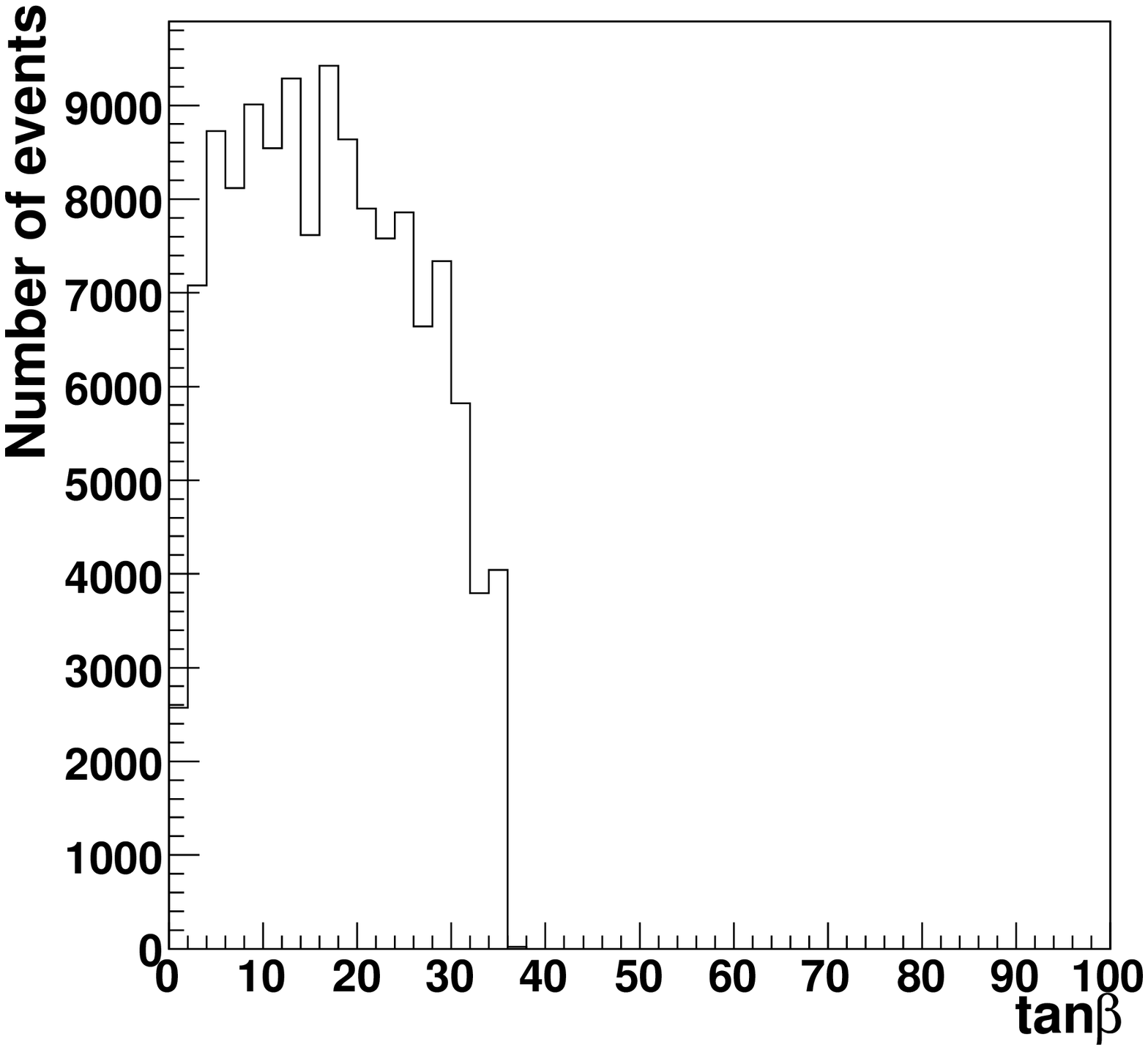}
\caption{The region of mSUGRA parameter space consistent with the
endpoint measurements obtained in section~\ref{sec:lala2}, for
positive $\mu$, with a 1 per cent jet energy scale error included.}
\label{plusmu_shift}
}

\FIGURE{
\twographst{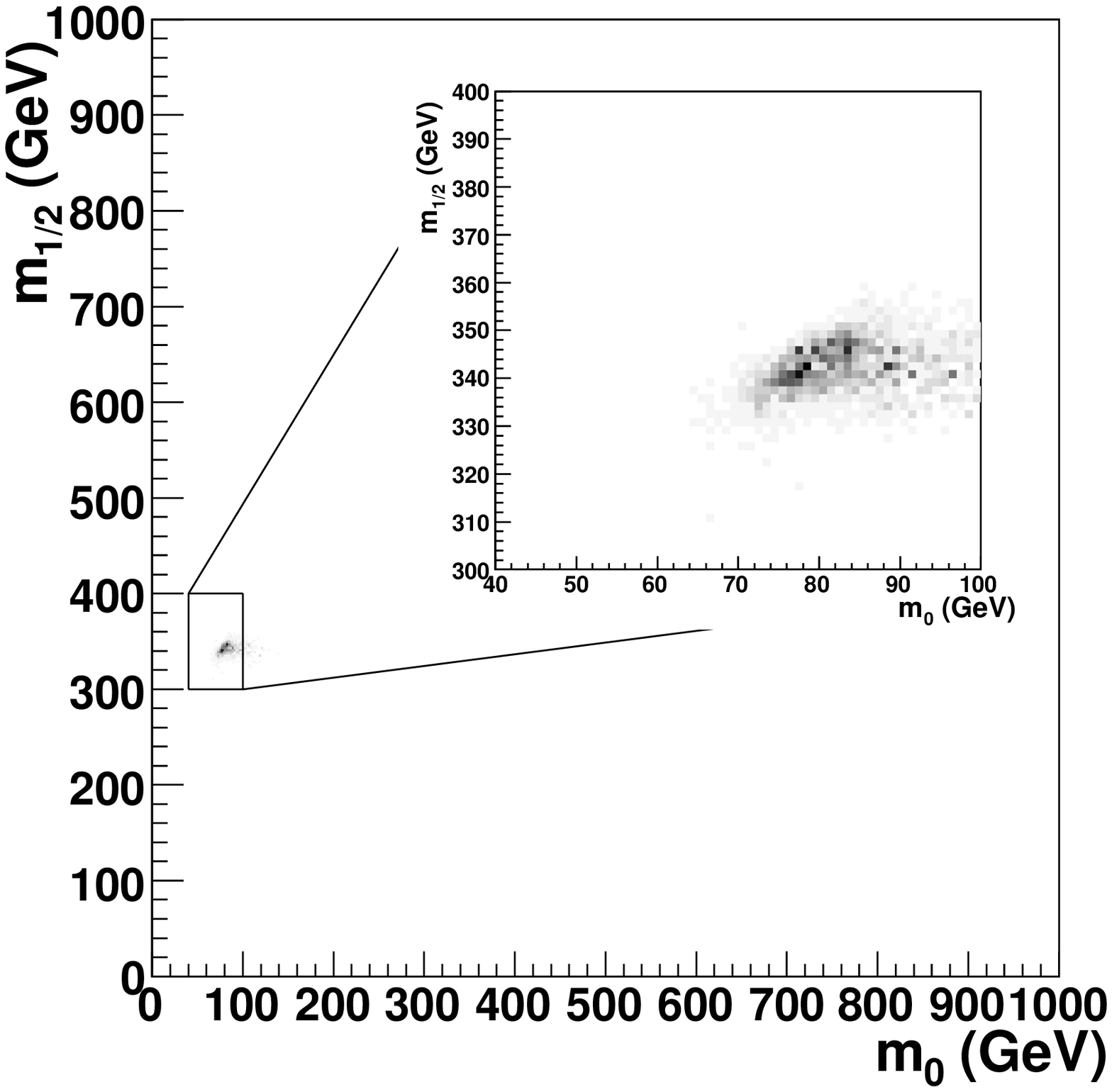}{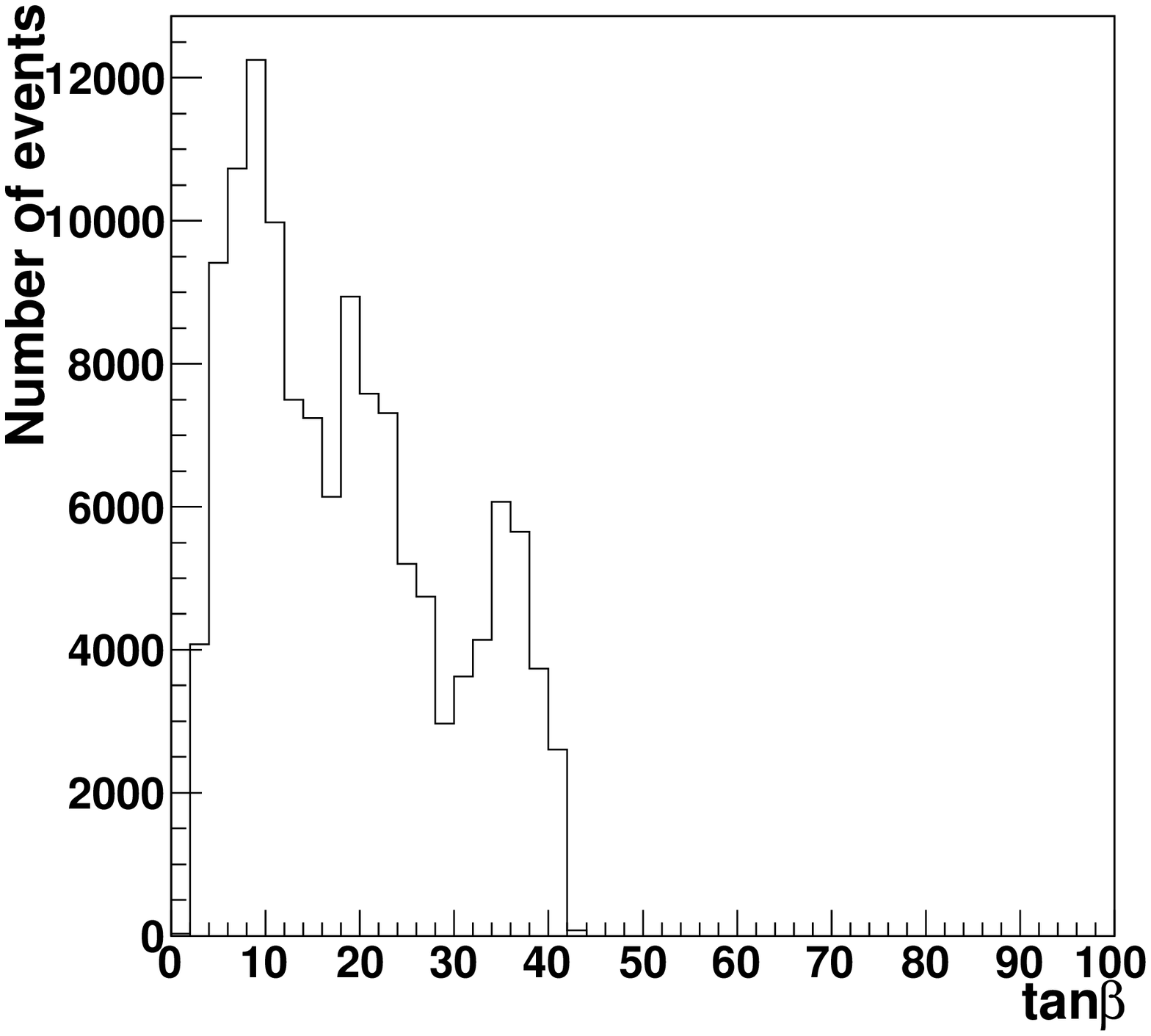}
\caption{The region of mSUGRA parameter space consistent with the
endpoint measurements obtained in section~\ref{sec:lala2}, for
negative $\mu$, with a 1 per cent jet energy scale error included.}
\label{minusmu_shift}
}

\subsection{Chain ambiguity in mSUGRA}
\TABLE{
\begin{tabular}{@{\extracolsep{-2mm}}|c|ccccccc|}
\hline
Name & \multicolumn{7}{c|}{Hieracrchy}\\
\hline 
$H_1$ & $\msq $&$ > $&$ \mnB $&$ > $&$ \mlL $&$ > $&$ \mnA $\\
$H_2$ & $\msq $&$ > $&$ \mnC $&$ > $&$ \mlL $&$ > $&$ \mnA $\\
$H_3$ & $\msq $&$ > $&$ \mnC $&$ > $&$ \mlL $&$ > $&$ \mnB $\\
$H_4$ & $\msq $&$ > $&$ \mnD $&$ > $&$ \mlL $&$ > $&$ \mnA $\\
$H_5$ & $\msq $&$ > $&$ \mnD $&$ > $&$ \mlL $&$ > $&$ \mnB $\\
$H_6$ & $\msq $&$ > $&$ \mnD $&$ > $&$ \mlL $&$ > $&$ \mnC $\\
$H_7$ & $\msq $&$ > $&$ \mnB $&$ > $&$ \mlR $&$ > $&$ \mnA $\\
$H_8$ & $\msq $&$ > $&$ \mnC $&$ > $&$ \mlR $&$ > $&$ \mnA $\\
$H_9$ & $\msq $&$ > $&$ \mnC $&$ > $&$ \mlR $&$ > $&$ \mnB $\\
$H_{10}$ & $\msq $&$ > $&$ \mnD $&$ > $&$ \mlR $&$ > $&$ \mnA $\\
$H_{11}$ & $\msq $&$ > $&$ \mnD $&$ > $&$ \mlR $&$ > $&$ \mnB $\\
$H_{12}$ & $\msq $&$ > $&$ \mnD $&$ > $&$ \mlR $&$ > $&$ \mnC $\\
\hline
\end{tabular}
\caption{The twelve mass hierarchies considered in Section~\ref{sec:hierarchysection}.}
\label{tab:hierarchies}
}

\label{sec:hierarchysection}
In order to investigate the effect of chain ambiguity on the mSUGRA
parameter space, the edge data from section~\ref{sec:lala2} are here
used in an mSUGRA fit {\em without} the assumption that the particles
in the decay chain have been identified correctly. It is still true
that there are few processes that can give the characteristic
endpoints associated with the squark cascade decay already described,
and it should be sufficient merely to include the possibility that any
of the neutralinos may be produced in the decay (provided of course
that the one further down the chain is lighter than that above it) and
that one has ambiguity over the slepton chirality. This gives twelve
possible mass hierarchies (see Table~\ref{tab:hierarchies}) and each
of these gives a series of possible endpoints in the mass spectra.
The issue of how to deal with parts of parameter space able to
generate the same final state through three- rather than two-body
decays (for example when the sleptons are too massive to produce
directly) is beyond the scope of this document but is ideal for
further study.

There can easily be points in parameter space at which almost all
sparticle production goes through one particular hierarchy (say
$H_1$), but in which a different hierarchy (say $H_2$) has
end-point locations which are a much better fit to the {\em positions}
of the ``observed'' edges.  This could be true even if the cross
section for $H_2$ was much less than for $H_1$.  Events from $H_2$
might not even be observable.  It is very costly to accurately
determine the observability (after realistic detector cuts and
consideration of backgrounds) of each of the hierarchies in
Table~\ref{tab:hierarchies} at every point in parameter space visited
by the Markov Chain.  For this reason, in this article we adopt the
following conservative position.  We choose not to consider the
(un)observability of end points at different points in parameter
space.  Instead we assume that {\em every} hierarchy consistent with
the masses of a given point is potentially visible.  This assumption
is conservative because in reality only a few hierarchies will be
visible.  The consequence of our assumption is that we will not reject
points of parameter space that a more in-depth analysis might be able
to reject.  It would be interesting for further work to pursue the
possibility of making stronger statements at each point in parameter
space based not only on the positions of the observed edges, but also
based on the number of event in them, and the number of events in
distributions which were {\em not} observed to have edges etc. How to
cope with points at which heavy sleptons force three- rather than
two-body neutralino decays should also be investigated.  In depth
analyses of this kind are beyond the scope of this paper, however.

If we label the $N_a$ different mass assignments with a tag $a_i$, the
likelihood for the $i$-th observed edge at each point $\textbf{p}$
in the mSUGRA parameter space $P$ now becomes:

\begin{eqnarray}
p(e_i^{obs}|\textbf{p}) & = & \sum^{N_a}_{j=1}{p(e_i^{obs}|\textbf{p},a_j)p(a_j)}  \nonumber \\
& = & \sum^{N_a}_{j=1}{p(e_i^{obs}|\textbf{m}_{a_j}(\textbf{p}))p(a_j)} \label{ambig}
\end{eqnarray}
where $p(a_i)$ is the prior for the mass assignments, and $N_a$ gives
the number of assignments open at that point in parameter space. If we
assume that each of the assignments is equally likely, the prior
$p(a_i)$ is simply $1/N_a$. The term
$p(e_i^{obs}|\textbf{m}_{a_i}(\textbf{p}))$ is calculated using
equation~(\ref{pof}) with the masses corresponding to the assignment
$a_i$.

Equation~(\ref{ambig}) makes the conservative assumption that any
observed edge could have come from any observed chain (i.e.\ not
necessarily from the same chain as that generating a different
observed edge).  Furthermore (but less realistically) it assumes that
there is no correlation between the chains generating each of the
\label{sec:whatasillnameforasec} edges, whereas in many parts of
parameter space it is highly likely that there is only one dominant
chain.  It is thus arguable that equation~(\ref{ambig}) should be
replaced by the stronger statement
\begin{eqnarray}
p(\textbf{e}^{obs}|\textbf{p}) & = & \sum^{N_a}_{j=1}{p(\textbf{e}^{obs}|\textbf{p},a_j)p(a_j)} \nonumber \\
& = & \sum^{N_a}_{j=1}{p(\textbf{e}^{obs}|\textbf{m}_{a_j}(\textbf{p}))p(a_j)} \label{ambiglessconservative}
\end{eqnarray}
which says that {\em all} the observed edges were the result of the
same (albeit unknown and unidentified) chain of sparticles.  We choose
to present results using (\ref{ambig}) rather than
(\ref{ambiglessconservative}).

The results for positive $\mu$ are seen in
figure~\ref{plusmu_edge_ambg}, whilst those for negative $\mu$ are in
figure~\ref{minusmu_edge_ambg}. The precision is worse than that
encountered previously, but not by much. It may be seen that there are
two favoured regions in each plot, rather than the single region
encountered previously.  The region at larger $m_0$ is one in which
hierarchy $H_1$ dominates the sum (\ref{ambig}).  The lower $m_0$
region has (\ref{ambig}) dominated by hierarchy $H_7$ in which the
right-slepton is substituted for the left-slepton.

The next course of action is to view the regions in the weak scale
mass space that correspond to the chosen mSUGRA points, and here we
have a problem. Since we are now assuming that we do not know exactly
which particles are in the decay chain, we can no longer take the
points in the mSUGRA plane and claim that they give us the masses of
the lightest two neutralinos and the left handed slepton. Instead, we
can merely say that we have measured a neutralino-like object and a
slepton-like object, but that we need some more facts before we can
say anything more.

We can, however, use some other information to tell us more about the
particles in the decay chain. For a start, we can look at the width of
the distribution for each mass (neutralino 1, neutralino 2, etc) that
results from the mSUGRA points and use these widths as a qualitative
guide. If the endpoints are really caused by a single mass hierarchy,
the masses in this chain should generally fit the data better than
other hierarchies, and this will manifest itself in a smaller spread
of masses for the masses involved in the correct hierarchy. In our
case, the endpoints should all be caused by a decay chain featuring
the lightest two neutralinos and the left handed slepton, so we expect
these masses to have narrower distributions. This is indeed the case
for the neutralinos, as seen in figure~\ref{mass-spread}, though the
selectron results are less different.

Note that figure~\ref{mass-spread} does not yet show mass
measurements.  The plots could only be interpreted as mass
measurements if further work were able to establish the identities of
the particles involved and confirm that they came predominantly from
just one chain.  Here we show them only to help get a hold on which
decay chains appear to be consistent with the results.

There are other things that can be done. Having had our attention
drawn to a small region of the mSUGRA parameter space, we can look
within that region at the branching ratios for the different possible
mass hierarchies, after which we might find that there are not enough
events of a given type to be consistent with the observed
endpoints. Therefore, although a decay chain featuring a neutralino 3
and neutralino 2 may fit a given endpoint slightly better than the
correct chain, it might be impossible for that chain to produce an
endpoint with the same number of events present as has been
observed. This, in conjunction with the width of the mass
distributions, might be enough to confirm the nature of the true decay
chain, but it would be foolish to assume that the true chain will
always be easy to identify.

Given that the region in mSUGRA space has not substantially increased
in size, we will not add the cross-section information at this
stage. Instead, we will investigate the effect of relaxing some of the
assumptions of the mSUGRA model.

\FIGURE{
\twographst{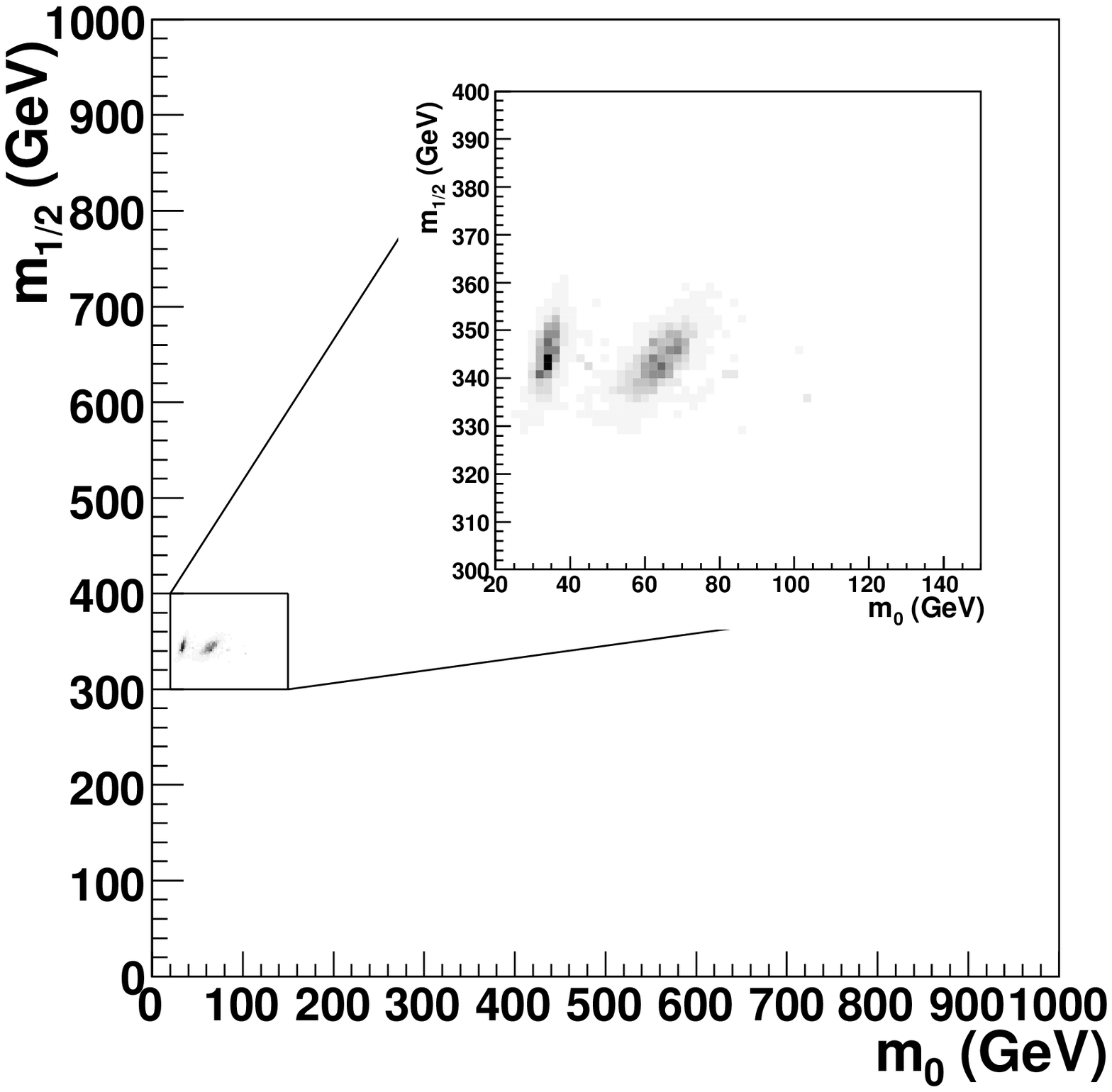}{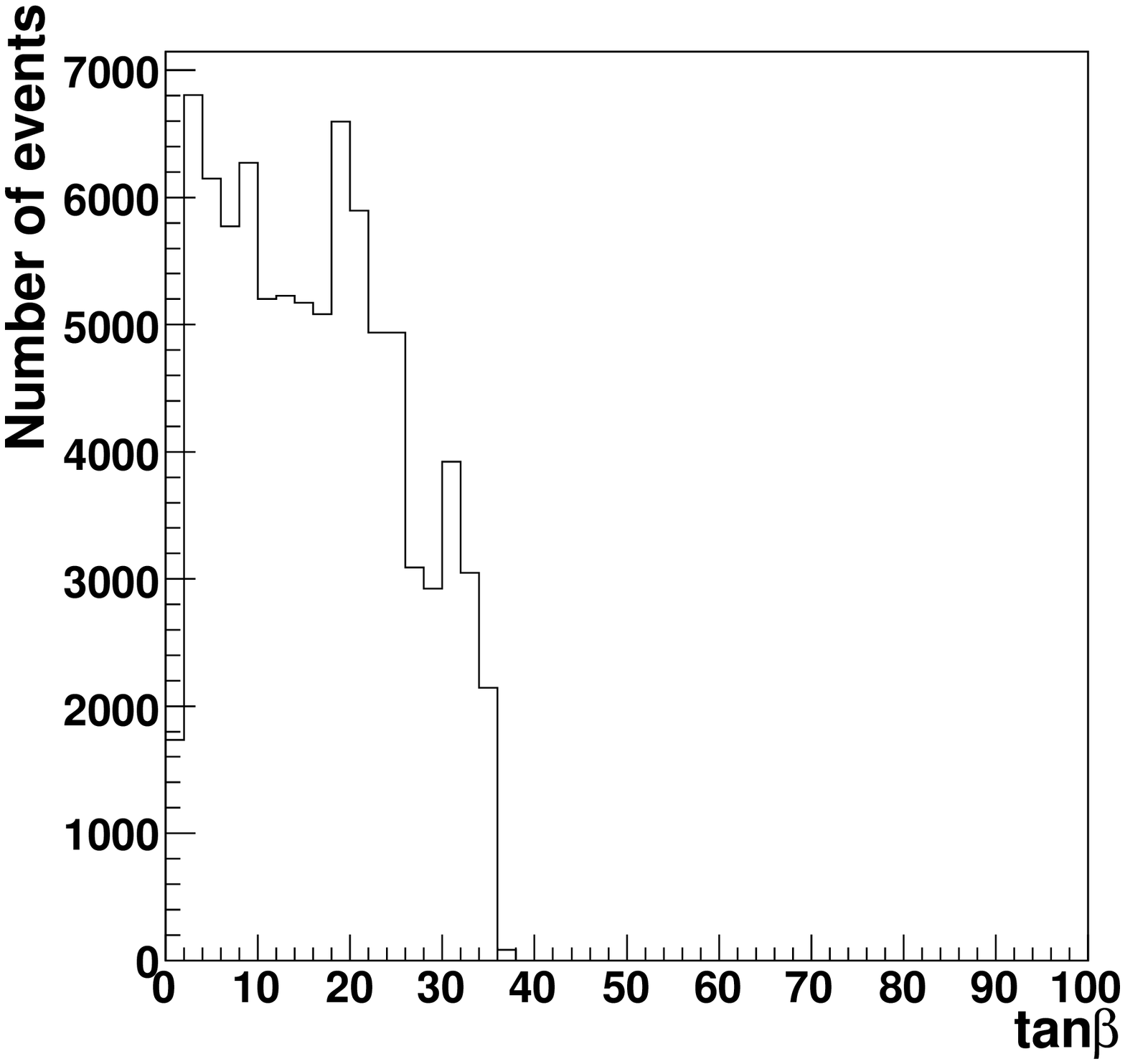}
\caption{The region of mSUGRA parameter space consistent with the
endpoint measurements of section~\ref{sec:lala2}, without the
assumption that the neutralinos and slepton in the squark decay chain
have been correctly identified. For full details, see text. Results
are shown for positive $\mu$.}
\label{plusmu_edge_ambg}
}

\FIGURE{
\twographst{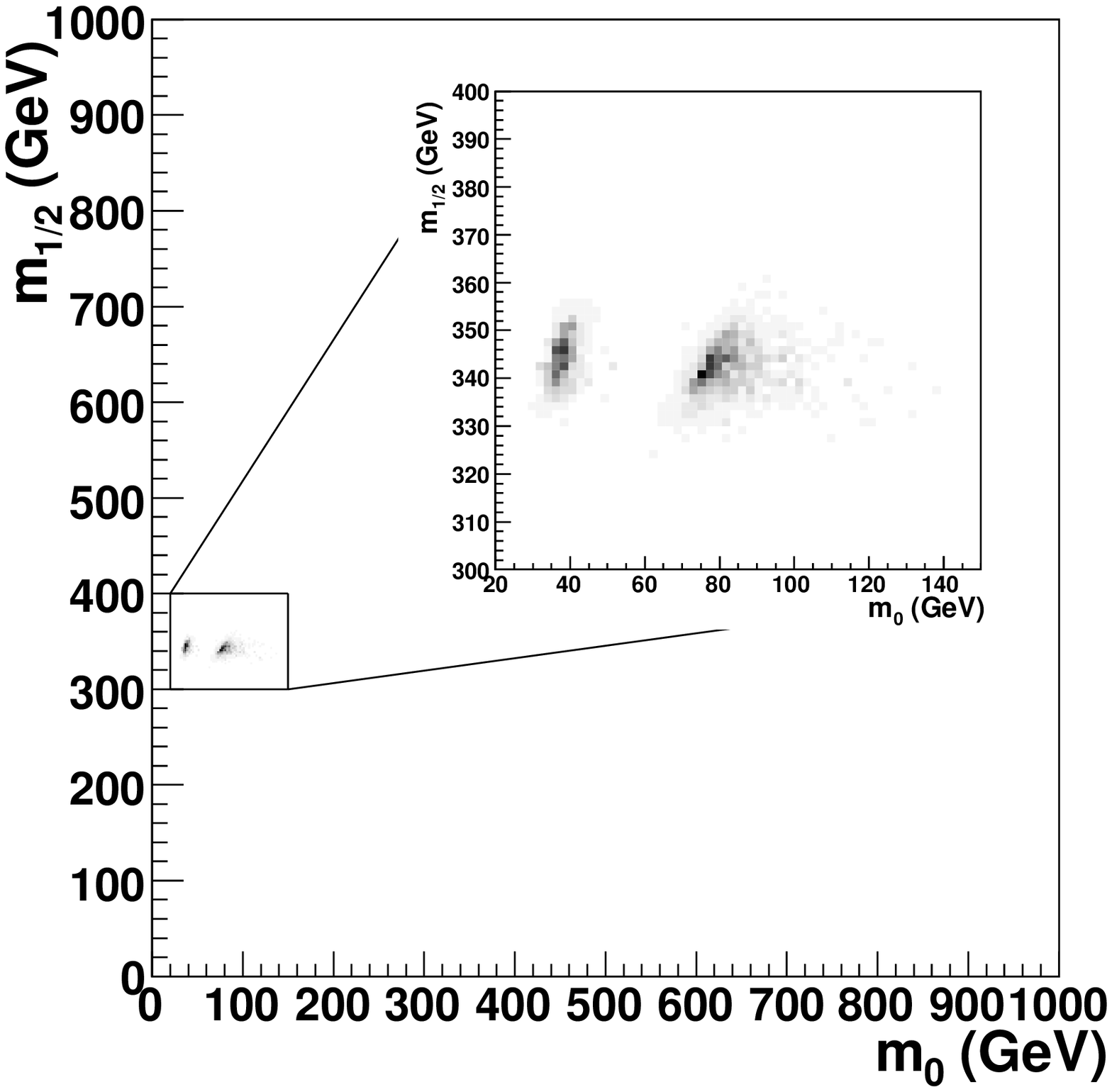}{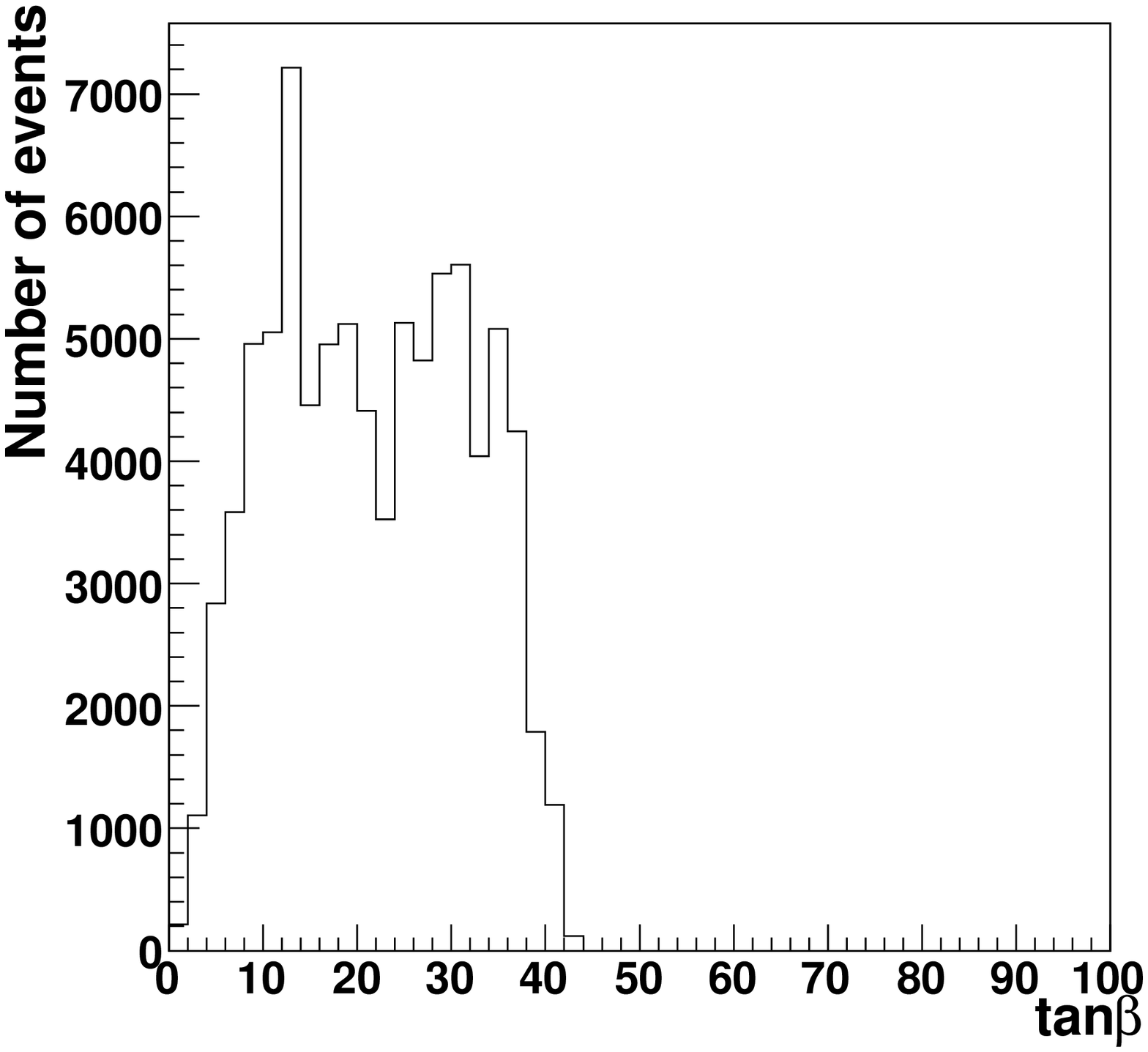}
\caption{The region of mSUGRA parameter space consistent with the
endpoint measurements of section~\ref{sec:lala2}, without the
assumption that the neutralinos and slepton in the squark decay chain
have been correctly identified. For full details, see text. Results
are shown for negative $\mu$.}
\label{minusmu_edge_ambg}
}

\FIGURE{
\sixgraphs{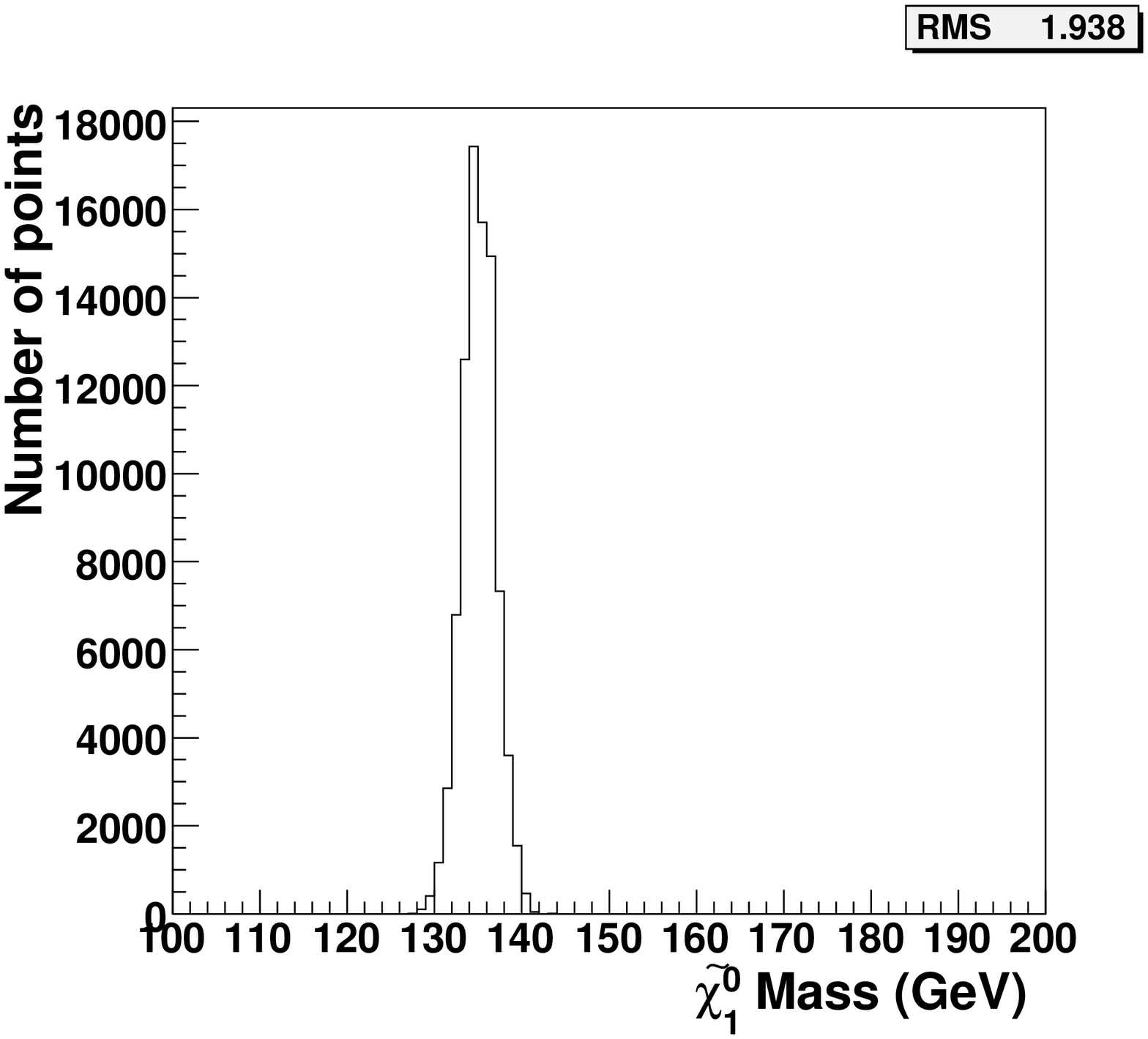}{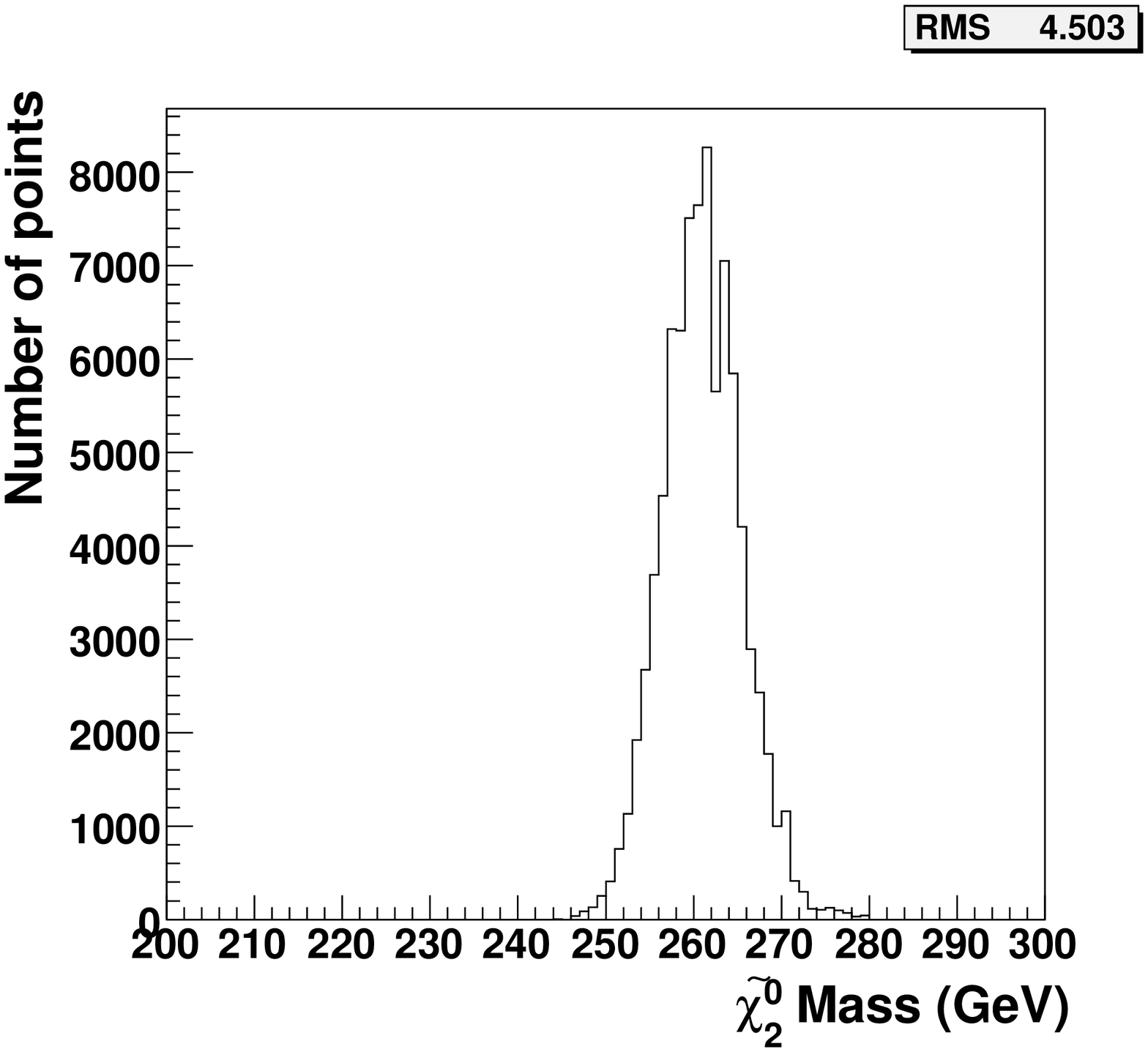}{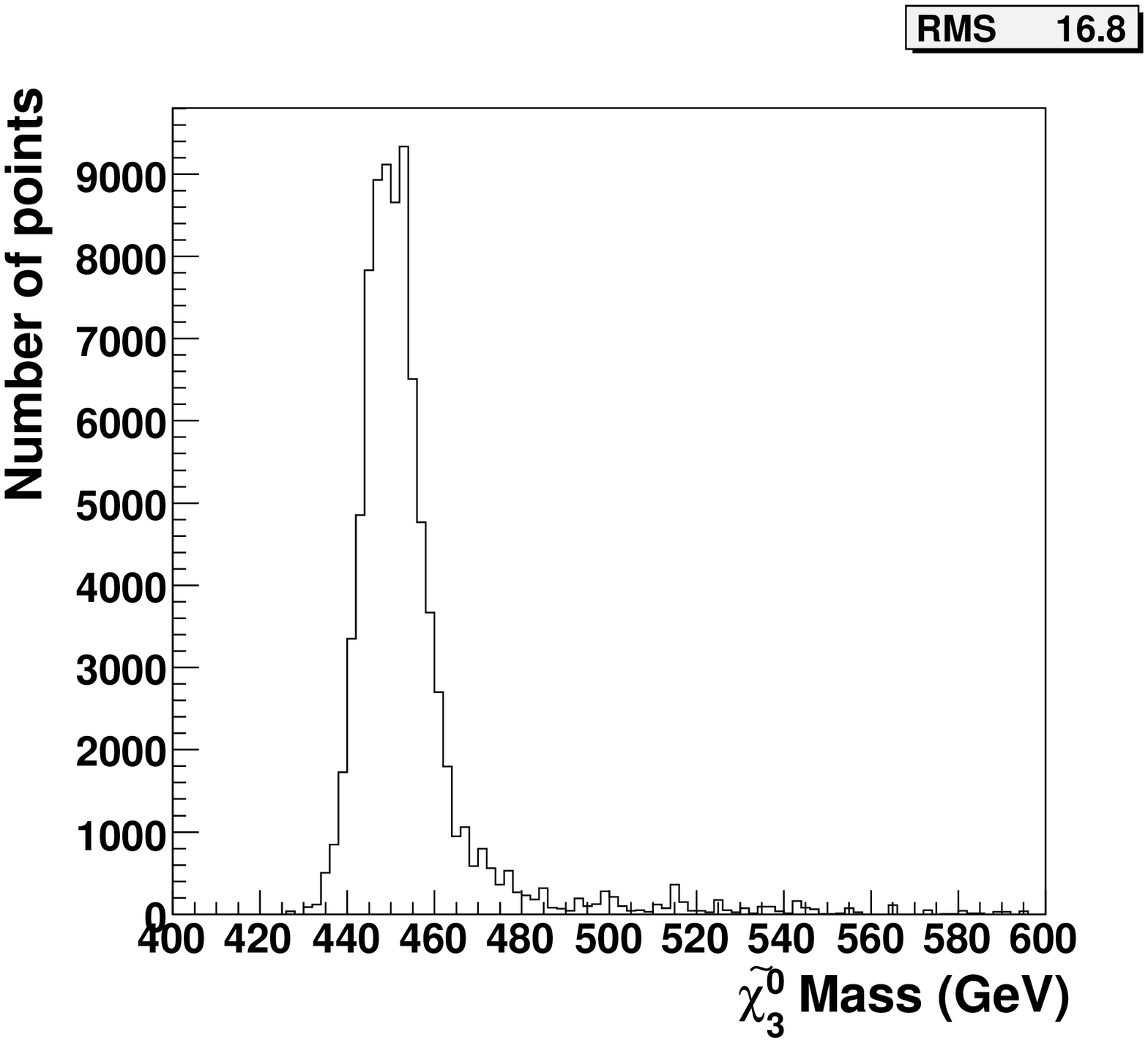}{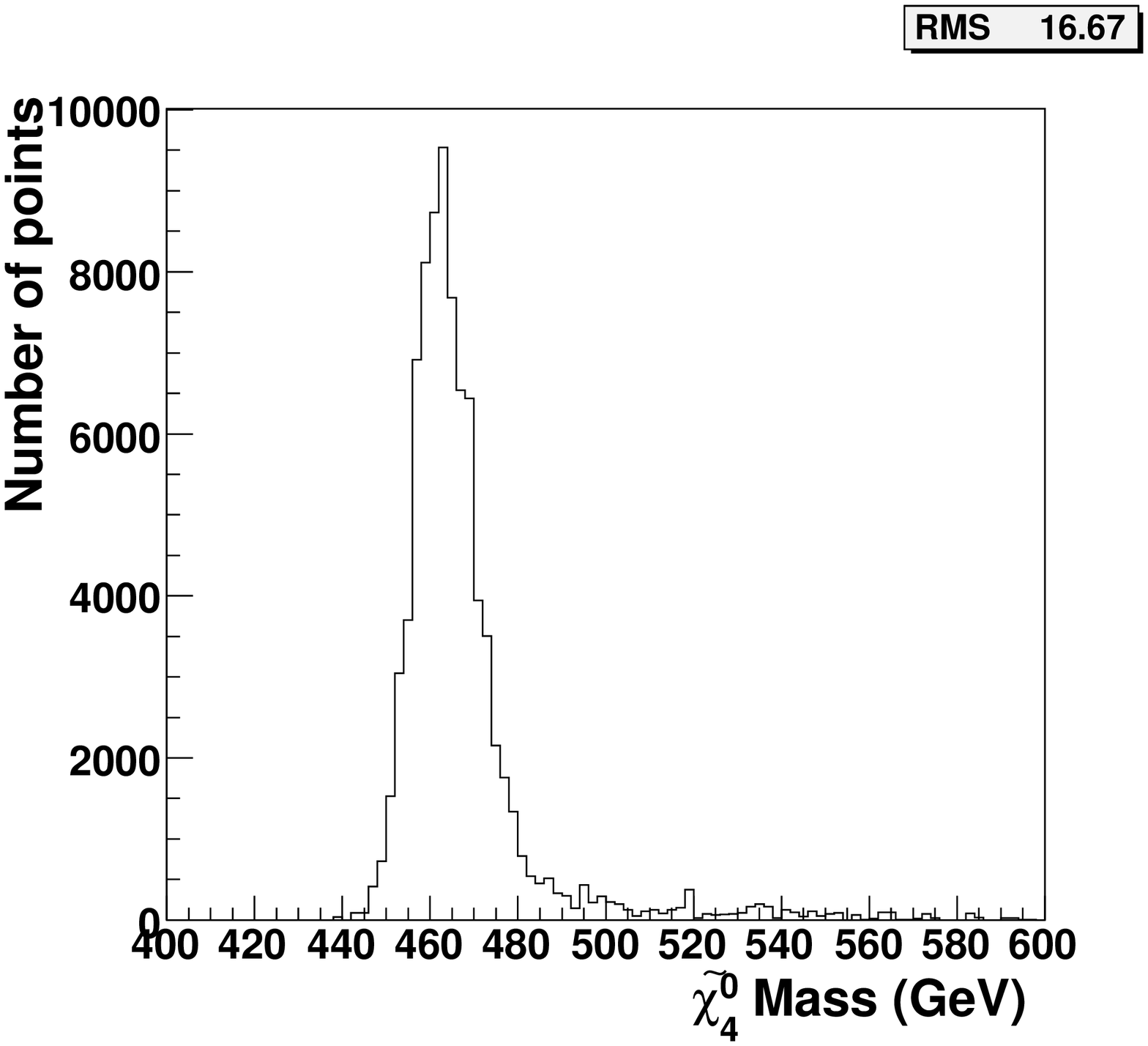}{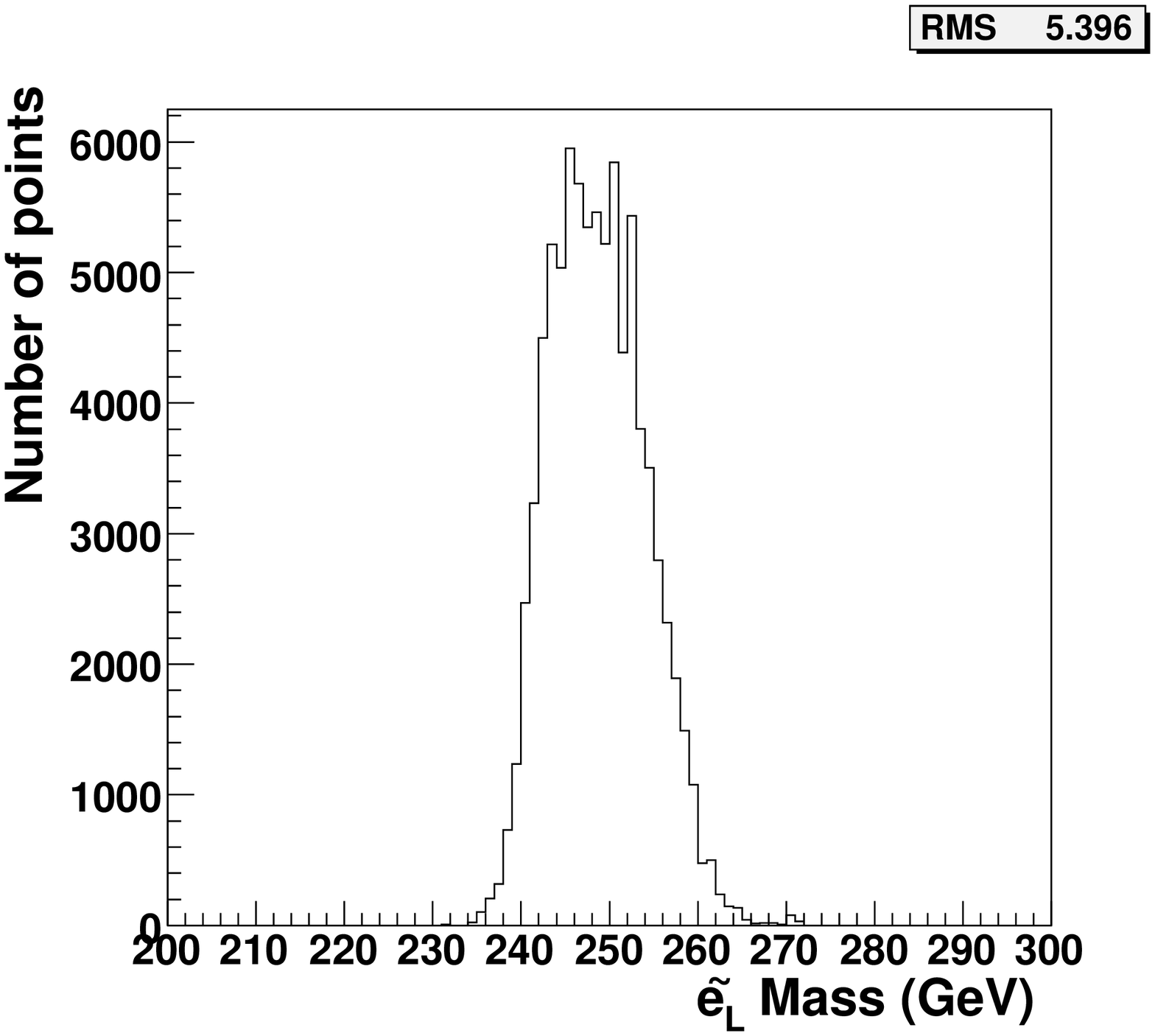}{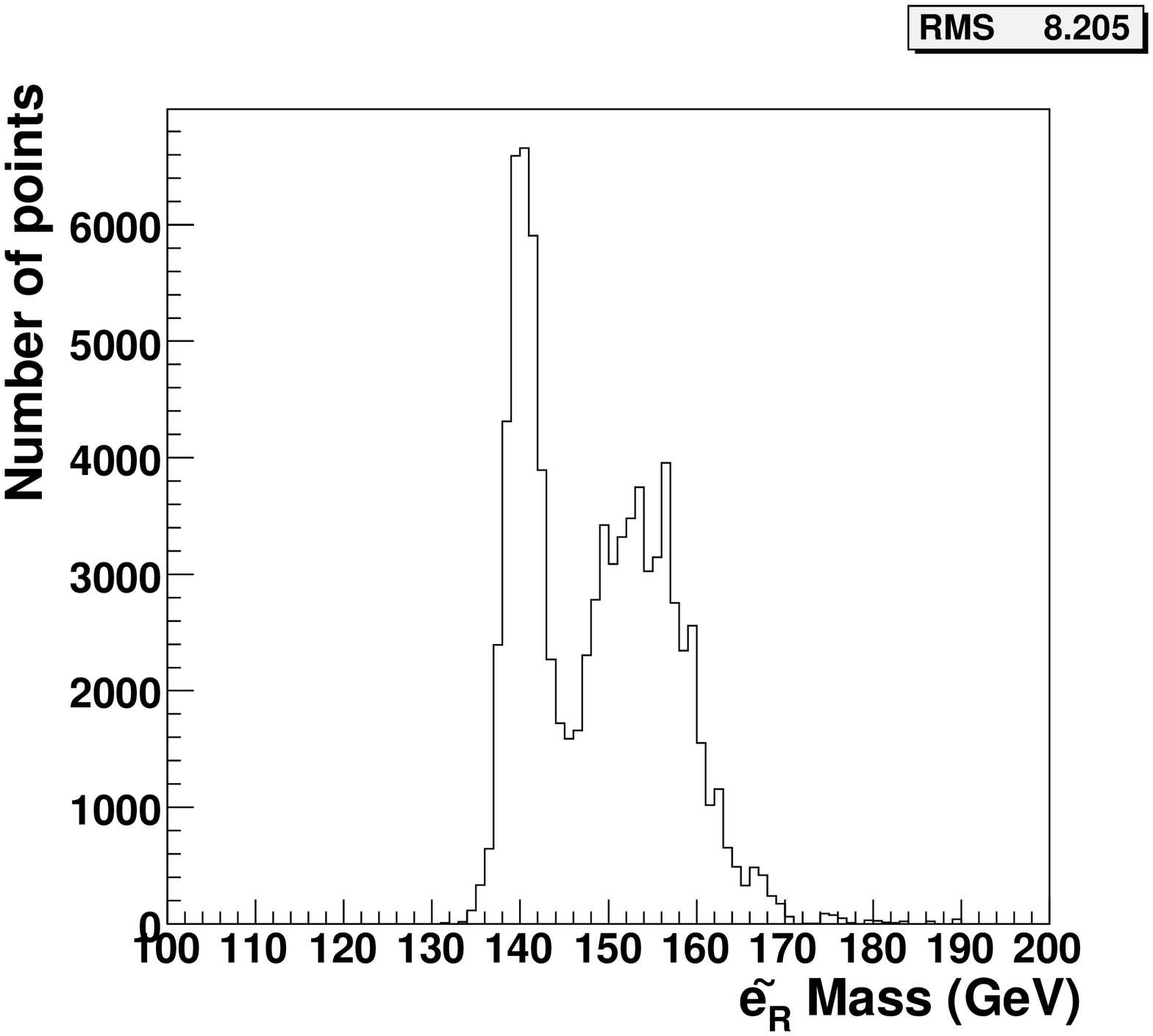}
\caption{The mass distributions obtained from the mSUGRA chain, in
which one has not assumed that the particles in the decay chain have
been identified. These are not to be confused with mass measurements!
The width of each plot (RMS about the mean) is recorded for each plot
in GeV.}
\label{mass-spread}
}

\subsection{A non-universal SUGRA model}
The mSUGRA model assumes universality of the scalar and gaugino masses
at the GUT scale, and also unifies the trilinear couplings at the GUT
scale. Although this helps in reducing the SUSY breaking parameter set
to a manageable level, reality may present a more complicated
case. Hence, there is a very strong motivation for developing
techniques that are either model independent or are at least able to
tackle some more general SUSY models.

In this subsection, we investigate the effect of relaxing the
assumption of universal GUT scale gaugino masses, whilst still
retaining the chain ambiguity and jet energy scale effects encountered
in the sections~\ref{sec:jetenscalesec} and
\ref{sec:whatasillnameforasec}. It is important to realise that this
is merely a first example of the use of the techniques developed here;
one could just as easily relax more of the mSUGRA assumptions provided
that one has made enough measurements to provide suitable constraints
on the resulting model.

\subsubsection{Kinematic edge constraints on non-universal SUGRA}
The parameter set for the SUGRA model now becomes $m_0$, tan$\beta$,
$A_0$, sgn($\mu$), $M_1$, $M_2$ and $M_3$. A Metropolis sampler was
used to sample from this parameter space (along with the jet energy
scale error $s$), with the mass spectrum of each point found using
{\tt ISAJET 7.69}. Chain ambiguity was incorporated in the same way as
described in section~\ref{sec:whatasillnameforasec}.  The results are
seen in figures~\ref{esugra_plusmu} and \ref{esugra_minusmu}: it
should be noted that the previous $m_0$ vs $m_{1/2}$ plot has been
superseded by three plots against the various GUT scale gaugino
masses. The plots shown contain 800,000 points, after which the
sampler was still clearly exploring new areas of the parameter space.
In these plots, the Markov Chain has not yet converged, and this lack
of convergence is sufficient to show that the endpoint data {\em
alone} do not provide sufficient information to adequately constrain
the non-universal SUGRA model, and so we have indeed reached a point
where we need to consider additional measurements -- such as the
cross-section.

\subsubsection{Kinematic edge data {\em and} cross-section constraints on non-universal SUGRA}
A further Metropolis sampler was used to explore the parameter space
of our non-universal SUGRA model using both the cross-section
information {\em and} the edge data in the definition of the
probability weight for each point. The results for positive $\mu$ are
seen in figure~\ref{esugrafinal_plusmu}, whilst those for negative
$\mu$ are seen in figure~\ref{esugrafinal_negativemu}, and the
difference from the plots described above is immediately apparent. The
system is much more tightly constrained, and it has not wandered too
far from the region corresponding to an mSUGRA model in which $M_1$,
$M_2$ and $M_3$ are degenerate. One can convert this GUT scale region
to a region in mass space as before (see figure~\ref{esugran1n2}),
though with the previous disclaimer that we have not yet identified
which of the particles are involved in the decay chain but merely the
range on the various masses that might be involved. Further work in
the form of Monte Carlo studies targeted in the selected region at the
GUT scale might possibly identify which masses are involved and hence
improve the precision further, a study that is perfectly feasible
given the relatively small extent of the region allowed by our data.

The results presented here are very encouraging, however, showing that
even with only one extra observable we can afford to be more honest
about our lack of information regarding decay processes whilst still
obtaining adequate precision within the framework of mSUGRA, and
reasonable precision in a more general model.

\FIGURE{
\fourgraphs{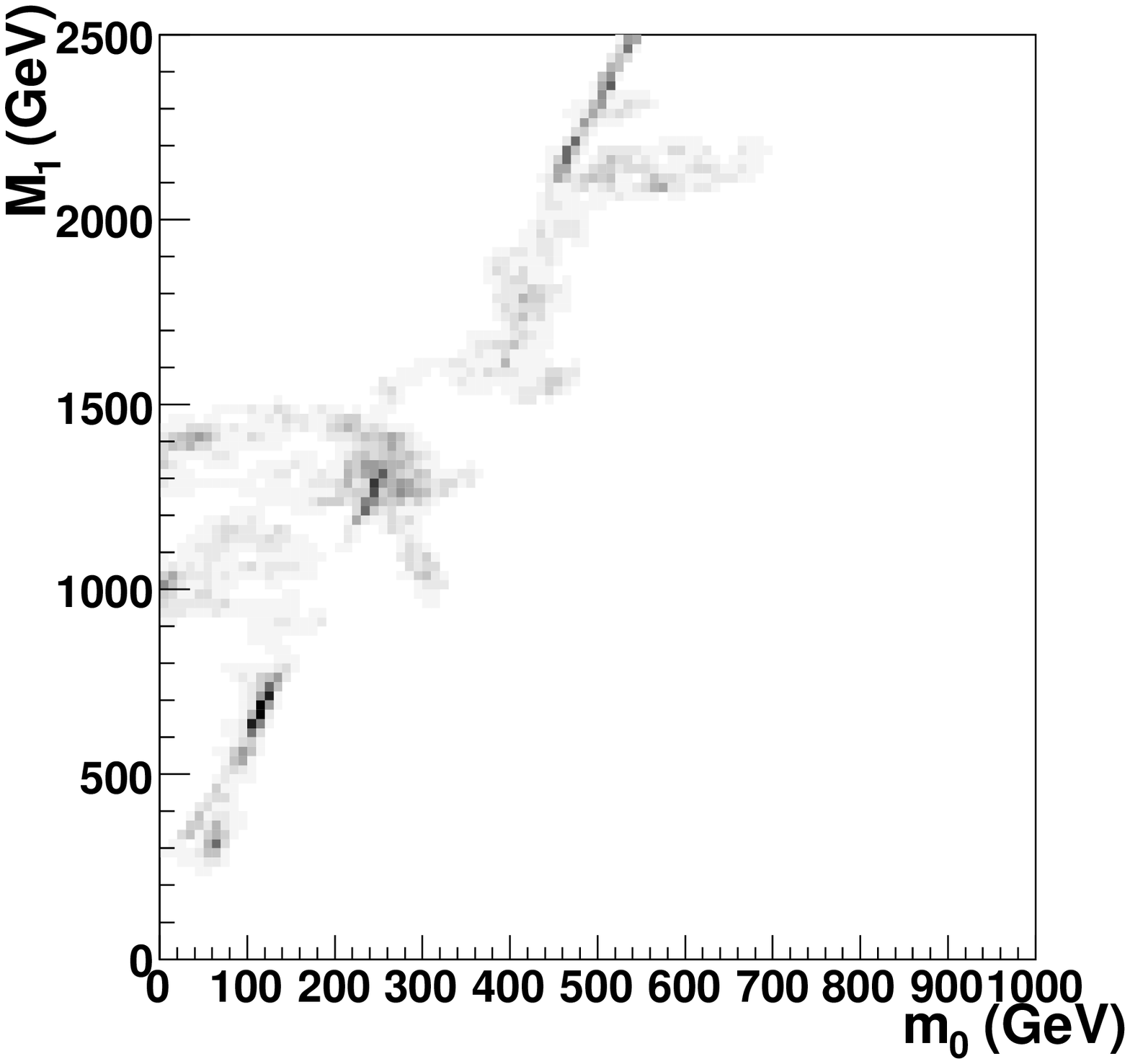}{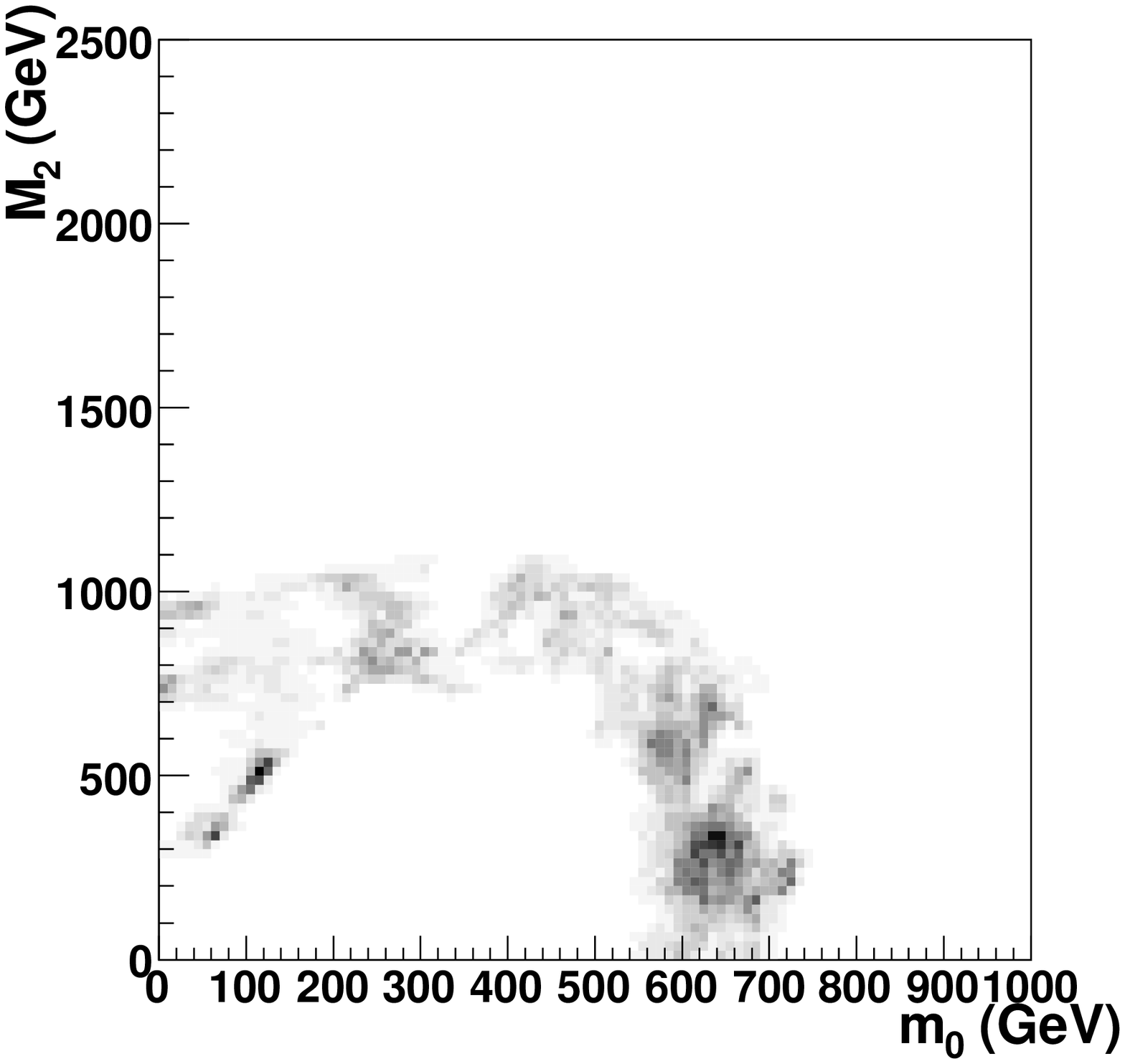}{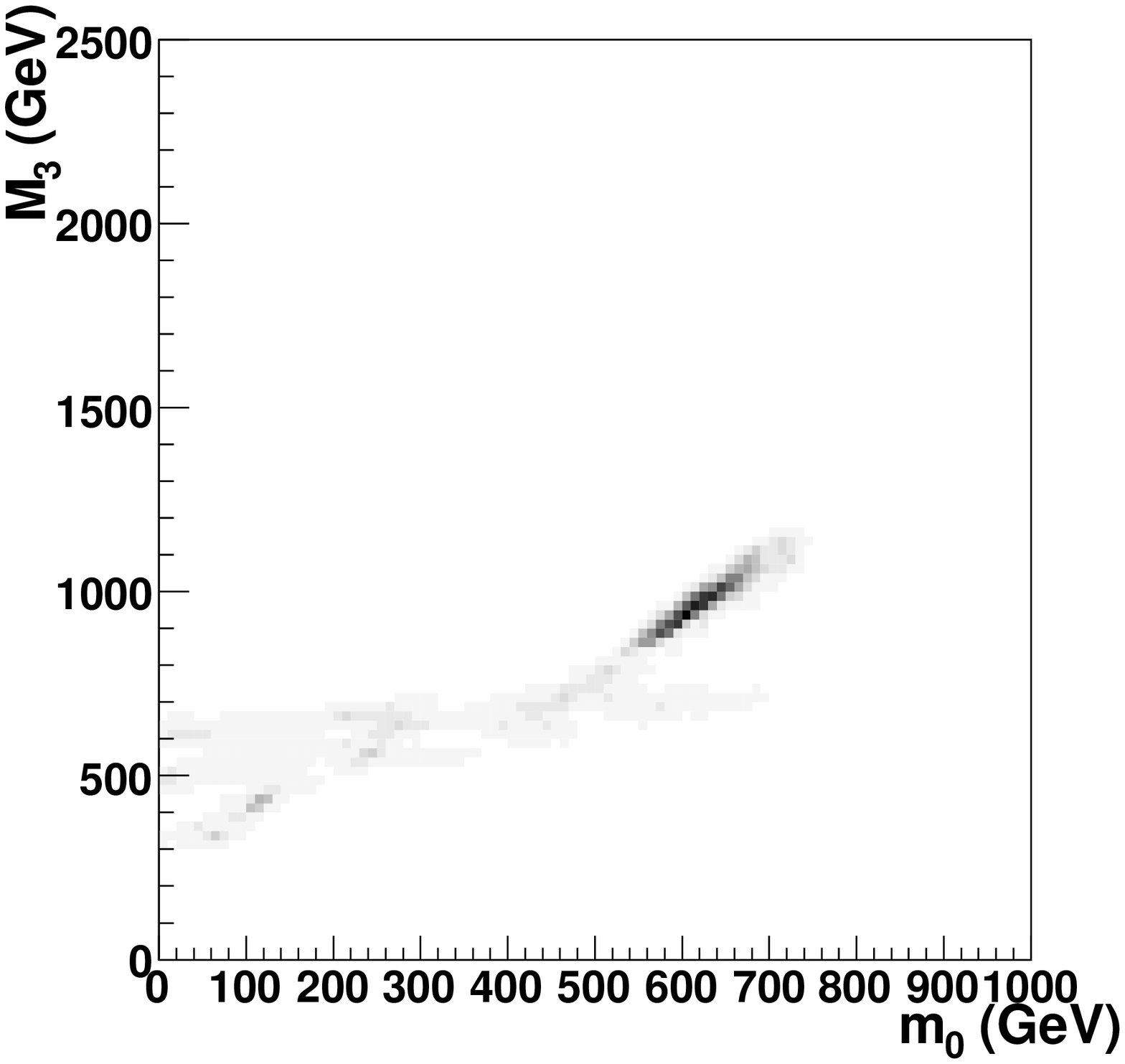}{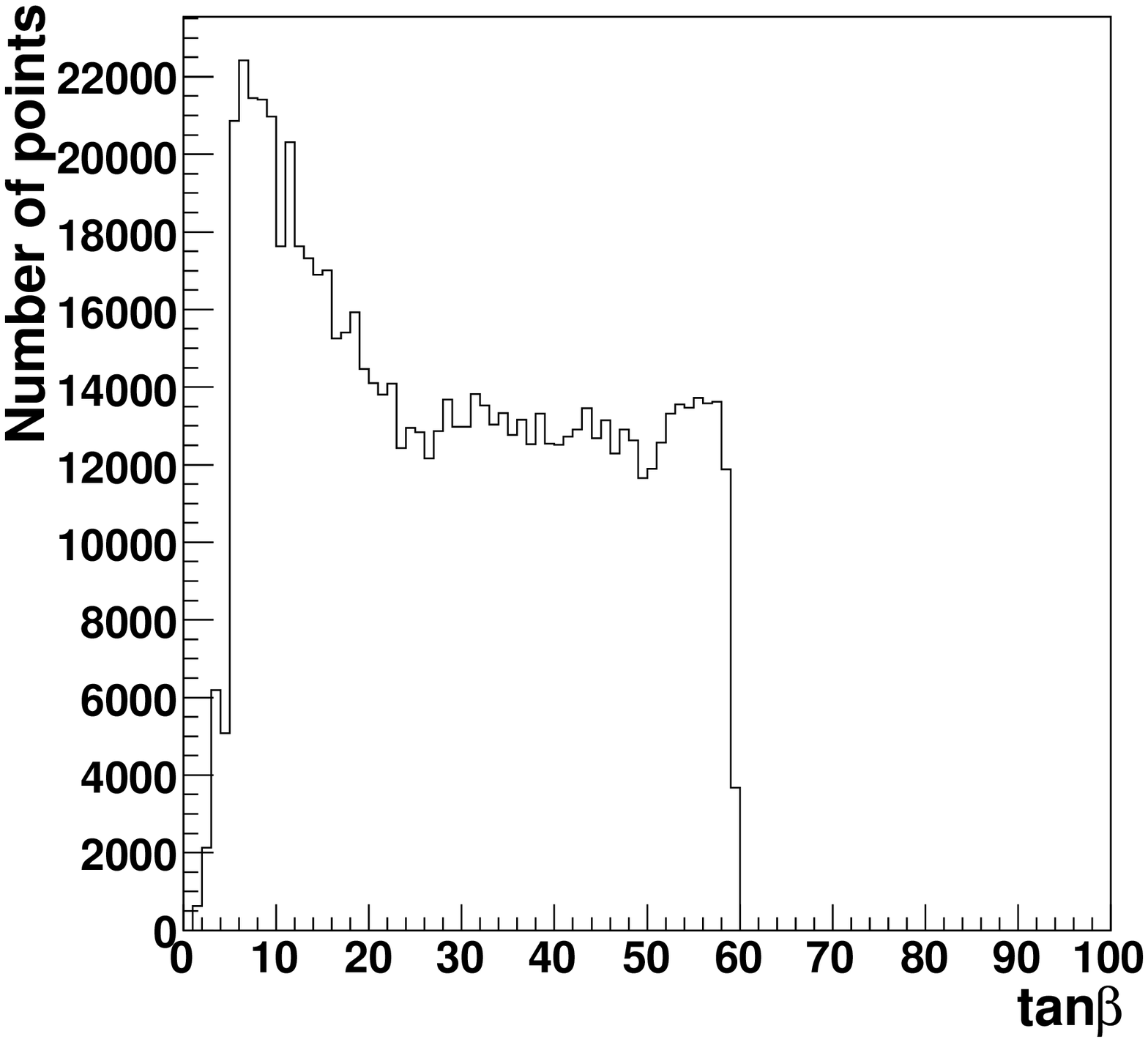}
\caption{The region of our non-universal SUGRA parameter space
consistent with the endpoint measurements of section~\ref{sec:lala2},
with chain ambiguity included. Results are shown for positive $\mu$.}
\label{esugra_plusmu}
}

\FIGURE{
\fourgraphs{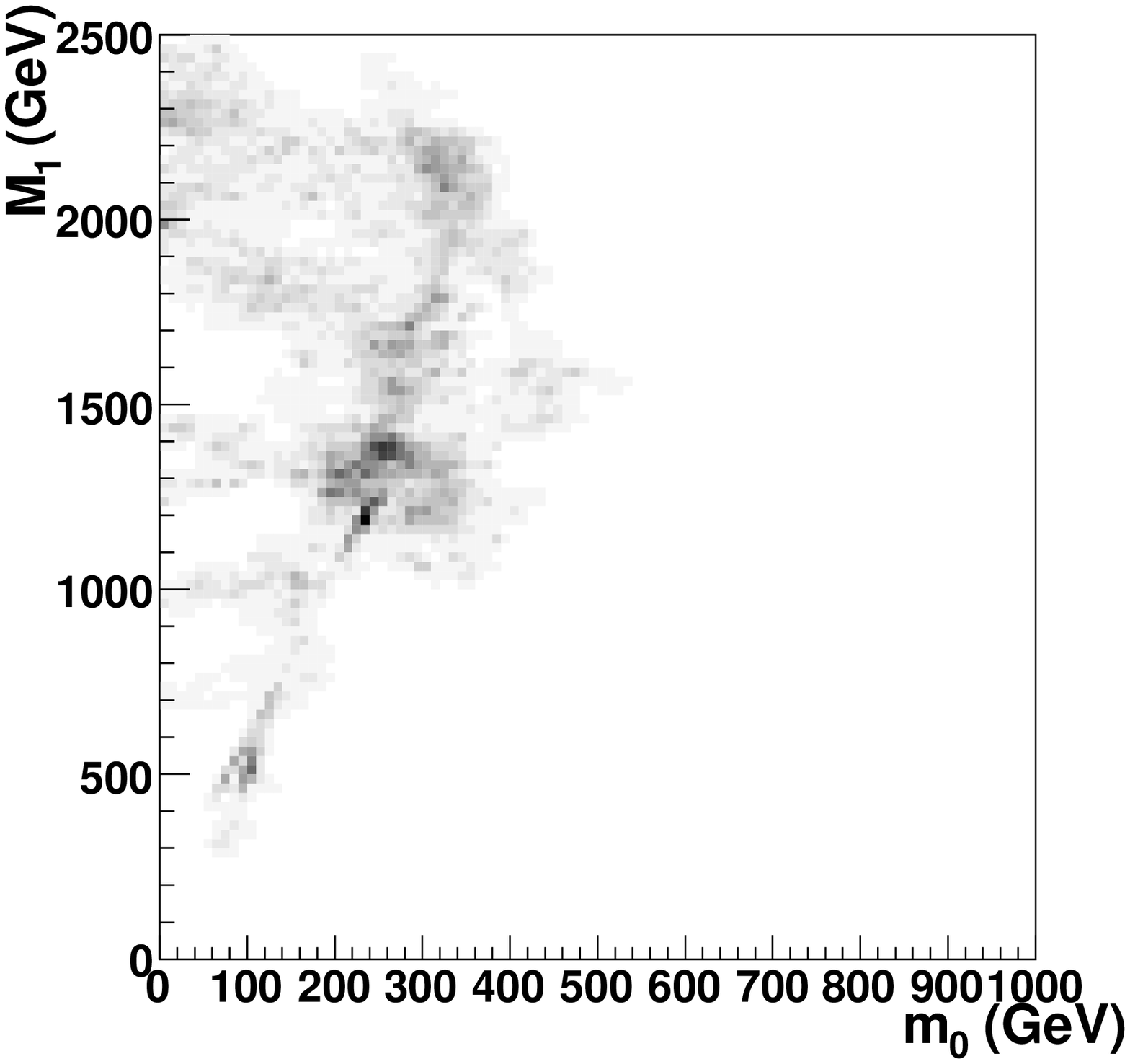}{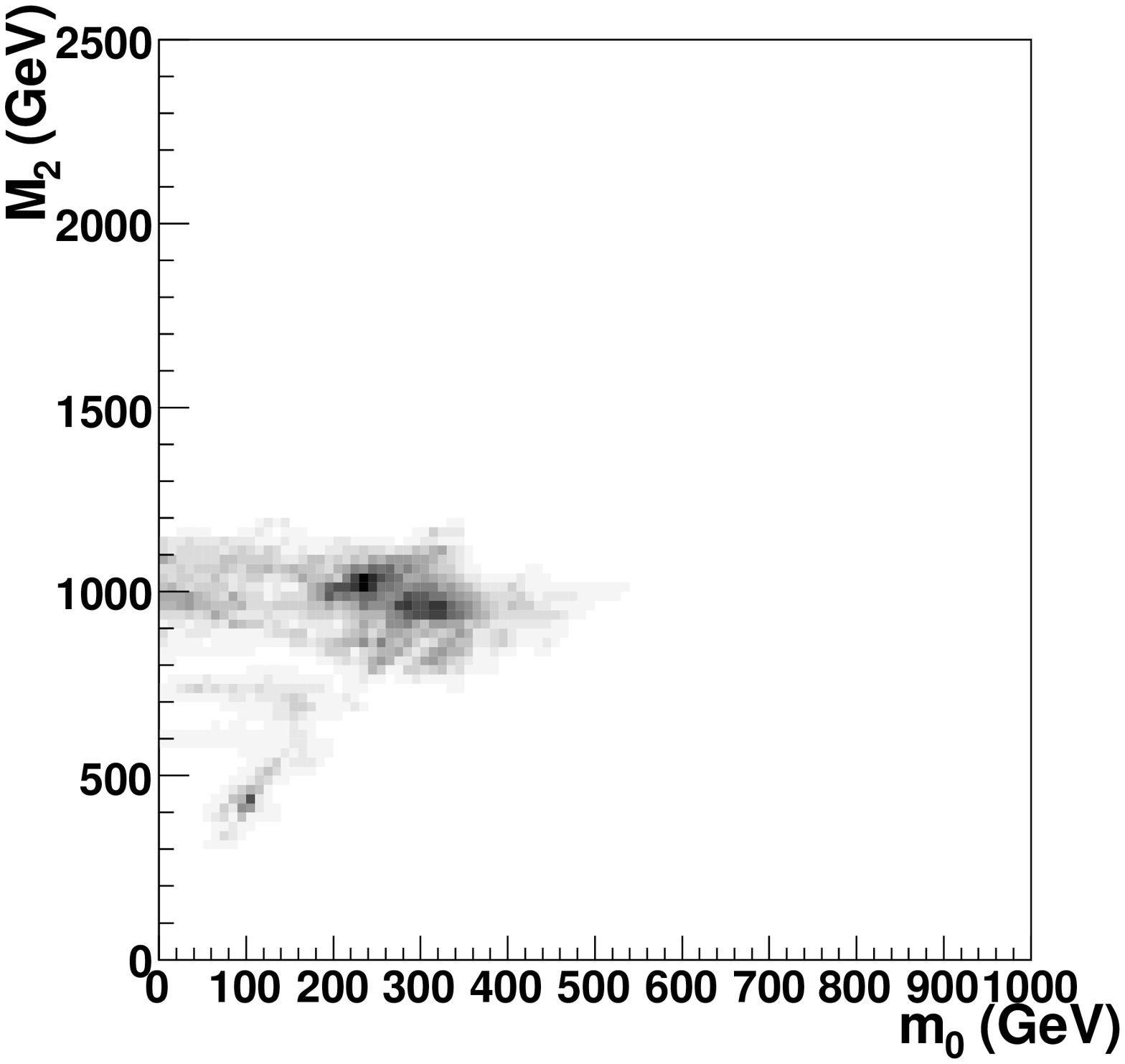}{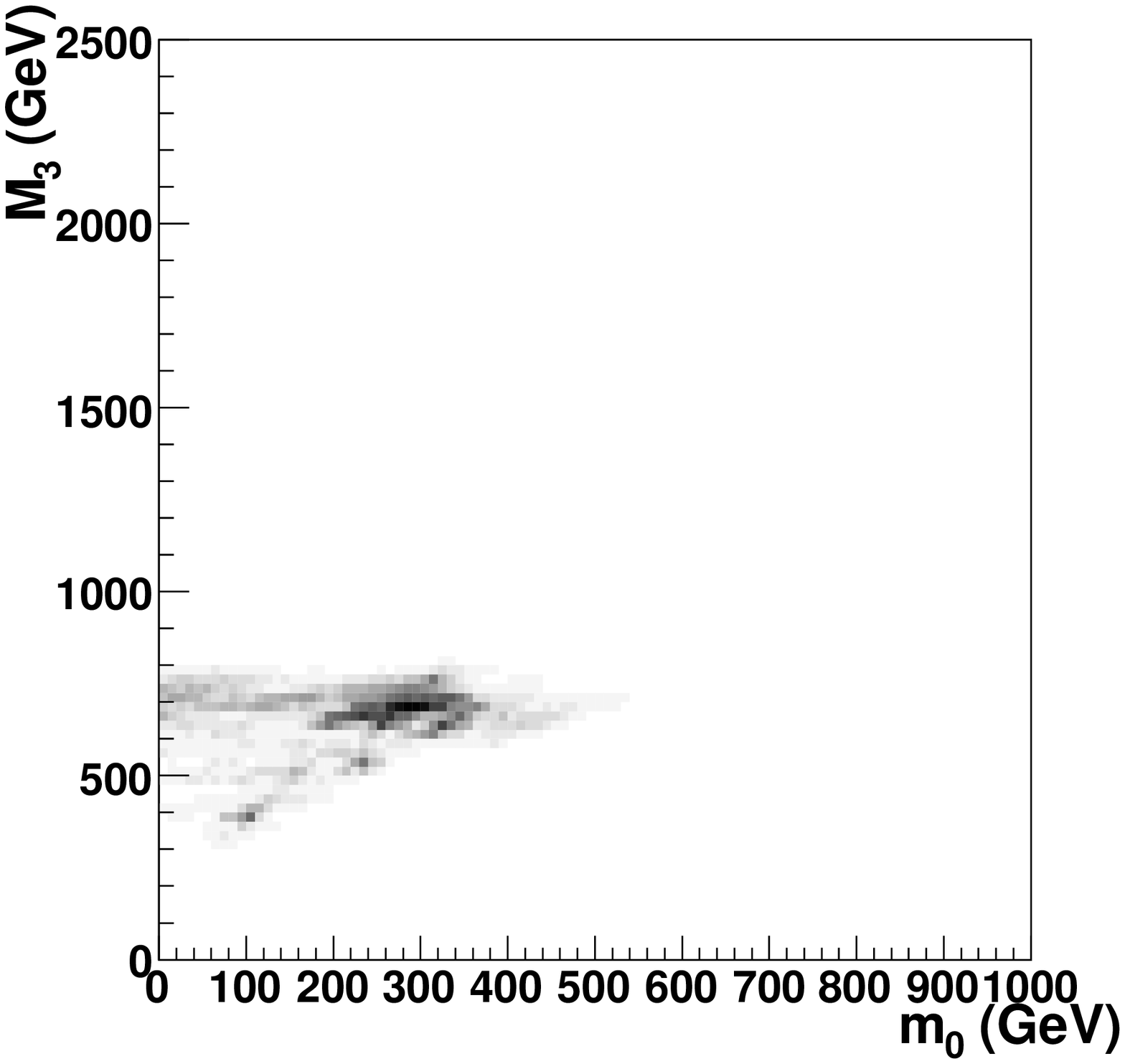}{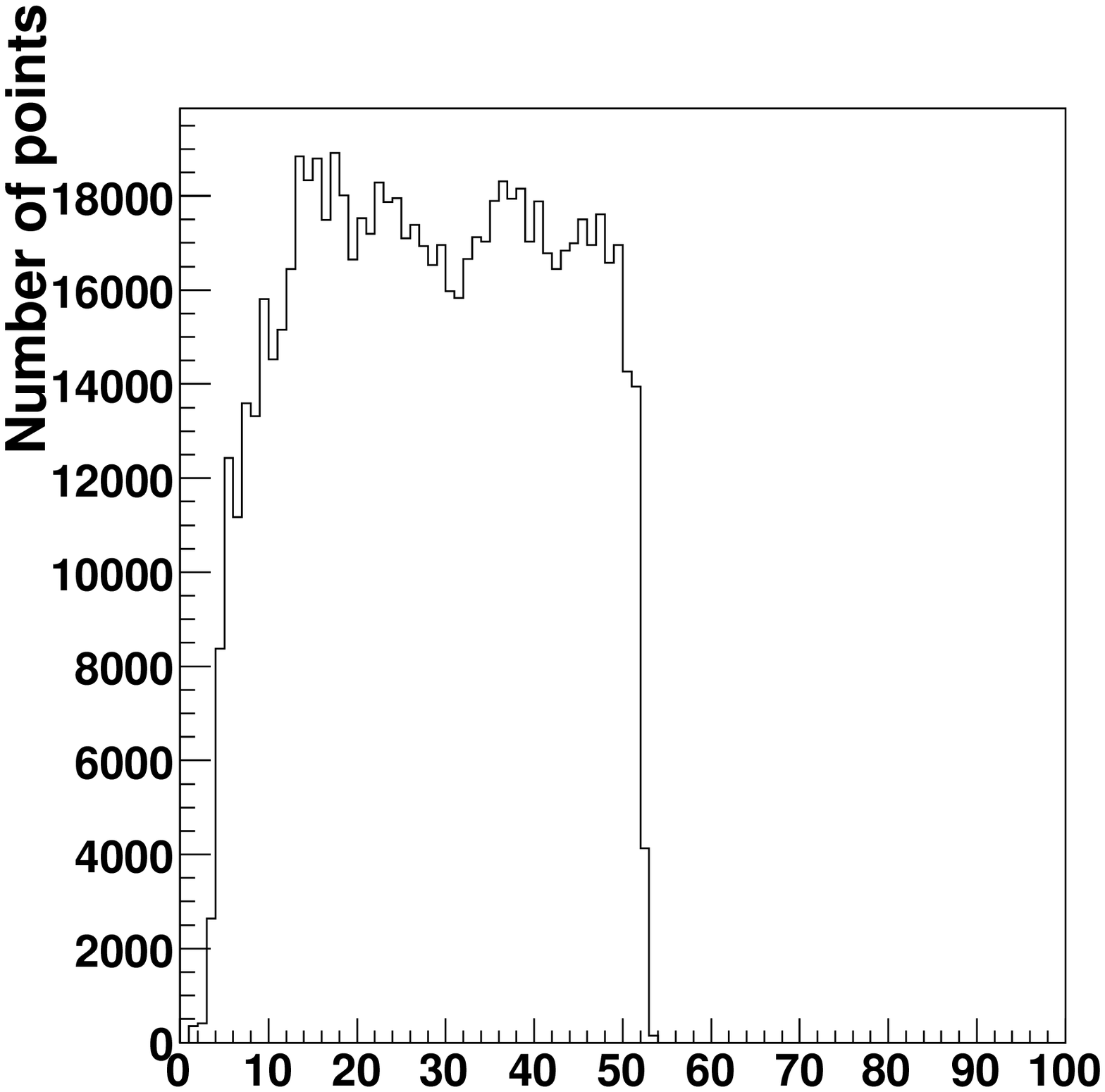}
\caption{The region of our non-universal SUGRA parameter space
consistent with the endpoint measurements of section~\ref{sec:lala2},
with chain ambiguity included. Results are shown for negative $\mu$.}
\label{esugra_minusmu}
}

\FIGURE{
\fourgraphs{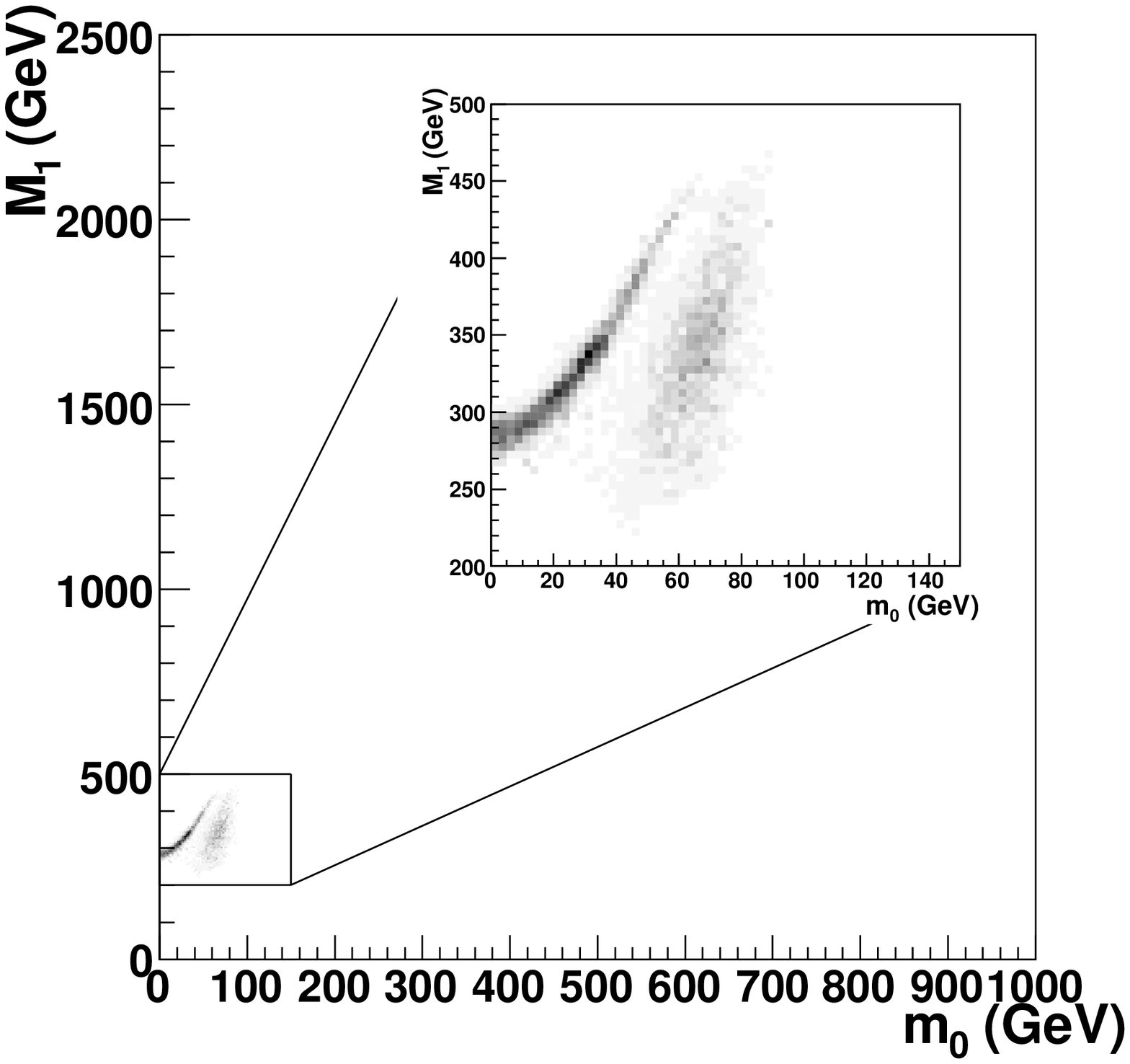}{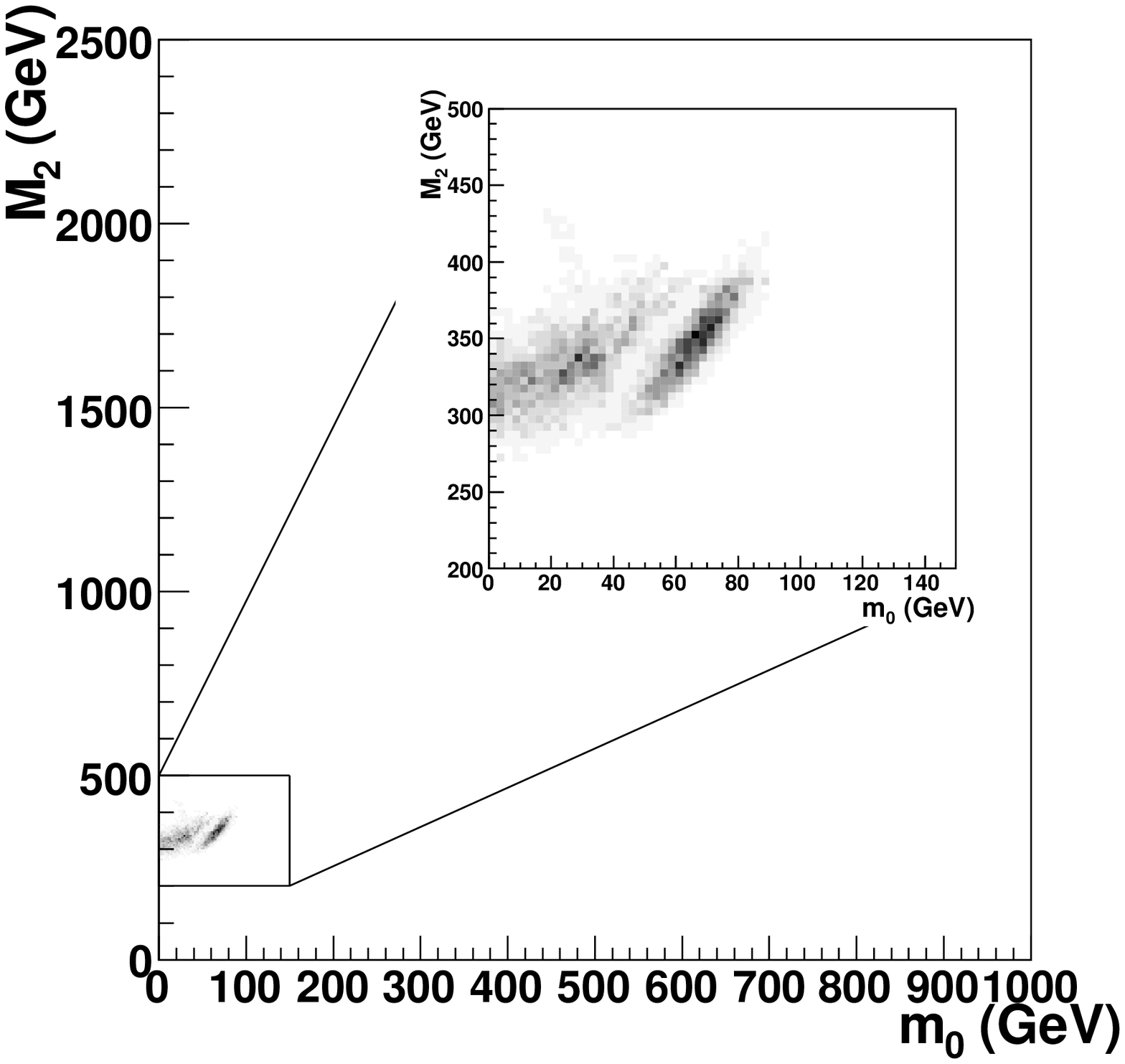}{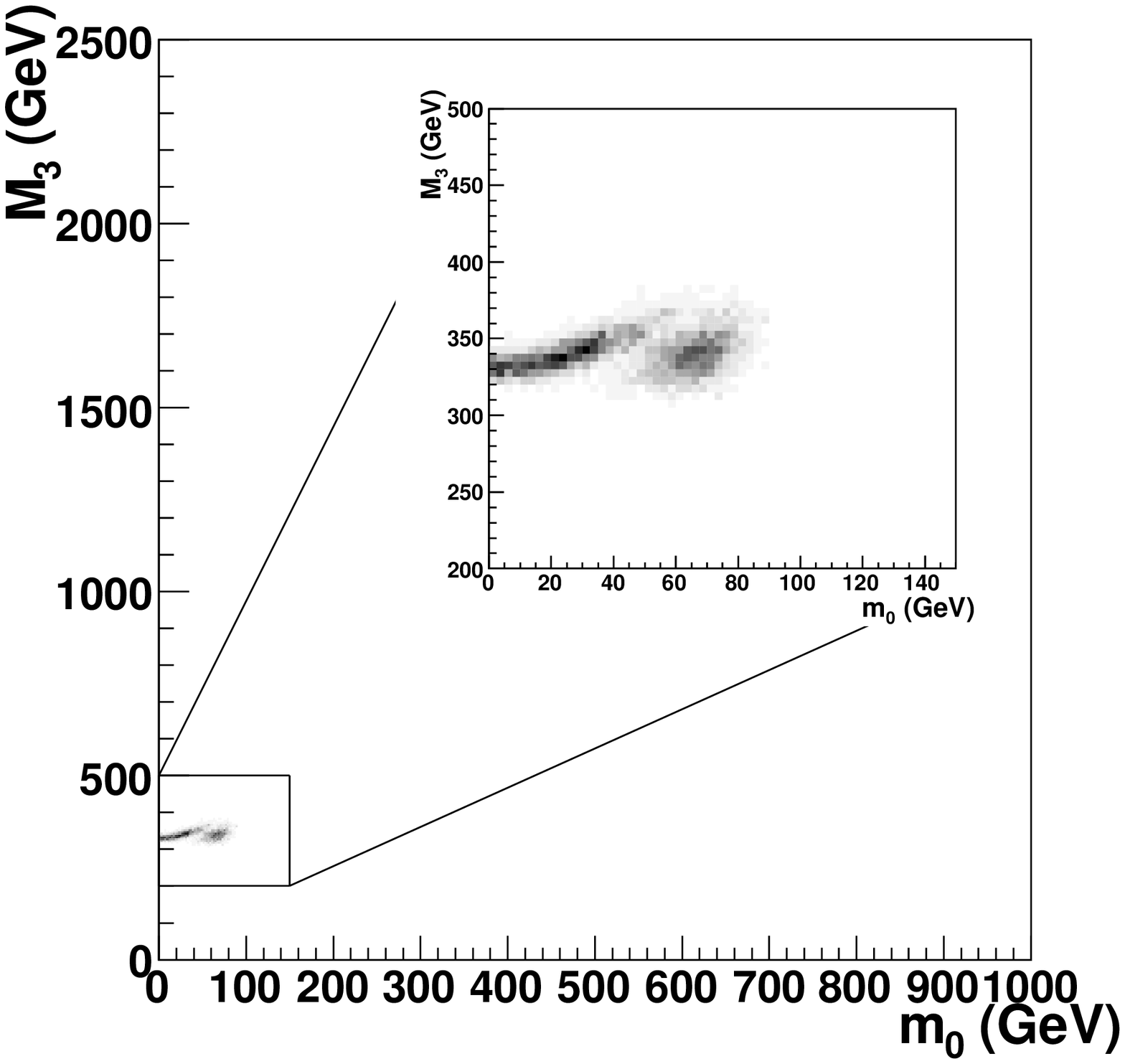}{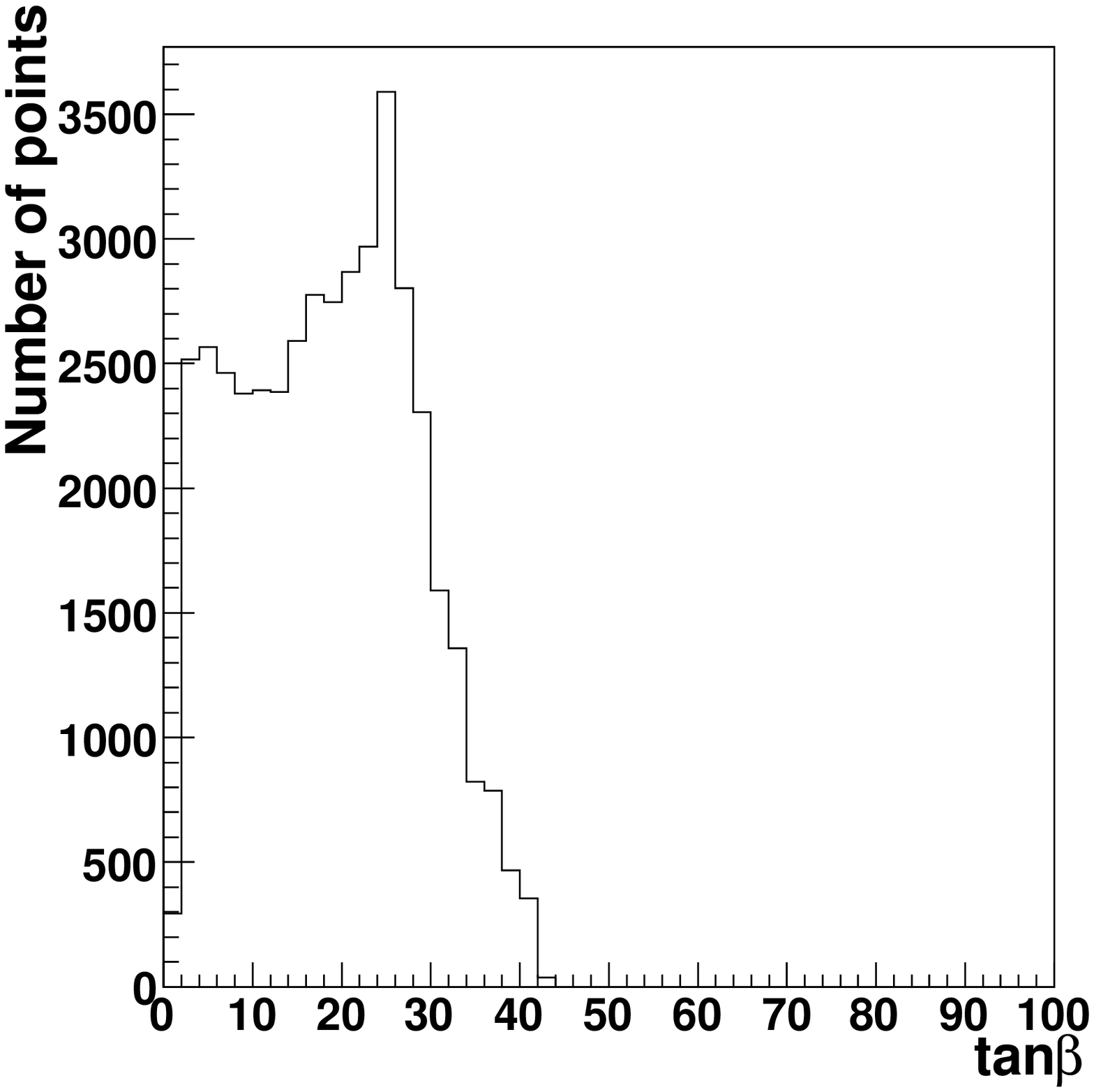}
\caption{The region of our non-universal SUGRA parameter space
consistent with the endpoint measurements of section~\ref{sec:lala2}
and the cross-section measurement, with chain ambiguity
included. Results are shown for positive $\mu$.}
\label{esugrafinal_plusmu}
}

\FIGURE{
\fourgraphs{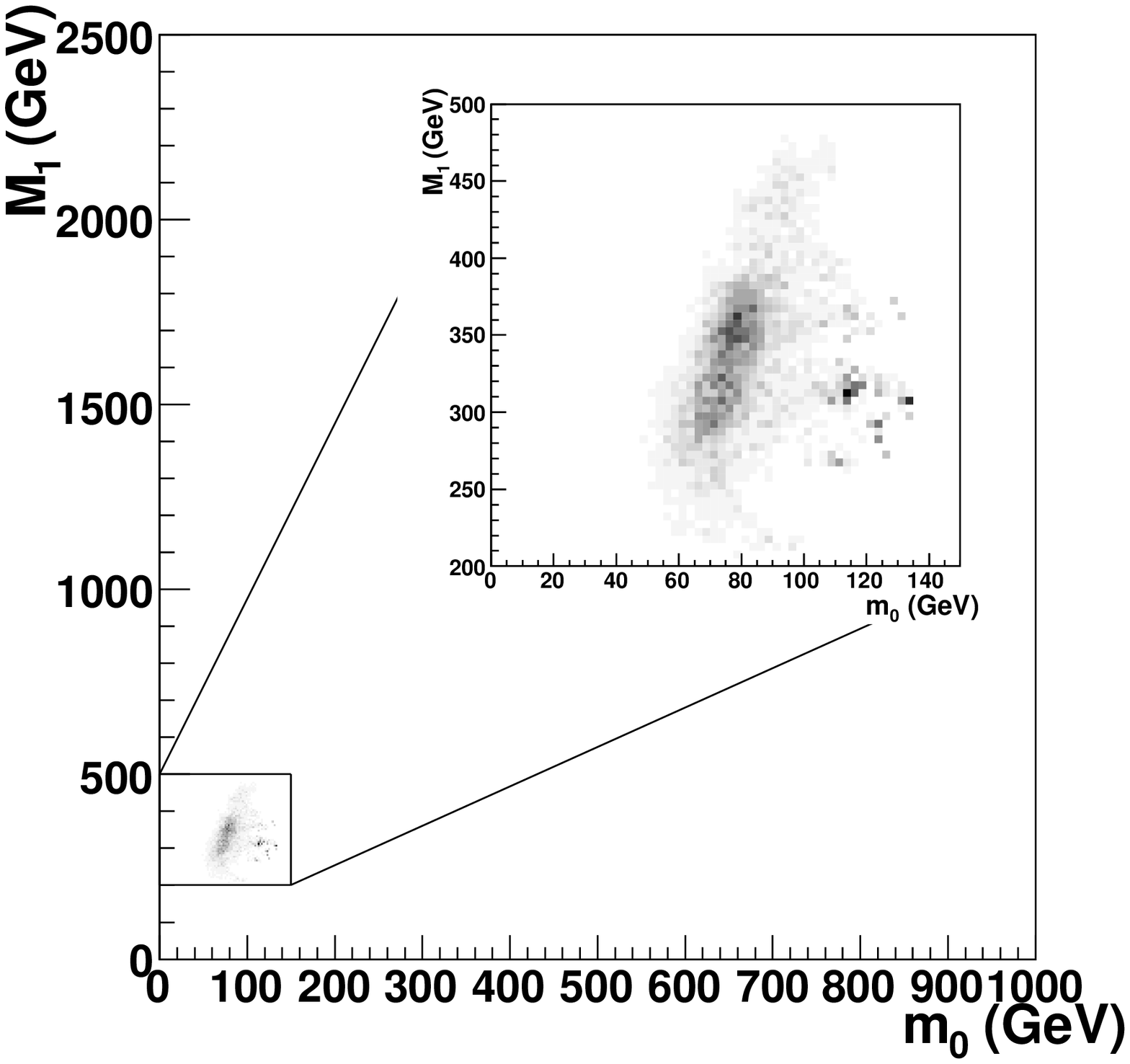}{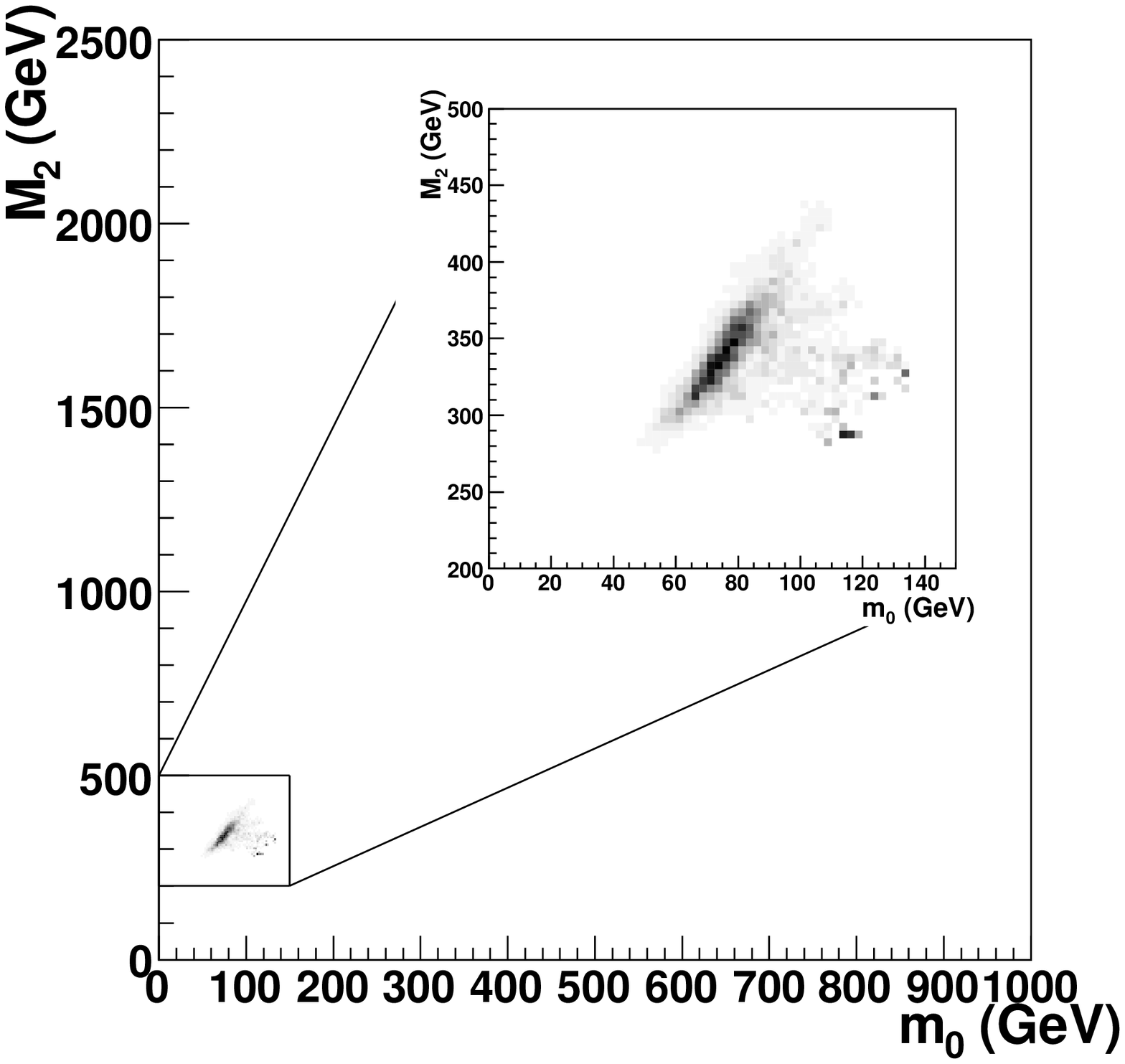}{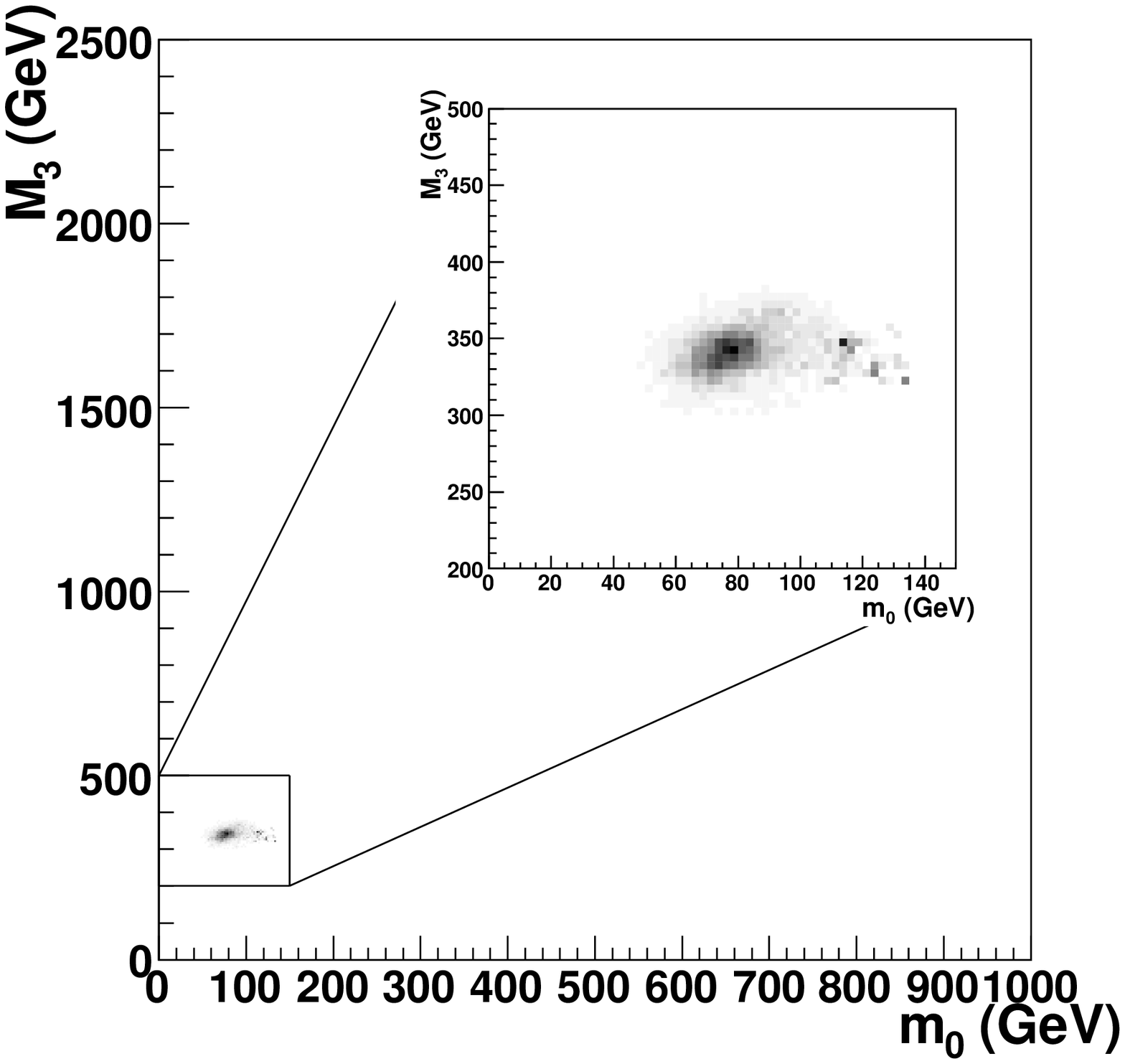}{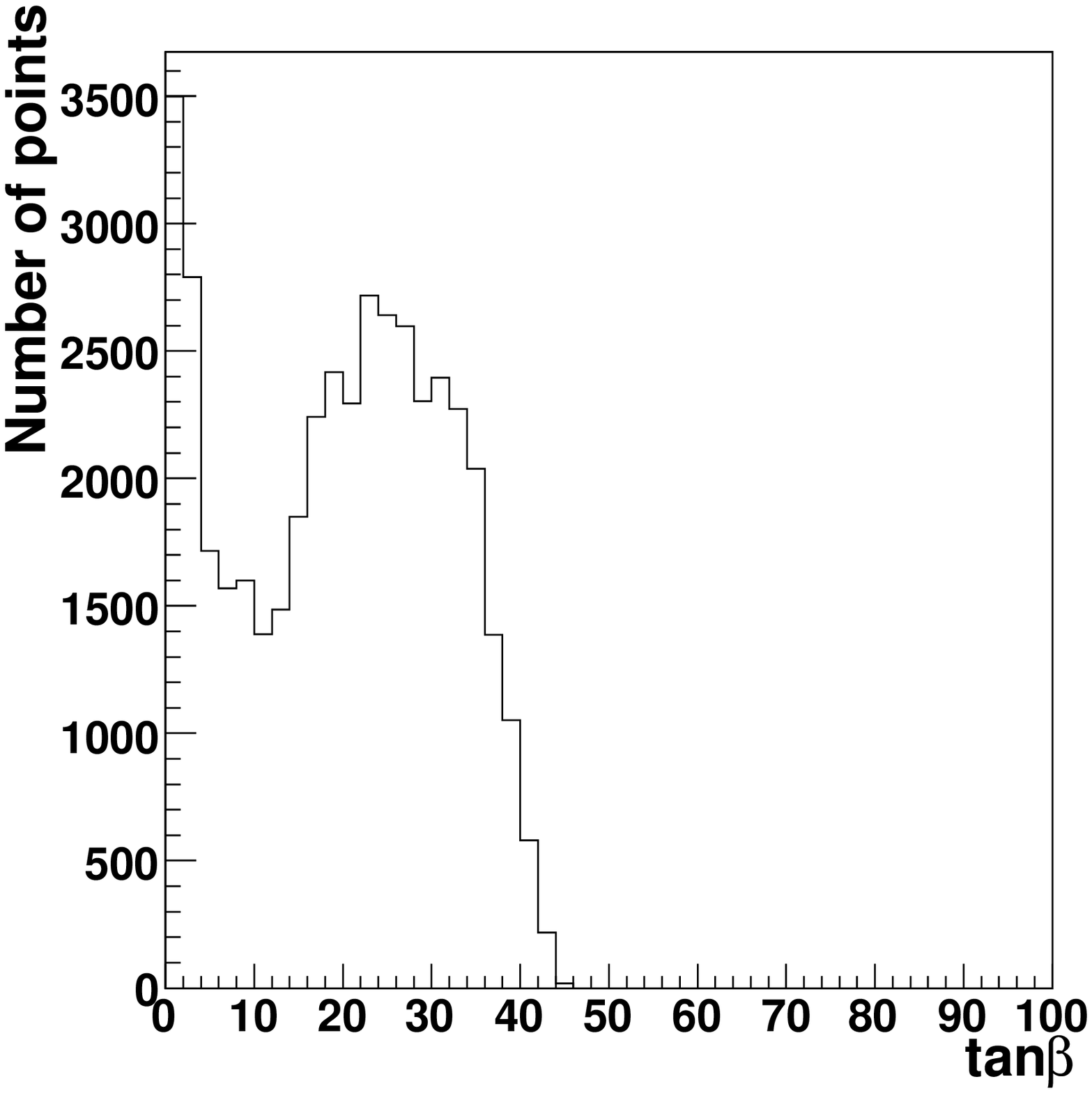}
\caption{The region of our non-universal SUGRA parameter space
consistent with the endpoint measurements of section~\ref{sec:lala2}
and the cross-section measurement, with chain ambiguity
included. Results are shown for negative $\mu$.}
\label{esugrafinal_negativemu}
}

\FIGURE{
\sixgraphs
{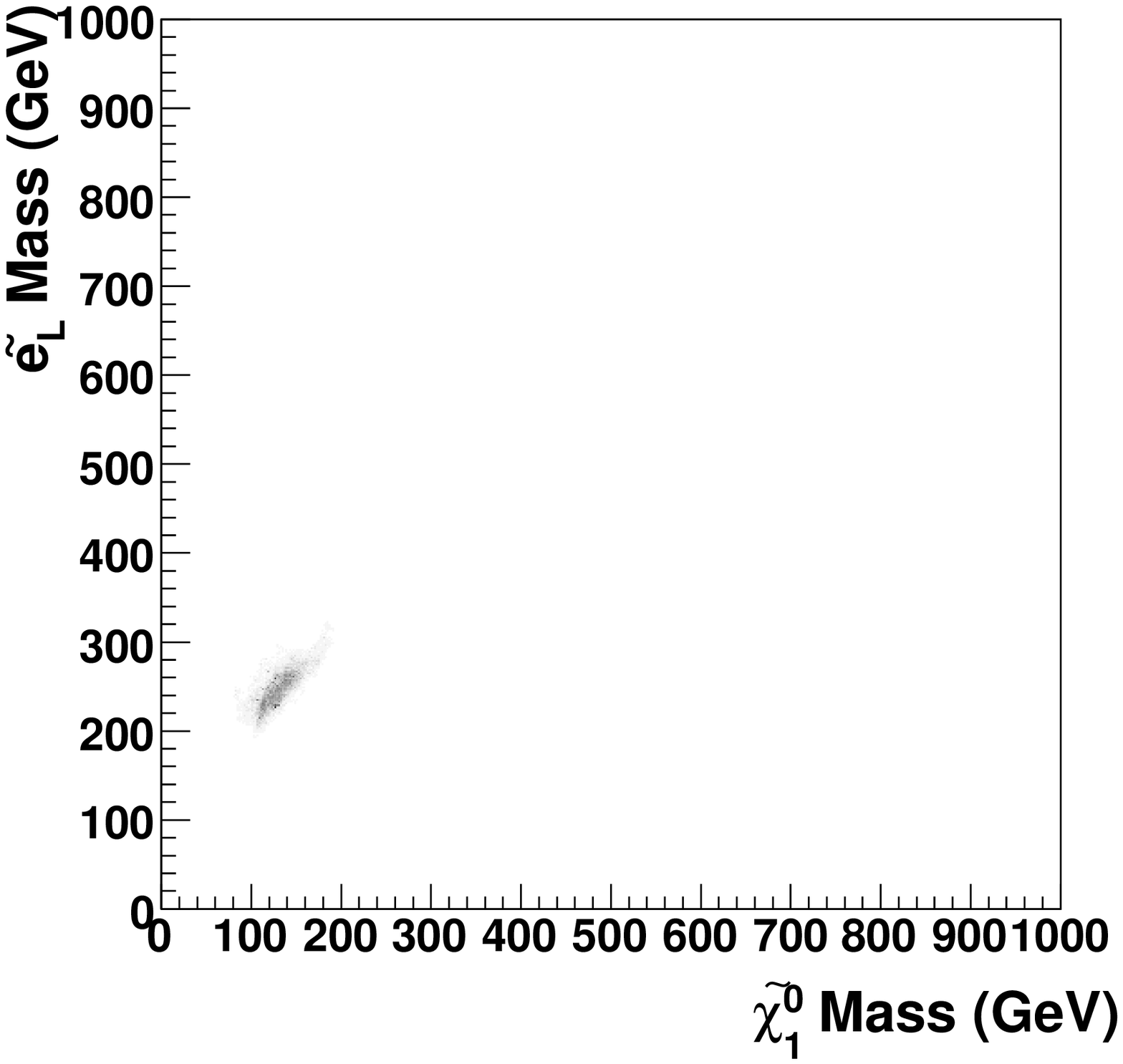}{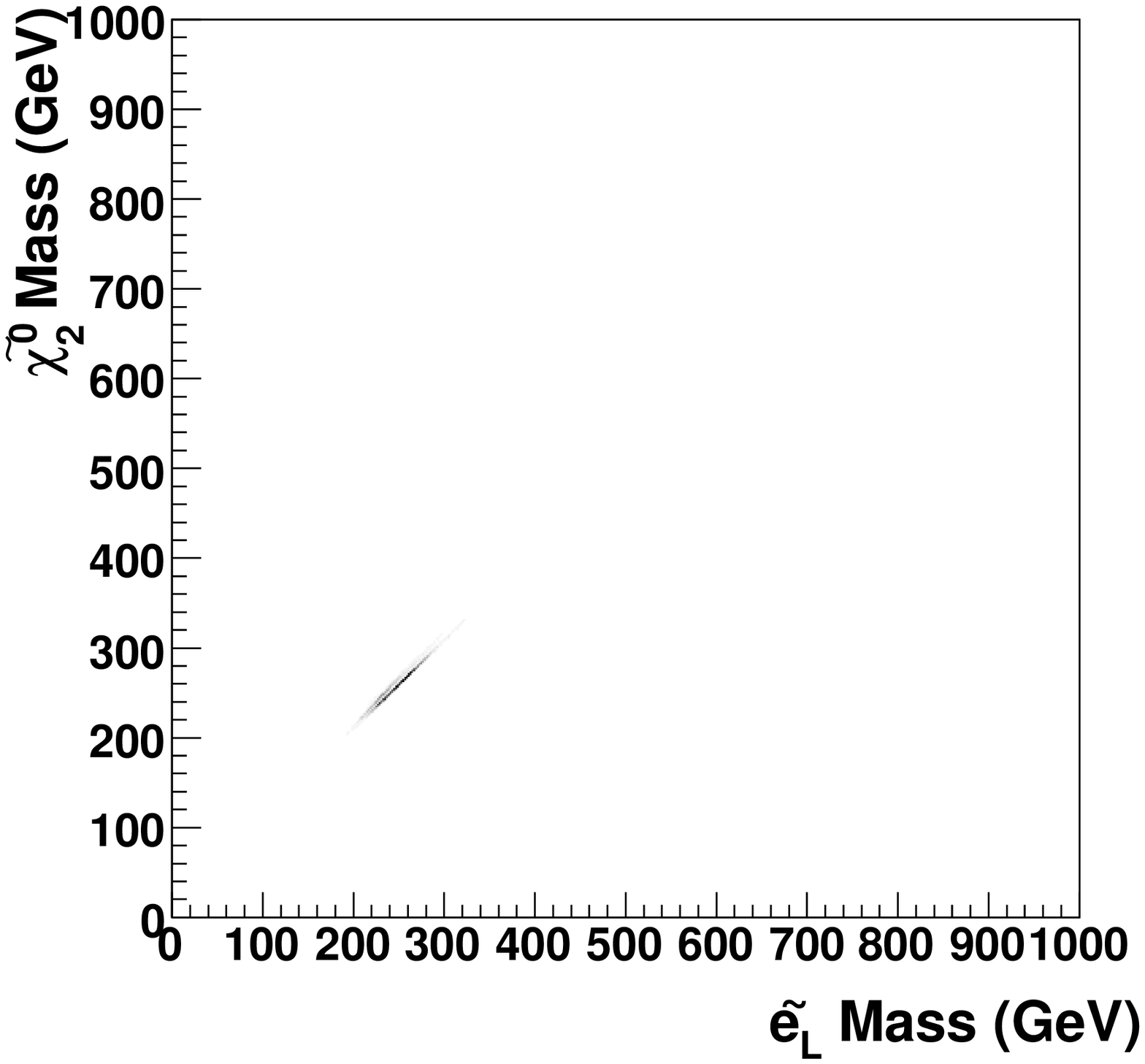}
{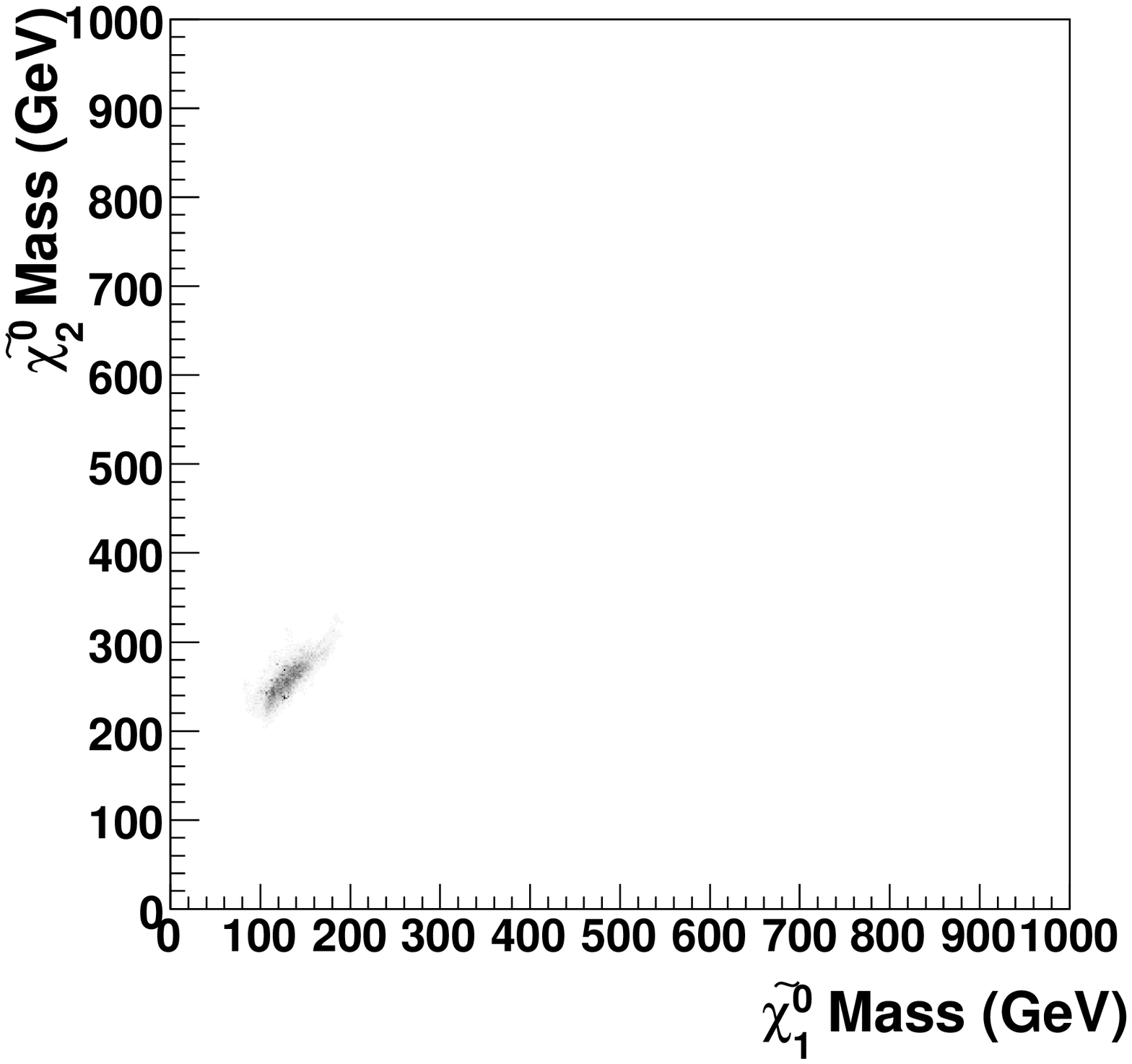}{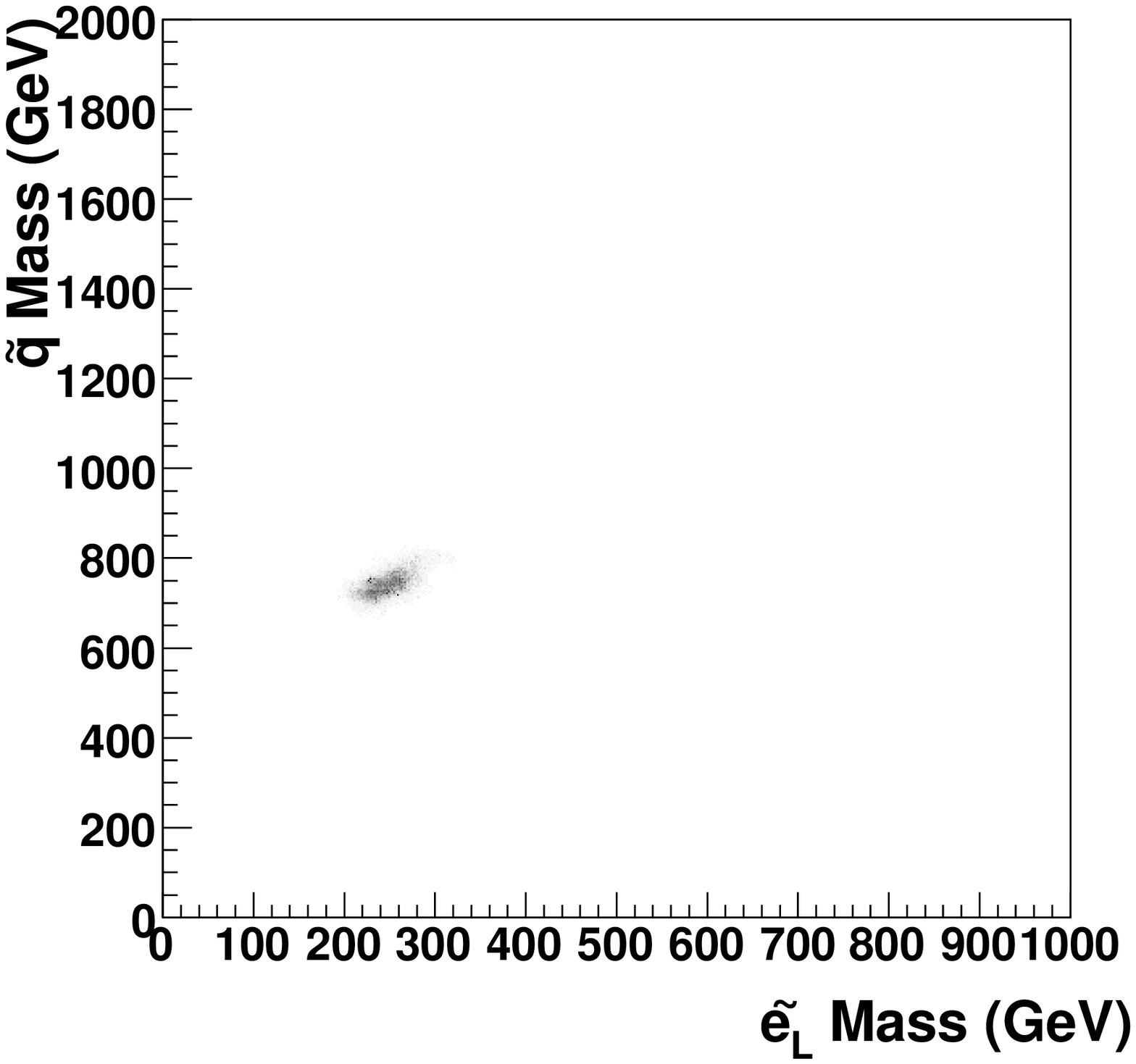}
{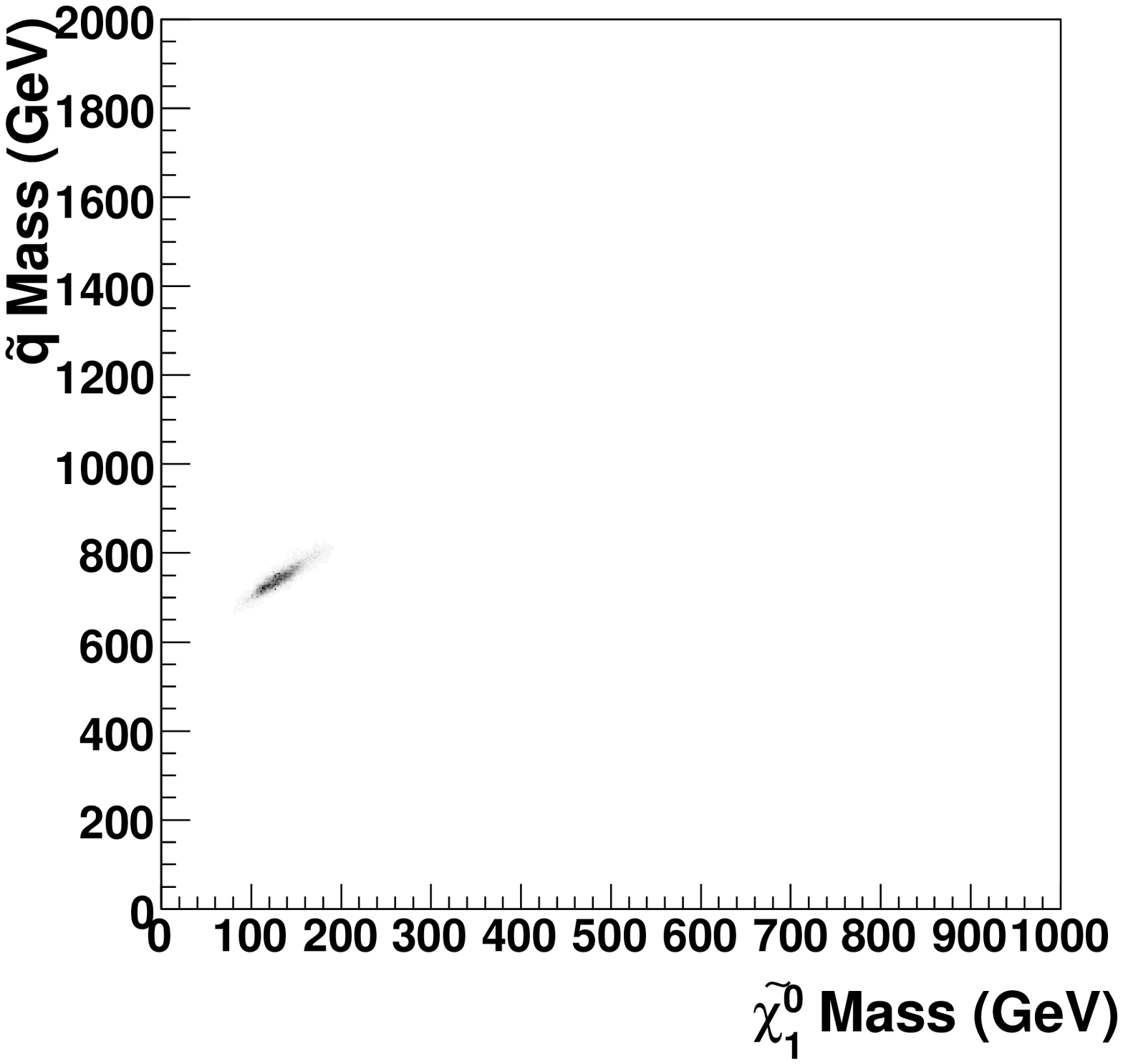}{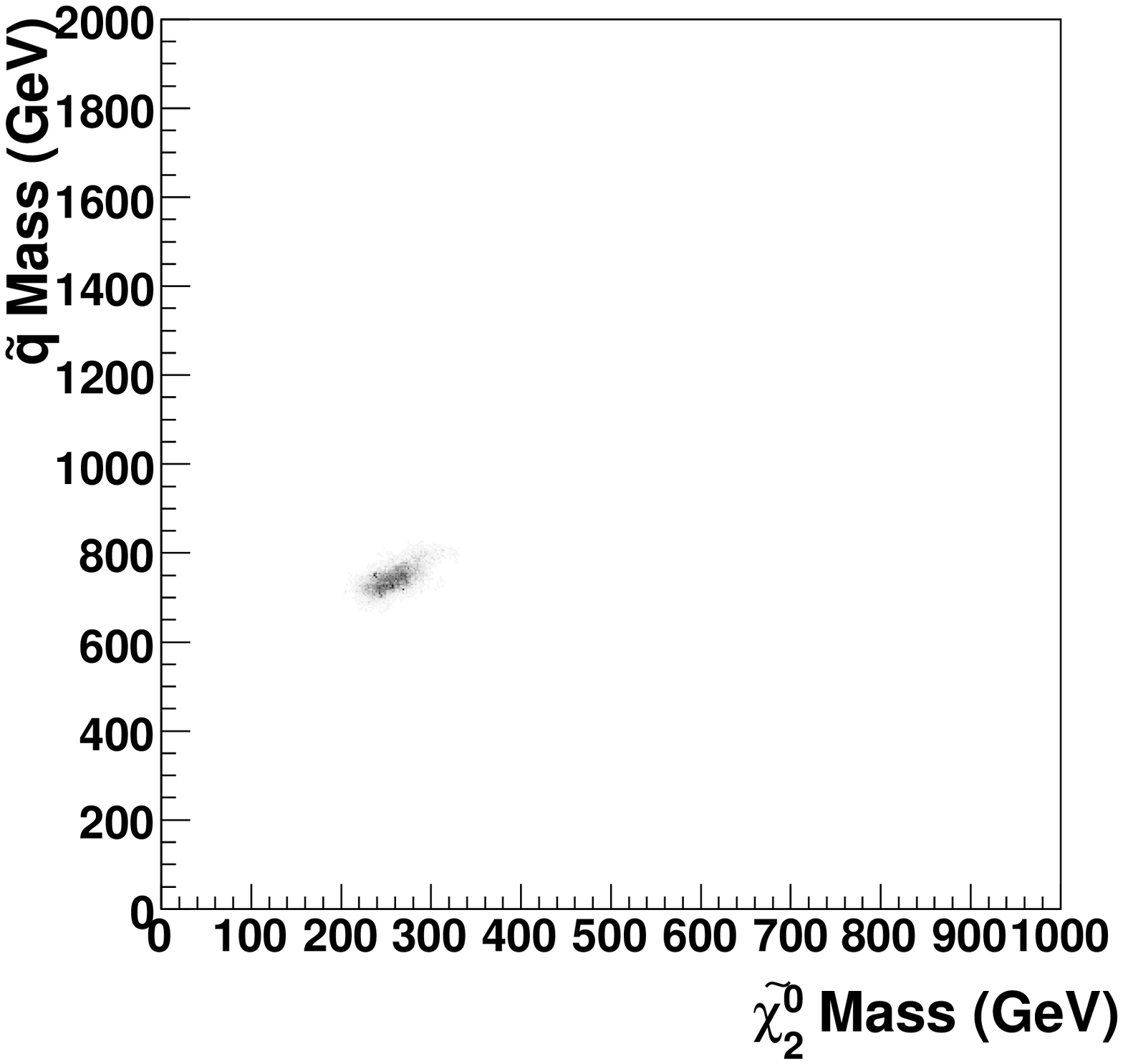}
\caption{The region of mass space corresponding to the non-universal
SUGRA parameter space region obtained in the text.}
\label{esugran1n2}
}

\section{Conclusions}
We have used Markov Chain sampling techniques to combine kinematic
endpoint measurements with a cross-section measurement in order to
obtain precision SUSY mass measurements in simulated ATLAS
data. Previous analyses have been extended to include ambiguity in the
decay chain responsible for the endpoints, and a preliminary study has
been made of a non-universal SUGRA model. Throughout it has been shown
that the precision of mass measurements is greatly improved through
the use of inclusive data, and the technique described offers a
rigorous and general approach to the problem of constraining SUSY at
the LHC. Reasonable precision has been obtained even with a fairly
conservative estimate of the errors on the endpoints themselves.

The work described here is the first step toward what is hoped will
be a powerful technique for future analysis. By collecting inclusive
observables, one can start to look at more and more general models,
with the final result limited only by the ability of physicists to
come up with new pieces of information. At the very least, the Markov
Chain approach is a powerful framework for combining information and
exploring multi-dimensional parameter spaces in an efficient manner.

As a final note, it is worth remarking that the technique is not
limited solely to data obtained at the LHC. Any piece of relevant data
is potentially useful, with obvious examples being cross-section
limits for rare decay processes, and dark matter measurements that are
currently already being used to set limits on theories. As we start to
explore models with greater numbers of parameters, this extra
knowledge could prove invaluable in providing a sufficient number of
constraints, and this will be the subject of future papers.

\section* {Acknowledgements}
We would like to thank members of the Cambridge SUSY working group for
helpful discussion and input, particularly Alan Barr and Bryan
Webber. Peter Richardson made invaluable contributions to the
parallelisation of {\tt HERWIG}. Computing time has been provided by
the Cambridge-Cranfield High Performance Computing Facility.  This
work has been performed within the ATLAS Collaboration, and we thank
collaboration members (in particular Dan Tovey) for helpful
discussions. We have made use of tools which are the result of
collaboration-wide efforts.

\clearpage
\appendix
\section {Markov Chain Sampling}
There follows a brief review of the relevant techniques involved in
the Markov chain methods used in our analysis. For a more
comprehensive explanation, see \cite{MacKay}).
\subsection{Markov Chains}
Let $X_i$ be a (possibly infinite) discrete sequence of random
variables. $X_1,X_2,$... is said to have the \emph{Markov property}
if:
\begin{equation}\label{markov}
P(X_{i+1}=x_{i+1}|X_i=x_i,X_{i-1}=x_{i-1},...,X_1=x_1)=P(X_{i+1}=x_{i+1}|X_i=x_i)
\end{equation}
for every sequence $x_1,...,x_i,x_{i+1}$ and for every $i\ge 1$. A
sequence of random variables with the Markov property is called a
\emph{Markov Chain}.

Suppose $i$ is a discrete step in a time variable. The Markov property
is then equivalent to stating that, given a present element of the
sequence $X_i$, the conditional probability of the next element in the
sequence is dependent only on the present. Thus, at each time $i$ the
future of the process is conditionally independent of the past given
the present.

\subsection{Sampling and probability distributions}
Suppose we wish to determine a probability distribution
$P(\mathbf{x})$; for example, the posterior probability of a model's
parameters given some data. It is assumed in general that $\mathbf{x}$
is an $N$-dimensional vector and that $P(\mathbf{x})$ can be evaluated
only to within a normalisation constant $Z$; i.e. we can evaluate the
function $P^*(\mathbf{x})$ where:

\begin{equation}\label{probdist}
P(\mathbf{x})=\frac{P^*(\mathbf{x})}{Z}
\end{equation}

Although $P(\mathbf{x})$ cannot be obtained analytically, we can in
theory solve the problem by sampling from $P(\mathbf{x})$ and plotting
the results. Two immediate problems present themselves; the first is
that $Z$ is in general unknown. The second, which holds true even if
we know $Z$, is that it is not obvious how to sample from
$P(\mathbf{x})$ efficiently without visiting every position
$\mathbf{x}$. We would like a way to visit places in
$\mathbf{x}$-space where $P(\mathbf{x})$ is large in preference to
places where $P(\mathbf{x})$ is small, thus giving a description of
the probability distribution with a minimum of computational effort.

\subsection{The Metropolis-Hastings Algorithm}
\label{app:metropmethod}
The above problem can be solved through the use of Markov Chain Monte
Carlo methods, one example of which is the Metropolis-Hastings
algorithm. This makes use of a proposal density $Q$ which depends on
the current state of a system, which we label
$\mathbf{x^{(t)}}$. (This state is really a point in a Markov Chain,
and may be, for example, a particular choice of the parameters in the
model whose probability distribution we are trying to sample). The
density $Q(\mathbf{x^{\prime}};\mathbf{x^{(t)}})$ (where
$\mathbf{x^{\prime}}$ is a tentative new state, or the next point in
the Markov chain) can be any fixed density from which it is possible
to draw samples; it is not necessary for
$Q(\mathbf{x^{\prime}};\mathbf{x^{(t)}})$ to resemble $P(\mathbf{x})$
for the algorithm to be useful, and it is common to choose a simple
distribution such as a Gaussian with a width chosen for the reasons
outlined in section~\ref{sec:whyQwaswhatitis}.

Assuming that it is possible to evaluate $P^*(\mathbf{x})$ for any
$\mathbf{x}$ as above, the first step in the Metropolis-Hastings
algorithm is to generate a new state $\mathbf{x^{\prime}}$ from the
proposal density $Q(\mathbf{x^{\prime}};\mathbf{x^{(t)}})$. The
decision on whether to accept the new state is made by computing the
quantity:

\begin{equation}\label{eq:aisdefined}
a=\frac{P^*(\mathbf{x^{\prime}})Q(\mathbf{x^{(t)}};\mathbf{x^{\prime}})}{P^*(\mathbf{x^{(t)}})Q(\mathbf{x^{\prime}};\mathbf{x^{(t)}})}
\end{equation}

Equation~(\ref{eq:aisdefined}) exists to ensure that the sampled
distribution {\em does not depend on the choice of $Q$}.

If $a \ge 1$ the new state is accepted, otherwise the new state is
accepted with probability $a$. It is noted that if $Q$ is a simple
symmetric density, the ratio of the $Q$ functions in
equation~(\ref{eq:aisdefined}) is unity, in which case the
Metropolis-Hastings algorithm reduces to the Metropolis method,
involving a simple comparison of the target density at the two points
in the Markov Chain.

If $Q$ is chosen such that $Q(\mathbf{x^{\prime}};\mathbf{x})>0$ for
all $\mathbf{x},\mathbf{x^{\prime}}$, the probability distribution of
$\mathbf{x^{(t)}}$ tends to $P(\mathbf{x})=P^*(\mathbf{x})/Z$ as $t
\to \infty$. Thus, by choosing points via the Metropolis algorithm and
then plotting them, we have achieved our goal of obtaining a
description of $P(\mathbf{x})$ in an efficient manner.

\subsection{Efficiency of the Metropolis-Hastings Algorithm}
Note that the presence of the caveat $t \to \infty$ implies that there
is an issue of convergence in the application of the
Metropolis-Hastings algorithm.  Each element in the sequence
{$\mathbf{x^{(t)}}$} has a probability distribution that is dependent
on the previous value $\mathbf{x^{(t-1)}}$ and hence, since successive
samples are correlated with each other, the Markov Chain must be run
for a certain length of time in order to generate samples that are
effectively independent -- at which point we say the chain has
``converged''.  The time it takes for the chain to converge depends on
the particular $P(\mathbf{x})$ being sampled, and on the details of
$Q$.  You cannot modify $P(\mathbf{x})$, but you are free to choose
the form of $Q$ so as to reduce the number of points which must be
sampled before convergence is reached.  Remember that
equation~(\ref{eq:aisdefined}) exists to ensure that the sampled
distribution {\em does not depend on your choice of $Q$}.

\label{sec:whyQwaswhatitis}

Finding a sensible $Q$ is a balance between choosing distributions
that are wide (and thus lead to successive samples being relatively
{\bf un}-correlated) and choosing distributions which are too wide
(and which then take a long time to random walk from one end of the
sample space to the other).  The widths of the proposal functions $Q$
used in this paper were chosen to be as large as possible, subject to
the Markov Chain's efficiency (the fraction of proposal points being
accepted) not falling much below one in twenty.  This choice only
affects the sampler's time to convergence and not the shape of the
resultant sampled distributions once convergence has been reached.


\bibliography{testbib}

\end{document}